\documentclass[useAMS,usenatbib]{mn2e}
\usepackage[totalwidth=480pt, totalheight=680pt]{geometry}
\input psfig.tex
\usepackage{latexsym}
\usepackage{natbib}
\usepackage{amssymb}
\usepackage{amsmath}

\title{Modelling galaxy spectra in presence of interstellar dust.\\
II. From the UV to the far infrared}

\author[L. Piovan, R.Tantalo \& C. Chiosi]{Lorenzo Piovan$^{1,2}$, Rosaria Tantalo$^1$ \& Cesare Chiosi$^1$\\
 $^1$Department of Astronomy, University of Padova,
       Vicolo dell'Osservatorio 2, 35122 Padova, Italy\\
 $^2$Max-Planck-Institut f\"ur Astrophysik, Karl-Schwarzschild-Str. 1, Garching bei M\"unchen, Germany\\
E-mail: {\tt piovan@pd.astro.it; tantalo@pd.astro.it;
chiosi@pd.astro.it}}

\date{\tt Submitted to MNRAS: September 2005;  Revised: March 2006}

\pubyear{2005}
\begin{document}
\maketitle
\title{Galaxies spectra with interstellar dust}

\begin{abstract}
In this paper, we present spectrophotometric models for galaxies of
different morphological type whose spectral energy distributions
(SEDs) take into account the effect of dust in absorbing UV-optical
light and re-emitting it in the infrared (IR). The models contain
three main components: (i) the diffuse interstellar medium (ISM)
composed of gas and dust whose emission and extinction properties
have already been studied in detail by \citet{Piovan06a}, (ii) the
large complexes of molecular clouds (MCs) in which new stars are
formed and (iii) the stars of any age and chemical composition.

The galaxy models stand on a robust model of chemical evolution that
assuming a suitable prescription for gas infall, initial mass
function, star formation rate and stellar ejecta provides the total
amounts of gas and stars present at any age together with their
chemical history. The chemical models are taylored in such a way to
match the gross properties of galaxies of different morphological
type. In order to describe the interaction between stars and ISM in
building up the total SED of a galaxy, one has to know the spatial
distribution of gas and stars. This is made adopting a simple
geometrical model for each type of galaxy. The total gas and star
mass provided by the chemical model are distributed over the whole
volume by means of suitable density profiles, one for each component
and depending on the galaxy type (spheroidal, disk and disk plus
bulge). The galaxy is then split in suitable volume elements to each
of which the appropriate amounts of stars, MCs and ISM are assigned.
Each elemental volume bin is at the same time source of radiation
from the stars inside and absorber and emitter of radiation from and
to all other volume bins and the ISM in between. They are the
elemental seeds to calculate the total SED.

Using the results for the properties of the ISM and the Single
Stellar Populations (SSPs) presented by \citet{Piovan06a} we derive
the SEDs of galaxies of different morphological type. First the
technical details of the method are described and the basic
relations driving the interaction between the physical components of
the galaxy are presented. Second, the main parameters are examined
and their effects on the SED of three prototype galaxies (a disk, an
elliptical and a starburster) are highlighted. The final part of the
paper is devoted to assess the ability of our galaxy models in
reproducing the SEDs of a few real galaxies of the Local Universe.

\end{abstract}



\section{Introduction}\label{intro}

In the early thirties of the past century, nearly at the same time
of the Hubble discoveries (external galaxies and expansion of the
Universe), another important discovery albeit less spectacular was
made by Trumpler, i.e. the existence of the interstellar dust, whose
implications could not be even imagined seventy years ago. He
suggested the existence of a solid component, made by particles of
many irregular shapes and dimensions, mixed and associated to the
ISM and absorbing the stellar radiation: the interstellar dust. It
was soon clear that the dust grains diffusing and absorbing the
light emitted by the stars, in particular in the UV-optical region,
cannot be neglected when measuring the light emitted by distant
stars and galaxies.

In the local universe a turning point was the IRAS survey that
discovered ten of thousands of galaxies, the major part of them too
weak to be included in the optical catalogs, emitting more energy in
the infrared than in all the other spectral regions. Galaxies with
IR luminosity as high as $L_{IR}>10^{11} L_{\odot}$, are the main
population of extragalactic objects in the Local Universe.

Which is the reason of this huge IR luminosity? What powers the IR
emission? Observations show that a great deal of the IR luminosity
is emitted by dust, which absorbs the UV-optical light emitted by
stars formed inside the MCs in huge bursts of activity and re-emits
it in the IR. About the $30\%$ or more of the stellar light in the
Local Universe is reprocessed by dust in the IR-submm range
\citep{Soifer87}.

An important new element in the study of high-redshift universe has
been the detection by the COBE satellite of a FIR and sub-mm
background radiation of extragalactic origin, the  so-called Cosmic
InfraRed Background (CIRB). This radiation implies that galaxies in
the past should have been much more active in the far-IR than in the
optical, and very luminous. It is likely that dust plays a
fundamental role in shaping the SED of these galaxies
\citep{Puget96,Guiderdoni97,Fixsen98,Hauser98,Pozzetti98}.
Observations with SCUBA \citep{Barger99,Hughes98}, ISOPHOT
\citep{Puget99,Dole99} and ISOCAM \citep{Elbaz98,Elbaz99} have
measured the CIRB at selected wavelengths trying to detect and
identify the sources of this radiation. Even if it is not easy to
identify these objects and to measure their redshift
\citep{Lilly99,Barger99}, from the observations it turns  out that the
IR luminosity seems to be emitted be the counterparts at high
redshift of the local LIRGs and ULIRGs, (ongoing high star
formation, obscuration and reprocessing of the stellar radiation by
a dusty environment).

Observations of high-$z$ galaxies at $z=3-4$
\citep{Steidel96,Madau96} confirm that these objects are
characterized by strong obscuration and emission by dust in the IR,
in such a way that only taking into account their emission  all over
the spectral range it is possible to study their properties. Going
to even higher redshifts, dust still plays a fundamental role even
in objects at redshift $z \gtrsim 4$, as indicated by the
observations of QSOs and galaxies
\citep{McMahon94,Omont96,Soifer98}.

It clearly appears how it is no longer possible to leave dust aside
in studies of the Milky Way, the Local and the High Redshift
Universe.

Stars and dust are therefore tightly interwoven, even if the
presence of dust is more spectacular in galaxies with strong star
formation. In disk galaxies, with active, mild star formation, dust
is partly associated to the diffuse ISM, partly to the molecular
regions with star formation, and, finally, partly to the
circumstellar envelopes of AGB stars. The contribution from all the
three kinds of source is evenly balanced. In starburst galaxies, the
situation is the same as above but now the key role is played by the
regions of intense star formation. In elliptical galaxies, which
show weak emission in the FIR (IRAS), dust is essentially associated
to AGB stars of small mass that continuously loose matter refueling
the ISM of gas and dust. The great importance of dust in relation to
the formation and evolution of galaxies is evident. Dust strongly
affects the observed SED of a galaxy, hampering its interpretation
in term of the fundamental physical parameters, such as age,
metallicity, initial mass function (IMF), mix of stellar
populations, star formation history (SFH). Determinations of
luminosity functions, number counts and many large-scale relations,
such as the Tully-Fisher just to mention one, are also affected by
the dust content of galaxies \citep{Calzetti01}. The strong effect
of dust, both for local \citep{Bell01} and high redshift galaxies
\citep{Madau96,Steidel99}, is the reason why the evolution with the
redshift of the galaxy star formation rate (SFR) is still a matter of vivid debate
\citep{Madau96,Steidel99,Barger00}.

To get precious informations on the process of galaxies formation
and evolution, we need to measure the SFR in galaxies at different
redshifts, to understand when and how galaxies form their stars. To
this purpose, many wide-field and all-sky surveys are currently
running or have been just completed, e.g. the Galaxy Evolution
Explorer [GALEX] \citep{Martin97} and the Sloan Digital Sky Survey
[SDSS] \citep{York00} in the UV-optical range; the Two Micron
All-Sky Survey [2MASS] \citep{Skrutskie97} and the Deep
Near-Infrared Survey of the Southern Sky [DENIS] \citep{Epchtein97}
in the near infrared. Recently, the Spitzer Space Telescope has open
the gate to dedicated studies aimed at improving our knowledge of
the Universe in the middle and far infrared (MIR and FIR), e.g. the
SIRTF Wide-Area Infrared Extragalactic Survey \citep[SWIRE
-][]{Lonsdale03} and the SIRTF Nearby Galaxies Survey \citep[SINGS
-][]{Kennicutt03}. The IR data will increase even more  with the
coming ASTRO-F mission \citep{Pearson04} and  the advent of the
James Webb Space Telescope (JWST). Combined with other astronomical
databases, they provide a huge amount of UV-optical and IR data for
millions of galaxies. Both spectral intervals are strongly affected
by dust.

To fully exploit the information on the physical properties of the
observed objects, this wealth of data must be accompanied by the
continuous upgrade of the  theoretical models of galaxy formation
and evolution. This means that the effects of dust must be included
in the SEDs of single stellar populations (SSP), assemblies of
coeval stars with the same initial chemical composition, and in the
SEDs of galaxies, composite systems  made of stars of any age and
chemical composition, gas and dust. Current theoretical models for
both SSPs and galaxies in presence of dust and still unsatisfactory
for many different reasons: the lack of theoretical spectra for very
cool likely dust-rich stars, the poor knowledge of the dust
properties in high metallicity environments and of the production of
dust itself, the relation between gas and dust content.

With a few exceptions
\citep[e.g][]{Bressan98,Mouhcine02a,Piovan03,Piovan06a}, the light
emitted by SSPs is modelled neglecting the presence of dust around
their stars, both for AGB and young stars
\citep[e.g.][]{Bertelli94,Tantalo98a,Girardi02,Bruzual03}. When we
fold many SSPs using the classical evolutionary population synthesis
technique (EPS) we simply convolve their fluxes with the SFH of the
galaxy. Many spectrophotometric models of galaxies are built in this
way: there is no dust at the level of SSPs and again no dust at the
level of the galaxy model \citep[see
e.g.][]{Arimoto87,Arimoto90,Bruzual93,Leitherer95,Leitherer99,Tantalo96,Kodama97,
Kodama98,Tantalo98a,Buzzoni02,Buzzoni05}. There are many reasons for
this lack of realistic dusty models. First, before IRAS (1983) and
COBE (1989), the important role played by dust in the galaxy SEDs
was not fully appreciated and consequently only  the  stellar
component was taken into consideration
\citep{Tinsley72,Searle73,Huchra77}. Second, the inclusion of the
dusty component and  the IR emission, in particular, require a
higher level of sophistication of the models. Indeed one has to
develop a 3D-model in which the sources of radiation and the
emitting/absorbing medium are distributed; one has to face and solve
the problem of the radiative transfer; one has to know optical
properties of the dust; one has to simulate in a realistic way the
interactions among the various physical components of a galaxy and
the computational cost is often very high.

Despite the above difficulties, many efforts have been made to
develop more and more complex and refined models, trying to take
into account both the effects of attenuation and emission by dust.
Considering the complexity of the problem, many even sophisticated studies
are limited to the UV-optical region of the spectrum and
consider or suggest and discuss only the attenuation of the stellar
radiation by dust at various levels of detail
\citep{Guiderdoni87,Bressan94,Tantalo96,Fioc97,Tantalo98a,Buzzoni02,Bruzual03,Buzzoni05}
. In some cases, the emission of dust in the IR/sub-mm range is
considered \citep[e.g.]{Rowan89,Guiderdoni98}, but no detailed model
of the stellar sources whose radiation is reprocessed by dust is
developed. In this case there is not a clear relationship between
the  sources of  UV flux and the reprocessed IR flux. There are also
models that include at different level of sophistication the effects
of obscuration and emission by dust, but many of them have been
developed to study particular objects or aspects of the radiative
transfer, or the effects of the space distribution of the dust
\citep{Efstathiou95,Bianchi96,Wise96,Cimatti97,Gordon97,Ferrara99,Gordon01,Misselt01,Dopita05a,Dopita05b}.
They cannot be applied to a general spectrophotometric study of
galaxies of different morphological type. The most recent models are
those by \citet{Devriendt99}, \citet{Devriendt00}, \citet{Silva98},
\citet{Silva99}, \citet{Takagi03a,Takagi04,Takagi03b}. The model of
\citet{Silva98} has been later updated and improved by including the
radio emission \citep{Bressan02}, the nebular emission
\citep{Panuzzo03} and the X-ray emission \citep{Silva03}. They
present some important differences in the way they approach the
problem. For instance, the models of \citet{Silva98} and
\citet{Takagi03a} are merely theoretical and all properties are
derived from a few important assumptions and/or ingredients, whereas
those of \citet{Devriendt00} rescale the SEDs to match the average
IRAS colors.

As in the meantime, much progress has been made in many aspects of
this complex problem, for instance better understanding of the dust
properties \citep[][]{Li01,Draine01}, more complete grids of stellar
evolutionary models and isochrones \citep{Salasnich00,Girardi00},
new libraries of stellar spectra at high resolution
\citep{Zwitter04}, better chemical models \citep{Portinari98}, new
SSPs with dust \citep{Piovan03,Piovan06a}, and finally more
detailed, multi-phase chemical models \citep{Dwek98}, we intend to
present here a new model of population synthesis in dusty conditions
trying to take advantage of some of these advancements. It follows
the theoretical approach of \citet{Silva98}, but yet improves upon
important aspects. Major changes are related to the chemistry and
optical properties of the dust (going from using updated cross
sections for absorption and emission processes, to including new
grain size distribution laws, to more accurate description of the
radiative transfer for star forming regions), to better models of
star formation and chemical enrichment histories in galaxies
including the most recent chemical yields and the effect of galactic
winds (whenever required), and finally to more recent libraries of
SSPs to calculate the photometric properties of the galaxies. The
resulting galaxy SEDs go from the far UV to the far IR.

\section{Strategy of the study}\label{strate}

The model we have adopted is shortly summarized in Sect.
\ref{gal_mod} where first we define the galaxy components we are
dealing with, i.e. bare stars, stars embedded in MC complexes, and
diffuse ISM (Sect. \ref{gal_comp}); second we outline the recipes
and basic equations for the gas infall, chemical evolution, initial
mass function and SFR (Sect. \ref{chem_mod}); third we describe how
the total amounts of stars, MCs and ISM present in the galaxy at a
certain age are distributed over the galaxy volume by means of
suitable density profiles, one for each component (Sect.
\ref{spatial}) that depend on the galaxy type: disk galaxies (Sect.
\ref{spatial_disk}), spheroidal galaxies (Sect. \ref{spatial_ell}),
and composite galaxies with both disk and bulge (Sect.
\ref{spatial_int}). In Sect. \ref{spe_syn_ism} we briefly summarize
some useful basic relationships/equations by \citet{Piovan06a} on
modelling the dusty ISM. In Sect. \ref{spe_syn_cal} we explain how
the SEDs of galaxies of different morphological type are calculated.
First the technical details of the method are described and basic
relationships/equations describing the interaction between the
physical components of the galaxy are presented. Second, we shortly
described the SSPs library in use \citep{Piovan03,Piovan06a} and the
space of the parameters. Third, three prototype galaxies (a disk
galaxy, an elliptical galaxy and a starburst galaxy) are used to
show the effects on the galaxy SED of varying the parameters. This
is presented  in Sect. \ref{par_examples_early} for an elliptical
galaxy, in Sect. \ref{par_examples_late} for a disk galaxy, and
finally \ref{par_examples_starburst} for a starburst galaxy. The
final part of the paper is devoted to assess the ability of our
model in reproducing the SED of a few galaxies  of different
morphological type belonging to the Local Universe. In Sect.
\ref{spiral_local}, we present two late-type spiral galaxies, in
Sect. \ref{early_local}, we show two early-type  galaxies and
finally in Sect. \ref{starburst_local} we examine two well studied
starburst galaxies. Some concluding remarks are drawn in Sect.
\ref{dis_concl5}. Finally, the mathematical derivation of some
expressions presented in this paper are given in Appendices A, B,
and C.

\section{Galaxy Models}\label{gal_mod}

\subsection{The main components and outline of the model }\label{gal_comp}

First, the galaxy models we are dealing with contain at least three
components:

\begin{enumerate}

\item[$1)$] The diffuse ISM, made of gas and dust. The physical properties of the
ISM with dust have been thoroughly discussed by \citet{Piovan06a}.
In Sect. \ref{spe_syn_ism}, we briefly summarize the main properties
and useful equations of the model we eventually choose for the ISM
of galaxies.

\item[$2)$] The large complexes of MCs in which active star
formation occurs. In our model we do not take HII regions and
nebular emission into account. Very young stars are totally or
partially embedded in these parental MCs and the SEDs emitted by
these stars are severely affected by the dusty MCs environment
around them and skewed toward the IR. \citep[see][for more
details]{Piovan06a}

\item[$3)$] The populations of stars that are no longer embedded in
their parental MCs. These stars can be subdivided in the
intermediate age AGB stars (from about 0.1 Gyr to a few Gyr) that
are intrinsically obscured by their own dust shells as described in
\citet{Piovan03}, and the old stars which shine as bare objects.

\end{enumerate}

Second, the amount of stars and gas (and its components) present in
a galaxy at any age must be the result of the star formation
activity framed in a suitable scenario.

Third, we have to adopt a suitable scheme for the 3D distribution of
the three components in the galaxy volume,  in order to describe
their interaction, and to calculate the transfer of the radiation
across the ISM of the galaxy.

\subsection{Gross history of star formation and chemical enrichment}\label{chem_mod}

The star formation and chemical enrichment histories of the model
galaxies are described by the so-called {\it infall-model}
developed by \citet{Chiosi80}, and ever since  used by many
authors among whom we recall  \citet{Bressan94},
\citet{Tantalo96}, \citet{Tantalo98a}, and \citet{Portinari98}. In
brief, within a halo of Dark Matter of mass $M_D$,  radius $R_D$,
and hence known gravitational potential, the mass of the luminous
matter, $M_{L} $, is supposed to grow with  time by infall of
primordial gas according to the law

\begin{equation} \label{rate_inf}
\frac{dM_{L}\left( t\right) }{dt}=M_{0}\exp \left( -\frac{t}{\tau
}\right)
\end{equation}

\noindent where $\tau $ is the infall time scale. The constant
$M_{0}$ is fixed by assuming that at the present  age $t_{G}$ the
mass $M_L \left(t \right)$ is equal to $M_{L}\left( t_{G}\right)$,
the luminous asymptotic mass of the galaxy (see also for more
details \citet{Tantalo96,Tantalo98a}):

\begin{equation}
M_{0}=\frac{M_{L}\left( t_{G}\right) }{\tau \left[ 1-e^{\left(
-t_{G}/\tau \right) }\right] }
\end{equation}

\noindent
Therefore, the time variation of the baryonic mass is

\begin{equation}
M_{L}\left( t\right) =\frac{M_{L}\left( t_{G}\right) }{\tau \left[
1-e^{\left( -t_{G}/\tau \right) }\right] }\left[ 1-e^{\left(
-t/\tau \right) }\right]
\end{equation}

\noindent Indicating with  $M_{g}\left( t\right)$ the mass of gas at
the time $t$, the corresponding gas fraction is $G\left( t\right)
=\frac{M_{g}\left( t\right) }{M_{L}\left( t_{G}\right) }$. Denoting
with  $X_{i}\left( t\right) $ the  mass abundance of the $i$-th
chemical species, we may write $G_{i}\left( t\right) =X_{i}\left(
t\right) G\left( t\right)$ where by definition $\sum_{i}X_{i}=1$.
The general set of equations governing the time variation of the
generic elemental species $i$ in presence of gas infall, star
formation, and gas restitution by dying stars has been introduced
and numerically solved by \citet{Talbot71} for closed models and
modified by \citet{Chiosi80} to include  the infall term for
open models:

\begin{eqnarray} \label{Talbot}
\frac{dG_{i}\left( t\right) }{dt} &=& -X_{i}\left(t\right) \Psi
\left( t\right) + \\ \nonumber &+& \int_{M_{l
}}^{M_{u}}\Psi(t^{\prime}_{M})R_{i}(t^{\prime}_{M}) \Phi\left(
M\right) dM +\left[ \frac{d}{dt}G_{i}\left( t\right) \right]_{\inf}
\end{eqnarray}

\noindent where $\Psi \left( t\right) $ is the SFR, $\Phi \left(
M\right) $ is the IMF (by mass), $M_{l}$ and $M_{u}$ are
respectively the lower and upper bounds of the IMF, $\tau _{M}$ and
$t^{\prime}_{M} = t-\tau_{M}$ are the lifetime and the birth time of
a star of mass $M$. $R_i(t^{\prime}_{M})$ is the fraction of a star
of initial mass $M$ that is ejected back in form of the species $i$.
The first term on right-end side represents the depletion of the
species $i$ from the ISM due to the start formation; the second term
represents the growth of the species $i$ ejected back to the ISM by
stars. The last term $\left[ \frac{d}{dt}G_{i}\left( t\right)
\right] _{\inf }$ is the gas accretion rate by infall. In the above
formulation only the ejecta of single stars are included.

It has been modified to include
the contribution of Type Ia supernovae assuming that the
originate in close binary systems \citep{Matteucci86,Portinari98} which are
supposed to obey the same IMF of single stars:

\begin{small}
\begin{eqnarray}
&& \large{\frac{dG_{i}\left( t\right) }{dt}} = -X_{i}\left(t\right)
\Psi \left( t\right) +\\ \nonumber &+& \int_{M_{l
}}^{M_{Bl}}\Psi(t^{\prime}_{M})R_{i}(t^{\prime}_{M}) \Phi\left( M\right) dM +\\
\nonumber &+&  \left(1-A\right)
\int_{M_{Bl}}^{M_{Bu}}\Psi(t^{\prime}_{M}) R_{i}(t^{\prime}_{M})
\Phi \left( M\right) dM +\\ \nonumber &+& \int_{M_{Bu}}^{M_{u
}}\Psi(t^{\prime}) R_{i}(t^{\prime}_{M}) \Phi \left( M\right) dM +\\
\nonumber &+&  A\int_{M_{Bl}}^{M_{Bu}} \frac{\Phi \left(M_{B}\right
)}{M_{B}} \int_{0}^{0.5} F\left( \mu \right)\Psi(t^{\prime}_{M_{1}})
R_{i}(t^{\prime}_{M_{1}}) d\mu
dM_{B} +\\
\nonumber &+&   E_{SNIi} \cdot A\int_{M_{Bl}}^{M_{Bu}} \frac{\Phi
\left(M_{B}\right )}{M_{B}} \int_{\mu_{l}}^{0.5} F\left( \mu
\right)\Psi(t^{\prime}_{M_{2}}) d\mu dM_{B} +\\
\nonumber &+& \left[ \frac{d}{dt}G_{i}\left( t\right) \right]_{\inf}
\end{eqnarray}
\end{small}

\noindent $ M_{Bl}$ and $M_{Bu}$ are the lower and upper limit for
the mass of the binary system ($M_{B}$ is the total mass of the
binary system, $M_{1}$ and $M_{2}$ represents the mass of the
primary and secondary star), $F\left(\mu \right)$ is the
distribution of the fractionary mass of the secondary $\mu
=M_{2}/M_{B}$, $\mu_{l}$ is the minimum value of this mass ratio and
$E_{SNIi}$ are the ejecta of SN{\ae} Ia, assumed to be independent
of $M_{B}$ or $\mu$ \citep{Portinari98}. In the range between
$M_{Bl}$ and $M_{Bu}$ of binary systems suitable to become a SN Ia,
the contribution of single stars (the fraction $1-A$ of the total) is
separated from that of binaries producing SN{\ae} Ia (the fraction $A$
of the total).

The first three integrals on the right-end
side represent the contribution of the ejecta of single stars. The
fourth integral represents the contribution of the primary star in a
binary system, assumed to be unaffected by its companion, as far as
the released chemical ejecta are concerned. The fifth term is the
contribution of SN{\ae} Ia exploding when the secondary star pours
all its ejecta on the primary star. The factor multiplying
$E_{SNIi}$ is $R_{SNI}\left(t\right)$, the rate of SN{\ae} Ia as
described in \citet{Greggio83}. Following \citet{Portinari98} we
have adopted $M_{Bl}=3M_{\odot}$, $M_{Bu}=12M_{\odot}$ and $A=0.2$.
The distribution function of the fractionary mass of the secondary
is $F\left(\mu\right)=24\mu^{2}$ \citep{Greggio83}.

The SFR, i.e. the number of stars of mass $M$ born in the time
interval $dt$ and mass interval $dM$, is given by $dN/dt=\Psi \left(
t\right) \Phi \left( M\right) dM$. \noindent The rate of star
formation $\Psi \left( t\right) $ is the \citet{Schmidt59} law
adapted to our model $\Psi \left( t\right) =\nu M_{g}\left( t\right)
^{k}$ which, normalized to $M_{L}\left( t_{G}\right) $, becomes

\begin{equation} \label{Schmidt_law}
\Psi \left( t\right) =\nu M_{L}\left( t_{G}\right) ^{k-1}G\left(
t\right) ^{k}
\end{equation}

\noindent
The parameters $\nu $ and $k$ are extremely important:
$k$ yields the dependence of the star formation rate on the gas
content. Current values are $k=1$ or $k=2$. The factor $\nu$
measures the efficiency of the star formation process.

\noindent In this type of model, because of the competition between
gas infall, gas consumption by star formation, and gas ejection by
dying stars, the SFR starts  very low, grows to a maximum and then
declines. The time scale $\tau$ (eqn. \ref{rate_inf}) roughly
corresponds to the age at which the star formation activity reaches
the peak value.

The chemical models are meant to provide the mass of stars,
$M_{*}\left(t\right)$, the mass of gas $M_{g}\left(t\right)$ and the
metallicity $Z\left(t\right)$ to be used as entries for the
population synthesis code.

We also introduce in the model composite galaxies made of a disk and
a bulge. In this case the mass of the galaxy is the sum of the two
components. The disk and the bulge are assumed to evolve
independently and for each component the evolution of its
$M_{*}\left(t\right)$, $M_{g}\left(t\right)$ and $Z\left(t\right)$
will be followed.

\subsection{The spatial distribution of stars and ISM}\label{spatial}

In the classical EPS, the structure of a galaxy has no effect on the
total SED which is simply obtained by convolving  the SSP spectra
with the SFH. The galaxy structure is mimicked by considering
different SFHs for the various morphological types and/or
componenents of a galaxy, e.g. bulge and disk \citep{Buzzoni02}.
This simple convolution is no longer suitable to be used when the
ISM and the absorption and IR/sub-mm emission by dust are taken into
account. In this case the spatial distribution of the ISM,  dust and
stars in the galaxy should be specified. To this aim we start from
the observational evidence that the spatial distribution of stars
and ISM depends on the galaxy type. In our models we consider three
prototype distributions, i.e. a pure disk, a spheroid and
combination of the two to simulate late-type (with no bulge),
early-type (classical ellipticals), and intermediate-type (with a
prominent bulge) galaxies, respectively.

\subsubsection{Disk galaxies}\label{spatial_disk}

The mass density distribution of stars, $\rho_*$, diffuse ISM,
$\rho_{M}$, and MCs, $\rho_{MC}$, inside a disk galaxy can be
represented by double decreasing exponential laws so that the mass
density decreases moving away from the equatorial plane and the
galactic center or both.

Taking a system of polar coordinates with origin at the galactic
center [$r,\theta,\phi$], the height above the equatorial plane is
$z=r\cos \theta $ and the distance from the galactic center along
the equatorial plane is $R=r\sin \theta $, where $\theta$ is the
angle between the polar vector $r$ and the z-axis perpendicular to
the galactic plane passing through the center. The azimuthal
symmetry rules out the
 angle $\phi$ and therefore the density laws for the three
components are:

\begin{equation}
\rho_{i}=\rho_{0i}\exp \left( -\frac{r\sin \theta
}{R_{d}^{i}}\right) \exp \left( -\frac{r\left| \cos \theta \right|
}{z_{d}^{i}}\right) \label{rho_stars_disk}
\end{equation}

\noindent where $``i"$ can be $``*",``M"$ or $``MC"$ that is stars,
diffuse ISM and star forming MCs. $R_{d}^{*}$, $R_{d}^{MC}$, and
$R_{d}^{M}$ are the radial scale lengths of stars, MCs and ISM,
respectively, whereas $z_{d}^{*}$, $z_{d}^{MC}$, $z_{d}^{M}$ are the
corresponding scale heights above the equatorial plane. To a first
approximation, we  assume the same distributions for stars and
MCs so that $R_{d}^{*}=R_{d}^{MC}$ and $z_{d}^{*}=z_{d}^{MC}$ thus
reducing the number of scale parameters. Distributions with
different scale parameters for the three components are,
however, possible.

The constants $\rho_{0i}$ vary with the time step. Let us indicate
now with $t_{G}$ the age of the galaxy to be modelled. For the
gaseous components  only the normalization constants
$\rho_{0MC}\left(t_{G}\right)$ and $\rho_{0M}\left(t_{G}\right)$ are
required as both loose memory of their past history and what we need
is only the amount and chemical composition of  gas at the present
time. This is not the case for the stellar component for which
$\rho_{0*}\left(t\right)$ is needed at each time  $0<t<t_{G}$. In
other words to calculate the stellar emission we need to properly
build the mix of stellar populations of any age
$\tau^{\prime}=t_{G}-t$ as result of the history of star formation.

The normalization constants are derived by integrating the density
laws over the volume and by imposing the integrals to equal the mass
obtained from the chemical model.  In general, the mass of each
component $M_i$ is given by

\begin{equation}
M_{i} =\int\nolimits_{V}\rho _{0i}\exp \left( -\frac{r\sin \theta
}{R_{d}^{i} }\right) \exp \left( -\frac{r\left| \cos \theta \right|
}{z_{d}^{i}}\right) dV
\end{equation}

\noindent The mass of stars born at the time $t$ is given by
$\Psi(t)$, and therefore $\rho_{0*}\left(t\right)$ will be obtained
by using $M_{*}\left(t\right)=\Psi(t)$. $M_{M}\left(t_{G}\right)$ is
the result of gas accretion, star formation and gas restitution by
dying stars. The current total mass $M_{MC}\left(t_{G}\right)$ is a
fraction of $M_M\left(t_{G}\right)$. The double integral (in $r$ and
$\theta$) is numerically solved for $\rho_{0i}$ to be used in eqn.
\ref{rho_stars_disk}.

The galaxy radius $R_{gal}$ is left as a free parameter of the
model, thus allowing for systems of different sizes and
distributions of the components.

There is a final technical point to examine, i.e. how to
subdivide the whole volume of a disk galaxy into a number of
sub-volumes so that the energy source inside each of these can be
approximated to a point source located in their centers. This
requires that the coordinates [$r, \theta, \phi$] are divided in
suitable intervals. As far as the radial coordinate is concerned,
test experiments carried out in advance have indicated that
subdividing the galaxy radius in a number of intervals $n_{r}$ going
from $40$ to $60$ would meet the condition, secure the overall
energy balance among the sub-volumes, speed up the computational
time and yield numerically accurate results.

The number of radial intervals is derived by imposing that the mass
density among two adjacent sub-volumes scales by a fixed ratio
$\rho_{j}/\rho_{j+1}=\zeta$, where $\zeta $ is a constant. Upon
simple manipulations the above relation becomes
$r_{j+1}=r_{j}+\zeta$. Therefore, the radial grid is equally spaced
in constant steps given by $R_{gal}/n_r$ \citep{Silva99}.

The grid for the angular coordinate $\theta $ is chosen in such a
way that spacing gets thinner approaching the equatorial plane, thus
allowing for more sub-volumes in regions of higher mass density. We
split the angle $\theta $ going from 0 to $\pi$ in $ n_{\theta}$
sub-values. We need an odd value for $n_{\theta} $ so that we have
$\left( n_{\theta}-1\right) /2$ sub-angles per quadrant. Following
\citet{Silva99}, the angular distance $\alpha _{1}$ between the two
adjacent values of the angular grid is chosen in such a way that
$R_{gal}$ subtends a fraction  $f\lesssim 1$ of the disk scale
height $\left( z_{d} \right)$; in other words $\alpha _{1}=\arcsin
\left( fz_{d}/R_{gal}\right)$. Logarithmic angular steps are then
imposed $\Delta \log \theta =\left( 2/\left(n_{\theta}-3\right)
\right) \log \left( \pi/2\alpha_{1}\right) $ where $n_{\theta}$ is
determined by the condition that the second angular bin near the
equatorial plan is $g\alpha_{1}$, with $ g \lesssim 3$. This implies
$ n_{\theta} =\frac{2\log \left(\pi /2\alpha _{1}\right) }{\log g}+3
$.

\noindent The grid for the angular coordinate $\phi $ is chosen to
be suitably finely spaced near $\phi =0$ and to get progressively
broader and broader moving away clockwise and counterclockwise from
$\phi =0$. The angular steps are spaced on the logarithm scale. As a
matter of fact, thanks to the azimuthal symmetry it is sufficient to
calculate the radiation field impinging on the volume $V\left(
r_{i},\theta _{i},\phi _{i}=0\right)$ from all other volumes
$V\left( r_{k},\theta _{k},\phi _{k}\right)$. A grid thinner at
$\phi \simeq 0$ than elsewhere guarantees the approximation of
point-like energy sources at the center of the volume elements and
the conservation of the total energy in turn.

\subsubsection{Early-type galaxies}\label{spatial_ell}

The luminosity distribution of early-type galaxies can be described by
the density profiles of Hubble, de Vaucouleurs and King
\citep{Kormendy77}, the most popular of which is  the King law
that yields a finite central density of mass \citep{Froehlich82}.
However, following \citet{Fioc97}, we use a density profile slightly
different from the King law in order to secure a smooth behavior at
the galaxy radius $R_{gal}$. We suppose that the mass density
profiles for stars, MCs, and diffuse ISM are represented by the
laws

\begin{equation}
\rho _{i}=\rho _{0i}\left[ 1+\left( \frac{r}{r_{c}^{i}}\right)
^{2}\right] ^{-\gamma _{i}}  \label{rhostar_ell}
\end{equation}

\noindent where again $``i"$ can be $``*",``M"$ or $``MC"$ (stars,
diffuse ISM and MCs) and $r_{c}^{*}$, $r_{c}^{MC}$, $r_{c}^{M}$ are
the core radii of the distributions of stars, MCs, and diffuse ISM,
respectively.

\noindent The density profile has to be truncated at the galactic
radius $R_{gal}$, which is a free parameter of the model. This would
prevent the mass $M\left( r\right) \rightarrow \infty $ for
$r\rightarrow \infty$. In analogy to what already made for disk
galaxies, the constants $\rho_{0*}\left(t\right)$, $\rho
_{0MC}\left(t_{G}\right)$ and $ \rho_{0M}\left(t_{G}\right)$ can be
found by integrating the density law over the volume and by equating
this value of the mass to the correspondent one derived from the
global chemical model

\begin{equation}
\rho _{0i} = \frac{M_{i}}{\displaystyle 4\pi \left( r_{c}^{i}\right)
^{3}\int_{0}^{R_{gal}/r_{c}^{i}}\frac{x^{2}}{\left( 1+x^{2}\right)
^{\gamma }}dx}  \label{rozero_normal}
\end{equation}

\noindent where $x = r/r_{c}^{i}$ while $\rho_{0i}$ and $M_i$ have
the same meaning as in  Sect. (\ref{spatial_disk}). The integral is
numerically evaluated and solved for $\rho _{0i}$.

Like
in the case of disk galaxies, the last step is to fix the spacing
of the coordinate grid [$r, \theta, \phi$]. The problem will be
simpler and the calculations will be faster than in the previous
case because of the  spherical symmetry. The spacing of the radial
grid is made keeping  in mind the energy conservation
constrain. Also in this case we take a sufficiently large number
of grid points $(n_r\simeq 40-60)$. The condition on the density
ratio between adjacent volumes, $\rho_{j}/\rho_{j+1} =\xi $ with
$\xi $ constant \citep{Silva99}, implies

\begin{equation}
r_{i}=r_{c}\sqrt{\left[ 1+\left( \frac{R_{gal}}{r_{c}}\right)
^{2}\right] ^{i/n_r}-1}
\end{equation}

\noindent where usually $r_{c} = r_{c}^{*}$. The coordinate $\theta
$ is subdivided into an equally spaced grid, with $ n_\theta $
elements in total, and $\theta_{1}=0,$ $\theta _{n_\theta }=\pi$.
Since the distribution is spherically symmetric, i.e. independent
from $\theta$, we do not need a thinner grid toward the equatorial
plane. For the azimuthal coordinate $\Phi $ we adopt the same grid
we have presented for disk galaxies.

\subsubsection{Intermediate-type galaxies}\label{spatial_int}

Intermediate-type galaxies are characterized by the relative
proportions of their bulge and disk: they go from the early
$\textrm{S0}$ and $\textrm{Sa}$ where the bulge is big, to the late
$\textrm{Sc}$ and $\textrm{Sd}$ where the bulge is small or
negligible. In our models, we take all this into account by means of
different SFHs for the disk and the bulge. Furthermore, both in the
bulge and disk we consider the three components (ISM, MCs and stars)
in a realistic way. In analogy to what already made  for purely disk
galaxies we adopt a system of polar coordinates with origin at the
galactic center [$r,\theta,\phi$].  Azimuthal symmetry rules out the
coordinate $\phi$.

In the disk, the density profiles for the three components are the
double decreasing exponential laws of eqn. (\ref{rho_stars_disk}).
Accordingly, we introduce the scale lengths $R_{d,B}^{*}$,
$R_{d,B}^{M}$, $R_{d,B}^{MC}$, $z_{d,B}^{*}$, $z_{d,B}^{M}$ and
$z_{d,B}^{MC}$, where the suffix $B$ indicates that now the scale
lengths are referred to the disk-bulge composite model. In the bulge
the three components are distributed according to  the King-like
profiles defined in eqn. (\ref{rhostar_ell}), where now the core
radii $r_{c,B}^{*}$, $r_{c,B}^{M}$ and $r_{c,B}^{MC}$ are referred
to the bulge. The constants of normalization are derived in the same
way described in Sects. \ref{spatial_disk} and \ref{spatial_ell} The
two SFHs of disk and bulge are assumed to be independent and are
simply obtained from the chemical models where the mass of disk and
bulge are specified. The content in stars, MCs and ISM in a given
point of the galaxy will be simply given by the sum of the
contributions for the disk and bulge.

Owing to the composite shape of the galaxy (a sphere plus a disk),
we have  to define a new mixed grid sharing the properties of both
components, i.e. those described in Sect. \ref{spatial_disk} and
Sect. \ref{spatial_ell}. Let us  indicate with $R_{B}$ the bulge
radius and with $R_{gal}$ the galaxy radius. The radial grid is
defined in the following way. We build two grids of radial
coordinates, $r_{B,i}$ and $r_{D,i}$, one for the disk and one for
the bulge, in analogy to what we did in Sects. \ref{spatial_disk}
and \ref{spatial_ell}. As the grid of the bulge is not equally
spaced, but thicker toward the center of the galaxy, we take the
coordinates $r_{i,B}$ of the bulge grid if $r_{i} < R_{B}$, while if
$r_{i} > R_{B}$, we take the coordinates of the disk $r_{D,i}$ until
$R_{gal}$. For the angular coordinate $\theta$ we proceed in the
same way. We build $\theta_{B,i}$ and $\theta_{D,i}$ as in Sects.
\ref{spatial_disk} and \ref{spatial_ell}. In this case the disk grid
is thinner toward the equatorial plane of the galaxy whereas the
bulge grid is equally spaced, so we take the coordinates
$\theta_{D,i}$ of the disk as long as $\theta_{D,i+1}-\theta_{D,i} <
\theta_{B,i+1}-\theta_{B,i}$ and
 $\theta_{B,i}$ elsewhere. For the azimuthal grid there is no problem
as it is  chosen in the same way both for the disk and the bulge.

\subsection{The elemental volume grid}\label{vol_grid}

Assigned the geometrical shape of the galaxies, the density
distributions of the three main components, and the coordinate grid
$\left( r,\theta ,\phi \right)$ (The number of grid points for the
three coordinates is $ n_r+1, n_\theta +1, n_\phi $), the galaxy is
subdivided into $\left(n_r, n_\theta, n_\Phi \right) $ small volumes
$V$, each one identified by  the coordinates of the center
$\left(r_{iV},\theta _{jV},\phi _{kV}\right)$ given by the mid-point
of the coordinate grid $r_{iV}=\left(r_{i}+r_{i+1}\right)/2$,
$\theta_{jV}=\left(\theta_{j}+\theta_{j+1}\right)/2$ and
$\phi_{kV}=\left(\phi_{k}+\phi_{k+1}\right)/2$. Thereinafter the
volume $ V\left( r_{iV},\theta _{jV},\Phi _{kV}\right)$ will be
simply indicated as  $V\left( i,j,k\right)$. The mass of stars, MCs,
and diffuse ISM, in each volume are easily derived from the
corresponding density laws $\rho _{i}\left( i,j,k\right) V\left(
i,j,k\right)$ where $i$ stands for stars, MCs, and ISM. By doing
this, we neglect all local gradients in ISM and MCs (gradients
inside each elemental volume). Since the elemental volumes have been
chosen sufficiently small, the approximation is fairly reasonable.

\section{Extinction and emission of the diffuse ISM}
\label{spe_syn_ism}

\citet{Piovan06a} presented a detailed study of the extinction and
emission properties of dusty ISMs. They take into account three dust
components, i.e. graphite, silicates and PAHs and the final global
agreement reached between theory and the ISM extinction and emission
data of MW, LMC and SMC has been very good. As we are now going to
include this dusty ISM model in our galaxies, it is wise to briefly
summarize here the basic quantities and relationships in usage.

First of all, the total cross section of scattering, absorption
and extinction is given by

\begin{equation}
\sigma_{p}\left( \lambda \right)
=\int_{a_{\min,i}}^{a_{\max,i}}\pi a^{2}Q_{p}\left( a,\lambda
\right) \frac{1}{n_{H}}\frac{dn_{i}(a)}{da} da \label{sigabs}
\end{equation}

\noindent where the index $p$ stands for absorption (abs),
scattering (sca), total extinction (ext), the index $i$ identifies
the type of grains, $a_{min,i}$ and $a_{max,i}$ are the lower and
upper limits of the size distribution for the i-type of grain,
$n_{H}$ is the number density of $H$ atoms, $Q_{p}\left( a,\lambda
\right)$ are the dimension-less absorption and scattering
coefficients \citep{Draine84,Laor93,Li01} and, finally
$dn_{i}(a)/da$ is the distribution law of the grains
\citep{Weingartner01a}. With the aid of the above cross sections it
is possible to calculate the optical depth $\tau_{p}(\lambda)$ along
a given path

\begin{equation}
\tau_{p}\left( \lambda \right) =\sigma _{p}\left( \lambda \right)
\int_{L}n_{H}dl=\sigma_{p}\left(\lambda \right) \times N_{H}
\label{tauabs}
\end{equation}

\noindent where $L$ is the optical path and all other symbols have
their usual meaning. In this expression for
 $\tau_{p}(\lambda)$ we have implicitly assumed that the
cross sections remain constant along the optical path.

Let us name $j_{\lambda}^{small}$, $j_{\lambda}^{big}$ and
$j_{\lambda}^{PAH}$  the contributions to the emission by small
grains, big grains and PAHs, respectively. How these quantities are
calculated is widely described  in \citet{Piovan06a} to whom the
reader should refer for more details. Let us summarize here just the
key relationships in usage.

The contribution to the emission by very small grains of graphite
and silicates is

\begin{eqnarray} \label{smallemission}
j_{\lambda }^{small}&=&\pi \int\nolimits_{a_{min
}}^{a_{flu}}\int\nolimits_{T_{min}}^{T_{max}} a^{2}Q_{abs}\left(
a, \lambda \right)B_{\lambda }\left( T\left(
a\right) \right)\nonumber \times \\
&& \times \frac{dP\left(a\right)}{dT} dT
\frac{1}{n_{H}}\frac{dn\left(a\right)}{da}da
\end{eqnarray}

\noindent where $dP \left(a\right)/dT $ is the distribution
temperature from $T_{min}$ to $T_{max}$  attained by grains with
generic dimension $a$ under an incident radiation field and
$B_{\lambda }\left( T\left( a\right) \right) $ is the Planck
function. $Q_{abs}\left( a, \lambda \right)$ are the absorption
coefficients, $dn\left(a \right)/da$ is the \citet{Weingartner01a}
distribution law for the dimensions, $a_{flu}$ is the upper limit
for thermally fluctuating grains, $a_{min}$ is the lower limit of
the distribution.

The emission by big grains of graphite and silicates is evaluated
assuming that they behave like  black bodies in equilibrium with the
radiation field. Therefore we have

\begin{equation} \label{bigemission}
j_{\lambda}^{big}=\pi \int_{a_{flu}}^{a_{max}}  a^{2} Q_{abs}
\left( a,\lambda \right) B_{\lambda} \left( T \left(a \right)
\right) \frac{1}{n_{H}}\frac{dn\left(a\right)}{da} da
\end{equation}

\noindent where $a_{max}$ is the upper limit of the distribution
and the meaning of the other symbols is the same as in eqn.
\ref{smallemission}.

The emission by PAHs is given by

\begin{eqnarray} \label{PAHemission}
j_{\lambda}^{PAH} &=& \frac{\pi }{n_{H}hc}\int\nolimits_{\lambda
_{min}}^{\lambda _{max}}I \left( \lambda^{^{\prime }} \right)
\lambda^{^{\prime
 }} \int\nolimits_{a_{PAH}^{low}}^{a_{PAH}^{high}} \frac{dn\left(a\right)}{da} \times \nonumber\\
&\times & a^{2} \left[ Q_{abs}^{IPAH} \left( a,
\lambda^{^{\prime}} \right)
S_{ION} \left( \lambda^{^{\prime}} ,\lambda ,a \right) \chi_{i} + \right. \\
& +& \left. Q_{abs}^{NPAH} \left( a, \lambda^{^{\prime}} \right)
S_{NEU} \left( \lambda^{^{\prime}} ,\lambda ,a \right) \left( 1-
\chi_{i} \right) \right] da  d\lambda^{^{\prime}}  \nonumber
\end{eqnarray}

\noindent where the ionization of PAHs is taken into account
\citep{Weingartner01b} and $\chi_{i}=\chi_{i}\left(a\right)$ is the
fraction of ionized PAHs with dimension $a$. $S_{ION} \left(
\lambda^{^{\prime}} ,\lambda,a \right)$ and $S_{NEU} \left(
\lambda^{^{\prime}} ,\lambda ,a \right)$ give the  energy emitted
at wavelength $\lambda$ by a molecule of dimension $a$, as a
consequence of absorbing a single photon with energy  $hc/\lambda
^{^{\prime }}$. $a_{PAH}^{low}$ and $a_{PAH}^{high}$ are the limits
of the distribution and $I \left( \lambda^{^{\prime }} \right)$ is
the incident radiation field.

\section{The galaxy SED}\label{spe_syn_cal}

Given the main components of a galaxy,  their spatial
distribution, the coordinate system, and the grid of elemental
volumes, to proceed further one has to model the interaction among
stars, dusty ISM and MCs to simulate the total SED emerging from
the galaxy.

Let us consider a generic volume $V^{\prime}=V\left( i^{\prime
},j^{\prime },k^{\prime}\right)$ of the galaxy: it will receive the
radiation coming from all other elemental volumes $V=V\left(
i,j,k\right)$. The radiation travelling from one volume to another
interacts with the ISM comprised between any two generic volumes.
Therefore one has to take into account the energy that is both
absorbed and emitted by the ISM under the interaction with the
radiation field. Two simplifying hypotheses are worth being made
here:

(i) The dust contained in a generic volume $V$ does not contribute
to the radiation field impinging on the volume $V^{\prime}$. This
approximation stands on the notion that, in first approximation,
owing to the low optical depths of the diffuse ISM in the MIR/FIR,
dust cannot effectively absorb the radiation it emits, except for
high density regions such as MCs, for which dust self-absorption has
already been taken into account. In other words, the dust contained in
$V^{\prime}$ will be transparent to the IR radiation emitted by the
dust contained in the volume $V$. Therefore, only stars and MCs will
contribute to the incoming radiation.

(ii) Following \citet{Rybicki79}, the radiative transfer from a
generic volume $V$ to $V^{\prime}$ is simply calculated by means of
the so-called  effective optical depth defined by

\begin{equation}
\tau_{eff}=\sqrt{\tau_{abs}\left( \tau_{abs}+\tau_{sca}\right) }
\label{tau_eff}
\end{equation}

\noindent The above relation stands on the following
considerations:  scattering increases the absorption, however as
photons are not destroyed, the effective optical depth must be
lower than the sum of the  optical depths by scattering and
absorption but higher than the one by sole absorption
\citep{Rybicki79}. Although this expression for $\tau_{eff}$ has
been derived for photons randomly travelling across an absorbing
 diffusive medium, so that it would strictly apply only to an
infinite, homogeneous, isotropically scattering  medium, we make use
of it here. The results from the above approximation have been
compared by \cite{Silva98} with those by \citet{Witt92} and
\citet{Ferrara99} using Monte-Carlo methods to solve radiative
transfer problems. The results fully agree with those by
\citet{Witt92} and \citet{Ferrara99} in the case of spherical
symmetry and partially  disagree  in the case of disk galaxies. For
these latter  the quality of agreement depends on view angle between
the galaxy and the observer \citep{Silva98}.

The total radiation field incident on $V^{\prime}$ is

\begin{eqnarray}
J\left( \lambda,V^{\prime } \right)
&=&\sum\limits_{i=1}^{n_r}\sum\limits_{j=1}^{n_\theta
}\sum\limits_{k=1}^{n_\Phi }   \frac{V \cdot \left[j^{MC}\left(
\lambda ,V\right) +j^{*}\left( \lambda ,V\right) \right] }
{r^{2}\left( V,V^{\prime}\right)}\\
&\times&  e^{\left[ -\tau _{eff}\left( \lambda ,V,V^{\prime}\right)
\right] } \nonumber
\end{eqnarray}

\noindent where the summations are carried over the whole ranges of
$i,j,k$ with $i\neq i^{\prime},j\neq j^{\prime}$ and $k\neq
k^{\prime}$; $r^{2}(V,V')$ is the value averaged over the volume of
the square of the distance between the volumes $V$ and $V'$; $\tau
_{eff}\left( \lambda ,V,V^{\prime}\right)$ is the effective optical
depth from $V$ to $V^{\prime}$ defined by eqn. (\ref{tau_eff}),
 which by means of eqn. (\ref{tauabs}) becomes

\begin{eqnarray}
\tau_{eff}\left( \lambda ,V,V^{\prime } \right) &=&
\sqrt{\sigma_{abs}(\lambda )\times \left( \sigma _{abs}\left(
\lambda \right) +\sigma _{sca} \left( \lambda \right) \right)}
\times \nonumber \\
&& \times \int\nolimits_{V(i,j,k)}^{V(i^{\prime },j^{\prime
},k^{\prime })}n_{H}(l)dl
\end{eqnarray}

\noindent This integral represents the number of H atoms contained
in the cylinder between $ V$ and $V^{\prime}$. All details on the
derivation of this quantity and $r^{2}(V,V')$ are given in Appendix
A and B.

The two terms $j^{MC}\left(\lambda,V\right)$ and
$j^{*}\left(\lambda,V\right)$ are the emission by MCs and stars
per unit volume of $V\left(i,j,k\right)$. They are evaluated at
the center of the volume element.

Let us now define two kinds of SSPs: those that are already free of
the parental cloud and are indicated as $ssp^{f}$ (classical free
SSPs), and those that are still embedded in their parental dusty
molecular clouds and are indicated as $ssp^{d}$ (dusty SSPs).

Let us denote with  $f_{d}$ the fraction of the SSP luminosity
that is reprocessed by dust and with $t_{0}$ the time scale for
this to occur, in such a way that

\begin{equation}
f_{d}=\left\{
\begin{array}{llll}
1          & & &\quad t\leq t_{0} \\
2-t/t_{0}  & & &\quad t_{0}<t\leq 2t_{0} \\
0          & & &\quad t\geq t_{0}
\end{array}
\right.
\label{fd}
\end{equation}

\noindent where $t_0$ is a suitable parameter determining the
evaporation time of the parental MCs. Accordingly, the fraction of
SSP luminosity that escapes without interacting with dust is
$f_{f}=1-f_{d}$.

\noindent The parameter $t_{0}$ will likely depend on the properties
of the ISM and type of galaxy in turn. Plausibly, $t_0$ will be of
the order of the lifetime of massive stars. It will be discussed in
more detail in Sect. \ref{par_chosen}.

The monochromatic luminosity of a free SSP of given age
$\tau^{\prime}=t_{G}-t$, born at $t$, and metallicity $Z$ for unit
of SSP mass is therefore

\begin{equation}
L_{\lambda}^{f}\left(\tau^{\prime},Z\right) =\frac{\displaystyle\int
\nolimits_{M_{L}}^{M_{U}}\Phi \left( M\right) f_{\lambda }\left(
M,\tau^{\prime},Z\right) dM}{
\displaystyle\int\nolimits_{M_{L}}^{M_{U}}\Phi \left( M\right) dM}
\label{sspdustnorm}
\end{equation}

Knowing the the monochromatic luminosity of the naked SSPs
$L_{\lambda}^{f}\left(\tau^{\prime}, Z\right)$, the monochromatic
luminosity of the dust enshrouded SSPs $L_{\lambda
}^{d}\left(\tau^{\prime}, Z\right)$ has been derived as described in
\citet{Piovan06a}. The emission of stars and MCs per unit  volume,
$j^{MC}\left(\lambda,V\right)$ and $j^{*}\left(\lambda,V\right)$
respectively, are given by

\begin{eqnarray}
&&j^{*}\left( \lambda ,V\right) = \int\nolimits_{2t_{0}}^{t_{G}}\rho
_{*}\left( t\right) L_{\lambda}^f
\left( \tau ^{\prime },Z\right) dt+    \nonumber \\
&&+\int\nolimits_{t_{0}}^{2t_{0}}\left(\frac{t}{t_{0}}-1 \right)
\rho_{*}\left( t\right) L_{\lambda}^f\left( \tau ^{\prime },Z\right)
dt
\end{eqnarray}

\noindent and

\begin{eqnarray}
&&j^{MC}\left( \lambda ,V\right) = \int\nolimits_{0}^{t_{0}}
\rho_{*} \left( t\right) L_{\lambda}^d \left( \tau ^{\prime
},Z\right) dt + \nonumber \\
&& + \int\nolimits_{t_{0}}^{2t_{0}}\left( 2-\frac{ t}{t_{0}}\right)
\rho_{*} \left( t\right) L_{\lambda}^d \left( \tau ^{\prime
},Z\right) dt
\end{eqnarray}

It is worth noticing that luminosity of the MCs depends only on the
luminosity of the young embedded stars and not on the mass of
molecular gas enclosed in the MCs. The  factors
$\left(2-t/t_{0}\right) $ and $ \left(1-\left(2-t/t_{0}\right)
\right) = \left(t/t_{0}-1\right)$ follows from relations (\ref{fd})
and the definition of $f_{f}$.

Once calculated the incident radiation field $J\left(\lambda,
V^{\prime}\right)$ we can obtain the emission per unit volume from
the dusty ISM. At this point the azimuthal and spherical
symmetries of the galaxy models become very important. The
emission per unit volume from the dusty ISM calculated at the
center of the volume elements is the same  everywhere,
independently of the coordinate $\Phi$. Therefore it is sufficient
to calculate the dust emission at $\Phi = 0$ for all possible
values of $r$ and $\theta$ on this "galaxy slice". It is obvious
that the symmetry will be lost when considering the total emission
from a certain volume element because owing to the adopted spacing
of the galaxy the elemental volumes are not equal. However, as
long as we refer to the emission per unit volume, the symmetry
properties above allows us to avoid  tedious and lengthy numerical
calculations. The total radiation field for unit volume emitted by
a single element is

\begin{equation}
j^{TOT}\left( \lambda ,V\right) =j^{MC}\left( \lambda ,V\right)
+j^{*}\left( \lambda ,V\right) +j^{ISM}\left( \lambda ,V\right)
\label{Geitot}
\end{equation}

\noindent where $j^{ISM}\left( \lambda ,V\right) $ is the radiation
outgoing from a unit volume of the dusty diffuse ISM and is given by
the sum of the contributions from silicates, graphite and PAHs
described by eqns. (\ref{smallemission}), (\ref{bigemission}), and
(\ref{PAHemission}). The total outgoing emission from the volume $V$
is therefore given by $j^{TOT}\left( \lambda ,V\right) \times V$
obviously different from volume to volume.

The radiation emitted by each elemental volume $\left(n_r, n_\theta,
n_\Phi \right)$ has to travel across a certain volume of the galaxy
itself before reaching the edge, escaping from the galaxy, and being
detected by an external observer. While finding its way out, the
radiation is absorbed and diffused by the ISM. The external observer
will see the galaxy along a direction fixed by the angle $\Theta $,
where $\Theta = 0$ means that the galaxy is seen face-on, whereas
$\Theta = \pi/2$ means that the galaxy is seen edge-on. To this aim,
we need to determine the properties of the cylinder of matter from
the center of each element $V$ to the edge of the galaxy, along the
direction $\Theta $ in order to calculate the amount of radiation
emitted by the galaxy along this direction. Therefore, the
monochromatic luminosity measured by an external observer is

\begin{eqnarray}
L\left(\lambda ,\Theta \right) &=& 4\pi
\sum\limits_{i=1}^{n_rr}\sum\limits_{j=1}^{n_\theta
}\sum\limits_{k=1}^{n_\Phi }V \times \nonumber  j^{TOT}\left(\lambda
,V\right) \\
&& \times e^{\left[ -\tau _{eff}\left( \lambda ,V,\Theta \right)
\right] }
\end{eqnarray}

\noindent where $\tau_{eff}\left(\lambda,V, \Theta \right)$ is the
effective optical depth between $V\left(i,j,k\right)$ and the
galactic edge along the direction $\Theta$. The details on the
derivation of the effective optical depth
$\tau_{eff}\left(\lambda, V, \Theta \right) $ are described in
Appendix C.

\subsection{SSPs in usage and parameters of the galaxy model}\label{par_chosen}

In this section first we shortly present the libraries of stellar
models, isochrones, and stellar spectra that are used to calculate
the SEDs of SSPs and galaxies. Second we summarize the results for
SSP intrinsically affected by dust calculated by
\citet{Piovan03,Piovan06a} and in the case of young SSPs we describe
the effect of the parameters of the model. Finally, we present the
other parameters governing the the ISM and chemical model and the
galaxy geometry.

\textbf{(1) Libraries of stellar models and stellar spectra}: We adopt
the set of isochrones by \citet{Tantalo98a} (anticipated in the data
base for galaxy evolution models by \citet*{Leitherer96}). The
underlying stellar models are those of the Padova Library calculated
with convective overshooting and are amply described by
\citet{Fagotto94a,Fagotto94b,Fagotto94c}, so that no detail is given
here.

The library of stellar spectra is from \citet{Lejeune98}, which
stands on the Kurucz (1995) release of theoretical spectra, however
with several important implementations \citep[see][for more
details]{Piovan03,Piovan06a}.

\textbf{(2) Dust enshrouded SSPs} As already mentioned, young stars
and AGB stars are both surrounded by their own dust. The young stars
because they are immersed in MCs, the AGB stars because they eject
dust on their own. In both cases the UV-optical light emitted by the
stars is absorbed and re-emitted in the FIR. But for old
ellipticals, in which star formation stopped at early epochs because
of the galactic wind so that no young stellar populations embedded
into parental MCs are present, for all the other morphological types
the impact of young dusty populations on the galaxy SED has to be
considered. This indeed is stronger than the one caused by dusty AGB
stars (see below). The reason of it can be reduced to the fact that
while for young SSPs all the emitted energy interacts with the dusty
environment, in the case of old SSPs only AGB stars interact with
the dust. However, as the ranges of wavelength interested in the
case of young SSPs and dusty AGBs are different, the accurate
description of the region around $1-2 \mu m$ requires that the dusty
shells around AGB stars are fully taken into account. SSPs for AGB
stars and young stars with this effect built in are  by
\citet{Piovan03} and \citet{Piovan06a}.

\textit{Intermediate age and old SSPs}: In this paper we decided not
to use the SSPs of \citet{Piovan03} where the effects of
circumstellar dusty shells around the AGB stars (of intermediate/old
ages) are included. The main reason is that dust enshrouded AGB
stars are available only for three metallicities. Work is in
progress to build a more complete set of SEDs in which the effects
of the dusty shells around AGB stars are included.

\textit{Young SSPs}: In general dust around young stars shifts the
UV/optical radiation into the MIR/NIR. However, the SED of young
SSPs \citep{Piovan06a} is governed by a number of parameters
describing the MC itself, chief among which are  the optical depth
$\tau_{V}$ in the $V$ band, the scale radius of the cloud $R$, the
$C$ abundance $b_{C}$ in the two log-normal populations of very
small carbonaceous grains according to  the distribution law of
\citet{Weingartner01a}, and finally the ionization state of PAHs. In
addition to these, there is another parameter related to young MCs,
namely the evaporation time $t_{0}$. Let us examine their effect in
some detail.

\noindent (i) The MC {\bf optical depth} $\tau_{V}$ in the $V$ band.
Up to now in our library of SSPs only two values are considered,
i.e. $\tau_{V} = 5$ and 35, corresponding to intermediate and high
optical depths. The effect of changing the optical depth is simple:
the higher the optical depth, the higher is the amount of energy
shifted toward longer wavelengths. A remark is worth here. Using a
certain set of SSPs (i.e. a set with given parameters) we implicitly
fix the  optical depth. Clearly, the optical depth of MCs not only
increases with the cloud mass, decreases with the cloud size, and in
general increases with the cloud density, but also increases with
the metallicity. Therefore the ideal situation would be the one in
which MCs in galaxy models cover the whole range of masses, radii,
densities and metallicities.

\noindent (ii) The {\bf  scale radius} of the MC. Two values are
allowed, $R = 1$ and 5. They correspond to MCs of different
compactness. Keeping constant all other parameters, the peak of
emission shifts to longer wavelengths at increasing scale radius
\citep[see also][]{Takagi03a}.

\noindent (iii) The {\bf carbon abundance} $b_{C}$ per H nucleus
fixes the abundance of this element in the two log-normal
populations of very small grains in the \citet{Weingartner01a} law.
As shown by \citet{Weingartner01a}, the sole extinction curve  is
unable to constrain $b_{C}$. It provides only an upper limit for
$b_{C}$, which at given ratio
$R_{V}=A\left(V\right)/E\left(B-V\right)$ is reached when the very
small carbonaceous grains and PAHs are able to account for  the
ultraviolet bump of the extinction curve. Values lower than the
upper limit are possible. Without other constraints,  for instance
the IR emission of the region of interest, it is not possible to
determine $b_{C}$, which plays an important role in the IR emission
of galaxies. Specifically, it drives the emission of PAHs in the MIR
generated by  young  MCs. The effect of varying $b_{C}$ is simple:
\textit{for higher values of $b_{C}$, the PAH emission in the MIR
reaches higher flux levels}. In contrast, the UV-optical  flux and
the shape of the absorbed stellar emission do not depend on $b_{C}$,
because the extinction curve remains the same for different values
of $b_{C}$. As the global abundance of $C$ is fixed, the
distribution of the grains has to compensate for the small number of
small grains with an higher number of big grains. Therefore, for low
values of $b_{C}$ we expect that the emission in the FIR slightly
increases (the total energy budget has clearly to be conserved).

\noindent (iv) The {\bf  ionization state} of PAHs. The optical
properties we have adopted are different for ionized and neutral
PAHs \citep[see][for more details]{Li01}. \textit{The ionization
state of PAHs affects the PAH emission profile in the MIR}. We
considered three ionization models. The first is the one by
\citet{Weingartner01b}. The  second model adopts for the MCs the
same ionization profile calculated by \citet{Li01} for the diffuse
ISM of the MW. The third model  simply takes into account only
neutral PAHs (very low ionization). In other words, we have a
sequence of models going from ionized to neutral PAHs \citep[see][
for more details]{Piovan06a}.

\noindent (v) $ \mathbf{t_{0}}$: Following the simple recipe by
\citet{Silva98}, this is the scale time required by a new generation
of stars to get rid of the parental MCs \citep[see
also][]{Piovan06a}. In \citet{Silva98} model, star formation is
reduced to a point source scheme, in which stars  born at the center
of  dusty MCs slowly evaporate  them to eventually shine free. As
pointed from \citet{Takagi03a}, if the time of the star forming
activity is shorter than the escaping time scale $t_{0}$, the light
coming from young stars is completely hidden at the UV-optical
wavelenghts. In general, in \citet{Silva98} model, for times shorter
than $t_{0}$, no light escapes from dusty star forming regions and
therefore the light observed in the UV-NIR should be negligible for
\textit{any aperture size} of the galaxies. This is not what we
observe in real galaxies especially in their central regions. The
discrepancy is cured by adopting the more realistic descriptions in
which stars are randomly distributed inside the MC, and solving the
radiation transfer problem with the ray tracing method
\citep[see][]{Piovan06a}. We can reproduce in detail in this way the
inner regions of starbursters like Arp$220$ and M$82$. Even with the
new scheme, the parameter $t_{0}$ drives the amount of energy
emitted by the young stars that is absorbed and reprocessed by the
local dust (i.e. in the region immediately around the stars
themselves). The evaporation is simulated by letting more and more
energy to escape without being reprocessed by dust. The time scale
$t_{0}$ likely goes from 3 to 100 Myr, i.e. the evolutionary
lifetime of a 100 and 5 $M_{\odot}$ respectively. High values of
$t_{0}$ mean that young stars are longer hidden by the parental
clouds and accordingly much of the light they emit  in the
UV-visible range is shifted to the FIR for a long period of time. In
such a case a large fraction of the IR light emitted by a galaxy
could be due to young stars still embedded in MCs. The opposite for
low values of $t_{0}$.

{\bf (3) Parameters of the ISM model} are the fractionary mass of
gas in the ISM $f_{M}$ and the metallicity $Z$.

\noindent (a) $\mathbf{f_{M}}$: The fraction of diffuse gas present
in a galaxy with respect to the total amount of gas can be inferred
from observational data on the ratio between the molecular and
atomic gas. In disk galaxies for instance most of hydrogen in MCs is
in form of $H_{2}$, whereas the inter-cloud medium is mainly made of
atomic $HI$. This parameter bears very much on emission and
extinction by the diffuse ISM. The larger $f_{M}$, the bigger is the
amount of gas present in the diffuse ISM. Therefore, the emission
(proportional to the number density of $H$ atoms) will be higher.
Furthermore, extinction of the diffuse ISM as well will increase
because of the bigger absorption by dust, which is also proportional
to the number density of $H$ atoms in the diffuse ISM.

\noindent (b) {\bf Metallicity $\mathbf{Z}$}: The metallicity
reached by the ISM of the galaxy fixes the properties of the dust,
drives the intensity of the ISM emission and its ability in
extinguishing the radiation field. We adopted as standard models for
$Z \lesssim Z_{\odot}$ the three compositions of MW, LMC and SMC,
that allow us to cover well this range of metallicities. The dust to
gas ratio scales with the metallicity: the higher  the metal content
of a galaxy, the higher is the abundance of grains per $H$ atom.
This simply implies that the ISMs of high metallicity galaxies will
both absorb and emit more energy. For metallicities higher than the
solar one ($Z \gtrsim Z_{\odot}$), we keep the relative proportions
of silicates, graphite and PAHs of the diffuse ISM of the MW, scaled
to the higher dust to gas ratio.

{\bf (4) Parameters of the chemical models}: Chief among others are
the exponent $k$ and the efficiency $\nu$ of the star formation law,
the infall time scale $\tau$, and limited to the case of elliptical
galaxies the parameters fixing the gravitational and thermal energy
of the gas and driving the onset of galactic winds. In the case of
galaxies composed by disk and bulge, the number of parameters
increases. There are $k_{B}$, $\nu_{B}$ and $\tau_{B}$ together with
$k_{D}$, $\nu_{D}$ and $\tau_{D}$, and the bulge to disk mass ratio
$M_B/M_D$. Finally, we will also consider the case in which galaxies
may undergo a burst of star formation. The burst is introduced
superimposing it to the star formation history law described by eqn.
(\ref{Schmidt_law}). The burst  is described by the following
parameters: the age $t_{BUR}$ at which the burst occurs, the
enhancement factor $\nu_{BUR}$  amplifying the current efficiency
$\nu$, and, finally, the duration $\Delta t_{BUR}$.

{\bf (5) Parameters of the geometrical models}: In the case of early
type galaxies (our spherical case) the parameters are $r_{c}^{*}$,
$r_{c}^{MC}$, $r_{c}^{M}$, i.e. the core radii for the distribution
of stars, MCs and ISM, respectively. In the case of disk galaxies
the spatial distributions of the three components are fixed by the
radial scale lengths $R_{d}^{*}$, $R_{d}^{MC}$, and $R_{d}^{M}$, and
the scale heights $z_{d}^{*}$, $z_{d}^{MC}$, and $z_{d}^{M}$, with
obvious meaning of the various symbols. For composite galaxies with
disk and bulge the number of parameters increases dramatically as we
are dealing with two distributions: the King law (as for early-type
galaxies) for the bulge, described with the parameters
$r_{c,B}^{*}$, $r_{c,B}^{MC}$, $r_{c,B}^{M}$ and a double decreasing
exponential law for the disk, described by the parameters
$R_{d,B}^{*}$, $R_{d,B}^{MC}$, $R_{d,B}^{M}$ and $z_{d,B}^{*}$,
$z_{d,B}^{MC}$, $z_{d,B}^{M}$. In all types of model, the scale
length of the young and old stellar populations is the same. Even if
this reduces the number of parameters, in reality it would be
interesting, for starburst galaxies in particular, to consider
different spatial distributions for  old and young stars
\citep[see][]{Calzetti01}. Work is in progress to improve upon this
type of model. The remaining geometrical parameters are: the radius
of the galaxy $R_{gal}$ and, if a bulge is present, the radius
$R_{B}$ of this, and finally the angle of inclination going from
edge-on to face-on galaxies, i.e. from $\Theta=0^{\circ }$ to
$\Theta=90^{\circ }$ with respect to the z-axis perpendicular to the
equatorial plane of the galaxy.

\subsection{Role of the parameters in template model galaxies}
\label{par_examples}

In this section we analyze in detail the effect of the various
parameters on galaxy SEDs. To this aim we calculate three template
galaxies, i.e. an old early-type galaxy, a late-type spiral galaxy,
and finally a starburst galaxy.

\subsubsection{Early-type galaxies}\label{par_examples_early}

\begin{figure}
\psfig{file=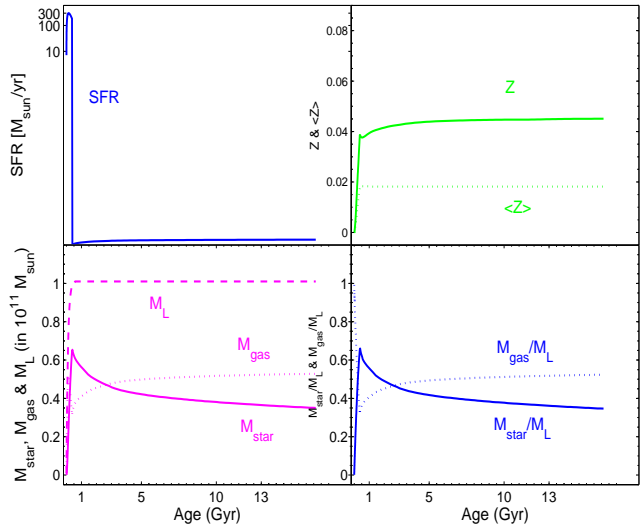,width=8.5truecm,height=7.0truecm}
\caption{Basic quantities of the chemical model for a prototype
early-type galaxy as function of the age: the top-left panel shows
the star formation rate in $M_\odot$/yr; the top-right panel
displays the maximum ($Z$, solid line) and mean metallicity
($\langle Z \rangle $, dotted line); the bottom-left panel shows the
mass of living stars $M_{star}$(solid line), the gas mass $M_{gas}$
(dotted line), and the total mass of baryons $M_{L}$ (dashed line);
finally the bottom-right panel displays the ratios $M_{star} /
M_{L}$ (solid line) and  $M_{gas} / M_{L}$ (dotted line). All masses
are in units of $10^{11}\,M_\odot$. Ages are in Gyr.}
\label{chimicaEll}
\end{figure}

Early-type galaxies are simulated by means of spherical models whose
total baryonic mass is  $10^{11} M_{\odot}$ and total dark matter
mass is $5$ times higher \citep[see][for details]{Tantalo96}. For
the purposes of this section we limit ourselves to consider the
typical SFH of an elliptical, characterized by a strong initial
burst of stellar activity, early onset of galactic wind, and
quiescence ever since. As the model stands on the infall scheme, the
SFR starts small, increases to a peak value and then declines ever
since. The time scale at which the peak occurs is of the order of
the infall time scale that is fixed at $\tau$=0.1 Gyr
\citep{Tantalo96} which means that the model tends to the closed-box
approximation but for the occurrence of galactic winds. The exponent
for the star formation law is $k$=1. The efficiency of star
formation is $\nu = 6$. The reason for the above choice and more
details can be found in \citet{Tantalo96,Tantalo98a}. The four
panels of Fig. \ref{chimicaEll} show a few relevant quantities
characterizing the galaxy.

To study the effect of the parameters we consider two stages of
the history of an early-type galaxy: the age of $0.15$ Gyr, when
the SFR reaches the peak and the present age of $13$ Gyr after
billion years of passive evolution.

In the very early stages of high star formation we can assume that
dust is mostly concentrated in the dense regions of star formation.
Following this idea we assign a  low gas content to the diffuse ISM
($f_{M} = 0.3$, which implies that  the mass  in dusty MCs is about
twice mass of the diffuse ISM). We adopt  $t_{0} \thicksim $ 40 Myr,
a long evaporation time, which sounds reasonable for a high density
star forming environment. Once the galactic wind has taken place and
star formation has ceased, we can adopt  $f_{M} =1$ and $t_{0} = 0$.

In the 13 Gyr model there is no longer star formation and the
evolution is merely passive. If the galaxy is free of gas and only
stars are present, the SED is expected to drop off long-ward of
about $2 \mu m$. However elliptical galaxies of the local universe
emit in the IR \citep{Guhathakurta86}. The origin of this flux in
the MIR/FIR may be due to diffuse dust which emits at those
wavelengths. Therefore to match this part of the spectrum one has to
allow for a small amount of diffuse ISM which is likely to exist. By
imposing $f_{M} = 1$ for our model of age $13$ Gyr, we assume that
all the gas is in form of diffuse ISM. Interesting questions to rise
are: how much gas can be present today in an elliptical galaxy? What
is the source of this gas?

To answer the above questions let us examine how the gas content of
an elliptical galaxy is expected to evolve with time
\citep{Gibson97a,Chiosi00}. According to the classical scenario
\citep{Larson74}, all the gas present in elliptical galaxies at the
onset of the galactic wind is supposed to be expelled from the
gravitational potential well of the galaxy into the
\emph{intracluster medium}. Despite the radiative losses, the energy
input from supernova explosions, stellar winds and dynamical
interactions overwhelms the gravitational potential and makes
galactic wind to occur in such a way to happen later and later at
increasing galaxy mass. Star formation suddenly ceases and
consequently the energy input by Type II (early on) and Type I
(afterwards) either stops (Type II) or get very small (Type I). All
this requires about $0.5$ Gyr. Subsequently low mass stars
($\lesssim 2 M_{\odot}$) loose mass by stellar wind (during the red
giant and the asymptotic giant branch phases) thus refueling the
galaxy of gas in amounts that are comparable to those before the
galactic wind \citep{Chiosi00}. What is the fate of this gas?

In the most plausible scenario \citep{Gibson97a} the phase of
galactic wind should last for about 0.5-1 Gyr thanks to the energy
input by Type Ia supernovae and then stop. The amount of gas lost by
low mass stars during this time interval turns out to be very large
\citep{Chiosi00} and it is not entirely clear how it can exceed the
gravitational energy. Nevertheless, there is general consensus that
on a rather short time scale the gas may escape the galaxy. Most
likely a sort of dynamical equilibrium is reached in which gas is
continuously ejected by stars and lost by the galaxy. It may happen
that a tiny amount of gas is always present in the galaxy, thus
accounting for the IR emission.

After the onset of the galactic wind, our chemical model is not able
to describe this complex situation. Dying stars emit lots of gas
whose fate is uncertain. Basing on the flux level observed in real
galaxies, the gas content of the galaxy is reduced by  a factor of
the order of $1000$.

Finally, we are left with the geometrical parameters to fix. For the
exponents $\gamma _{*}$ and $\gamma _{MC}$ a value around $1.5$
could be taken for ellipticals. The value for the ISM is more
uncertain. Following \citet{Froehlich82}, \citet{Witt92}, and
\citet{Wise96} we choose $\gamma _{M}\simeq 0.5-0.75$, that is the
ISM is less concentrated toward the center than the stars.

Before galactic wind, at the ages of 0.15, the gas is made of
molecular clouds and diffuse ISM. Therefore, we need the scale
lengths of MCs, dusty ISM and stars for which we assume $r_{c}^{MC}
= r_{c}^{*} = r_{c}^{M} = 0.5$ kpc. Afterwards, we need only the
scale lengths for ISM and stars. Both are kept unchanged (0.5 kpc).
In general we will assume $r_{c}^{*}=r_{c}^{MC}$ in spheroidals to
reduce the number of parameters.

In Table \ref{table_par}, columns (2) and (3), we summarize the set of
parameters we have used to model our test ellipticals.

\begin{figure}
\psfig{file=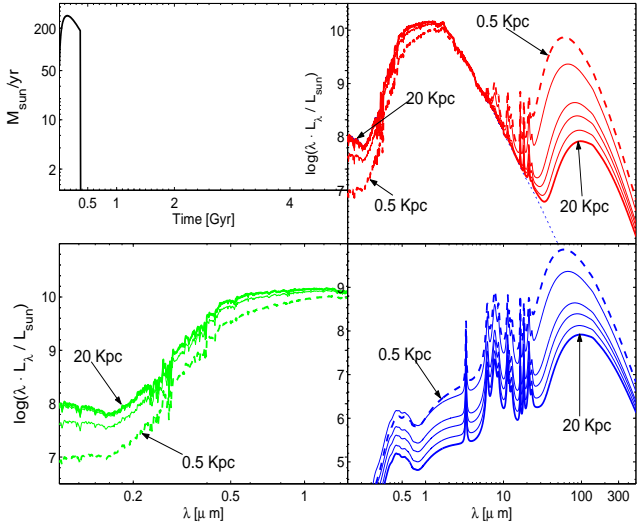,width=8.5truecm,height=7.0truecm}
\caption{\textbf{Top-left panel}: star formation history of the
prototype elliptical galaxy with $M_{L} = 10^{11}M_{\odot}$ at the
age of $13$ Gyr. \textbf{Top-right panel}: SEDs of elliptical
galaxies with radii of $20$, $10$, $5$, $3$, $1$ and $0.5$ kpc. The
full range of wavelengths is represented, from $0.1$ to $500 \mu m$.
\textbf{Bottom-left panel}: details on the UV-optical/NIR region of
the SEDs. \textbf{Bottom-right panel}: contribution to the total
emission of the diffuse ISM for the different radii.}
\label{RGALELLITTICA}
\end{figure}

At the age of $13$ Gyr only the geometrical parameters need to be
discussed. First we consider the galaxy radius $R_{gal}$, which is
let vary from $20$ to $0.5$ kpc, i.e. from very expanded to very
compact systems. In Fig. \ref{RGALELLITTICA}, we show the resulting
SEDs together with the emission in the UV-optical region
(bottom-left panel) and the contribution to the total flux  by the
diffuse ISM (bottom-right panel). At decreasing size of the galaxy,
the optical depth increases: the dimension of the galaxy scales
linearly, the density of  matter and the numerical density of $H$
atoms $n_{H}$ increase as $\varpropto r^{-3}$ and, therefore, the
optical depth $\tau$ increases  as $\varpropto r^{-2}$. From the
top-right and bottom-left panels of Fig. \ref{RGALELLITTICA} we can
see  the effect of it on the UV-optical/NIR part of the spectrum:
the smaller the dimension, the stronger is the extinction of the
UV-optical light. For the same reason the emission of dust in the
FIR becomes stronger at decreasing  dimensions because of the higher
density (bottom-right panel of Fig. \ref{RGALELLITTICA}). The
emission, in fact, linearly depends  on the numerical density of $H$
atoms.

As the mass of the physical components of the galaxy and all their
scale lengths are fixed (in our example all equal to $0.5$ Kpc), in
models of smaller size more gas, stars and dust are stored in the
outer regions with respect to the larger size models. In such a case
the  matter distribution tends to become uniform, a limit situation
reached when $r_{c}\gg R_{gal}$. For compact objects, the peak of
emission in the FIR also shifts toward shorter wavelengths. This is
due to the higher temperature of the grains, thanks to the combined
effect of the higher local emission and the presence of more sources
in the outer regions (bottom-right panel of Fig.
\ref{RGALELLITTICA}).

\begin{table*}
\centering \caption{Columns (2) through (5): parameters  for  a sample of model
galaxies, namely two ellipticals (E) at the ages of $13$ and $0.15$ Gyr, a disk
(D) at the age of $13$ Gyr and a starburst  (SB) at the age of $13$ Gyr. Columns (6)
through (11) best-fit parameters of  the models we have adopted to match the SEDs
of the spiral
galaxies M$100$ and NGC $6946$, the elliptical galaxies
NGC$2768$ and NGC$4494$, and finally  the local starburst galaxies
Arp$220$ and M$82$.} \vspace{0.2truecm}
\begin{tabular*}{150mm}{l|cccc|cccccc}
  \hline
  (1)                                         &(2)  &(3)  &(4)  &(5)       &(6)            &(7)            &(8)    &(9)  &(10)  &(11)\\
  \hline
  Model -- Galaxy                             &   E & E   & D   & SB       & M$100$        & NGC$6946$     & NGC$2768$ & NGC$4494$ & Arp$220$ & M$82$  \\
  \hline
  Age\footnotemark[1]                                         &  13 & 0.15& 13  & 13       & 13            & 13            & 13    & 13  & 13   & 13\\
  \hline
  $D$\footnotemark[2]                         & --  & --  & --  & --       & 19            & 5.5           & 21.5  & 20  & 76   & 3.25 \\
  $M_{L}$\footnotemark[3]                   & 1   &  1  & 1   & 1        & 2             & 1.2           & 2.2   & 2   & 1.35 & 0.18 \\
  $t_{0}$\footnotemark[4]                     & 0   & 40  & 5   & 15 or 30 & 5             & 3             & 0     & 0   & 80   & 100  \\
  $f_{M}$\footnotemark[5]                     & 1   & 0.3 & 0.5 &  0.5     & 0.35          & 0.55          & 1     & 1   & 0.5  & 0.9  \\
  $r_{c}^{*}/r_{c}^{M}$\footnotemark[6]       & 0.5 & 0.5 & --  & 0.5      & --            & --            & 0.5   & 0.5 & 1.0  & 0.2  \\
  $r_{c}^{*} = r_{c}^{MC}$\footnotemark[7]    & 0.5 & 0.5 & --  & 0.5      & --            & --            & 0.5   & 0.5 & 0.5  & 0.35 \\
  $R_{d}^{M}$\footnotemark[8],                & --  & --  & 5   & --       & 5             & 5             & --    & --  & --   & --   \\
  $R_{d}^{*} = R_{d}^{MC}$\footnotemark[9]    & --  & --  & 5   & --       & 5             & 5             & --    & --  & --   & --   \\
  $z_{d}^{M}$\footnotemark[10]                 & --  & --  & 0.4 & --       & 0.5           & 1             & --    & --  & --   & --   \\
  $z_{d}^{*} = z_{d}^{MC}$\footnotemark[11]   & --  & --  & 0.4 & --       & 0.5           & 1             & --    & --  & --   & --   \\
  $R_{gal}$ \footnotemark[12]                 & 10  & 10  & 10  & 10       & 19            & 13            & 20    & 12  & 17   & 10   \\
  $\tau$\footnotemark[13]                     & 0.1 & 0.1 &  4  & 9        &  4            & 5             &  0.1  & 0.1 & 9    & 9    \\
  $\nu$\footnotemark[14]                      & 6   & 6   & 0.7 & 1        &  0.7          & 0.7           &  2    & 2   &  0.4 & 1.2  \\
  $t_{BUR}$\footnotemark[15]                  & --  & --  & --  &  12.95   & --            & --            & --    & --  &12.95 & 12.95\\
  $\Delta t_{BUR}$\footnotemark[16]           & --  & --  & --  &  0.05    & --            & --            & --    & --  &0.05  &  0.05\\
  $\nu_{BUR}$\footnotemark[17]                & --  & --  & --  &  30      & --            & --            & --    & --  & 60   &  5   \\
  $\Theta$\footnotemark[18]                   & --  & --  & 60$^{\circ }$  & --       & 27$^{\circ }$ & 35$^{\circ }$ & --    & --  & --   & --   \\
  \hline
\end{tabular*}
\vspace{0.2truecm}
\begin{minipage}{\textwidth}
\footnotesize$^{1}${Age in Gyr.}
\footnotesize$^{2}${Distance of the galaxy in Mpc.}
\footnotesize$^{3}${Baryonic mass of the galaxy in units of $10^{11} M_{\odot}$.}
\footnotesize$^{4}${Evaporation time in Myr of young SSPs from dusty
parental regions. In the $13$ Gyr models and the galaxies NGC$2768$
and NGC$4494$ with no ongoing star formation,   the parameter
is set equal to $0$.}
\footnotesize$^{5}${Fractionary gas content in
the diffuse ISM.}
\footnotesize$^{6}${Ratio between the core radii
of stars and ISM in the spherical models.}
\footnotesize$^{7}${Scale
radii in kpc of stars and MCs in the spherical models.}
\footnotesize$^{8}${Radial scale in kpc of the diffuse ISM in the
disk models.}
\footnotesize$^{8}${Radial scales in kpc of stars and
MCs in the disk models.} \footnotesize$^{9}${Vertical scale in kpc
of the diffuse ISM in the disk models.}
\footnotesize$^{10}${Vertical scales in kpc of stars and MCs.}
\footnotesize$^{11}${Radius of the galaxy in kpc.}
\footnotesize$^{12}${Infall time scale in Gyr.}
\footnotesize$^{13}${Efficiency of star formation.}
\footnotesize$^{14}${Age of the beginning of the burst for starburst
models in Gyr.} \footnotesize$^{15}${Length of the burst for
starburst models.} \footnotesize$^{16}${Multiplying factor for the
star formation for starburst models.}
\footnotesize$^{17}${Inclination angle.}
\end{minipage}
\label{table_par}
\end{table*}

\begin{figure*}
\centerline{
\psfig{file=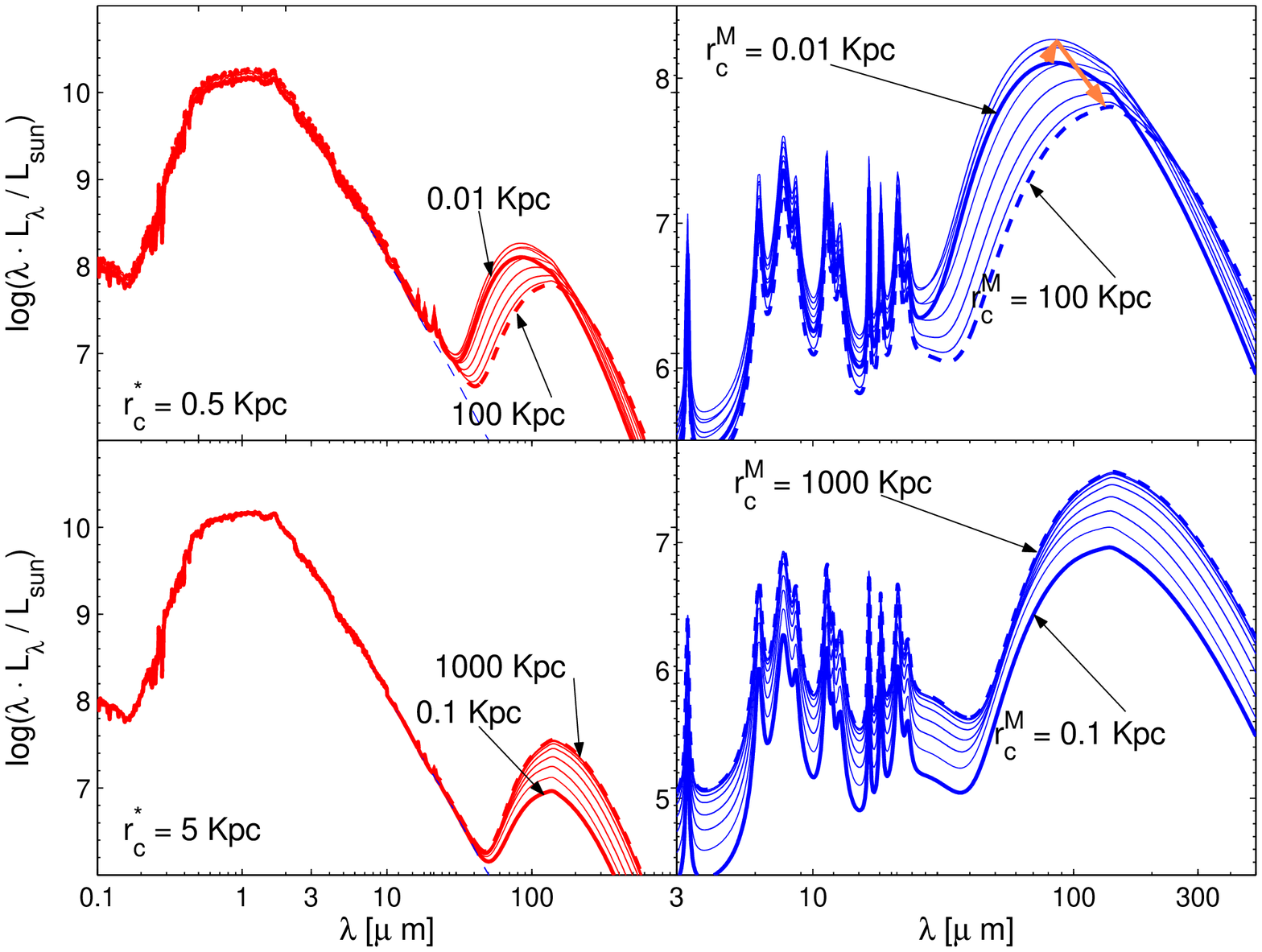,width=8.5truecm,height=6.0truecm}
\psfig{file=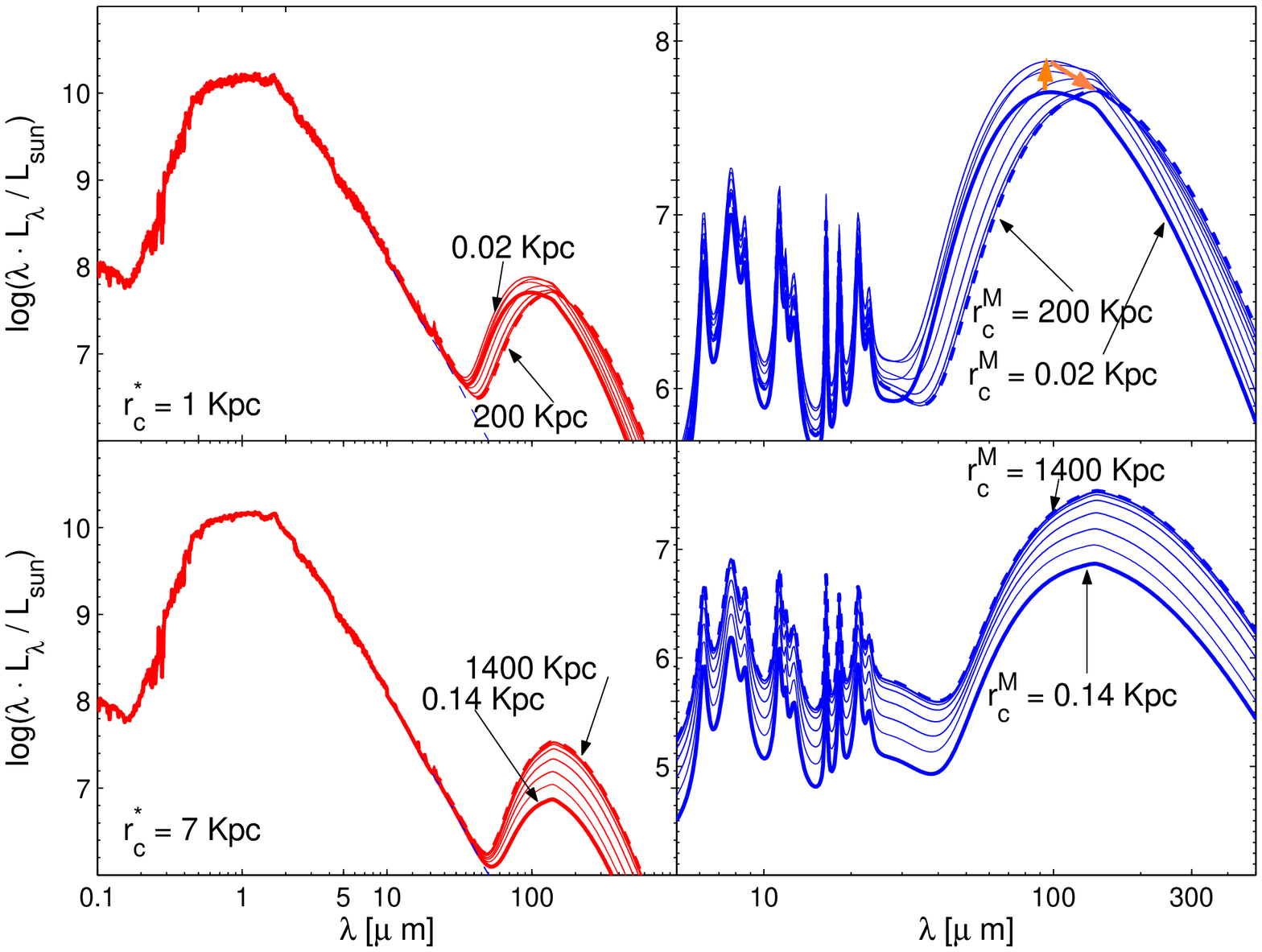,width=8.5truecm,height=6.0truecm}}
\caption{{\bf Left Panels}: the SEDs of the test elliptical galaxy
at the age of $13$ Gyr for  $r_{c}^{*}=0.5, 1, 5$ and $7$ kpc. For
each of these the ratio $r_{c}^{*}/r_{c}^{M}$ is varied  from
$0.005$ to $50$. The corresponding values of $r_{c}^{M}$ are also
shown. {\bf Right Panels}: the contribution to the total emission by
the diffuse ISM. The arrows show the evolution of the cirrus
emission going from the smaller to the higher values of $r_{c}^{M}$.
All the other parameters are kept fixed.} \label{RCSTARVARIARCC}
\end{figure*}

\begin{figure}
\psfig{file=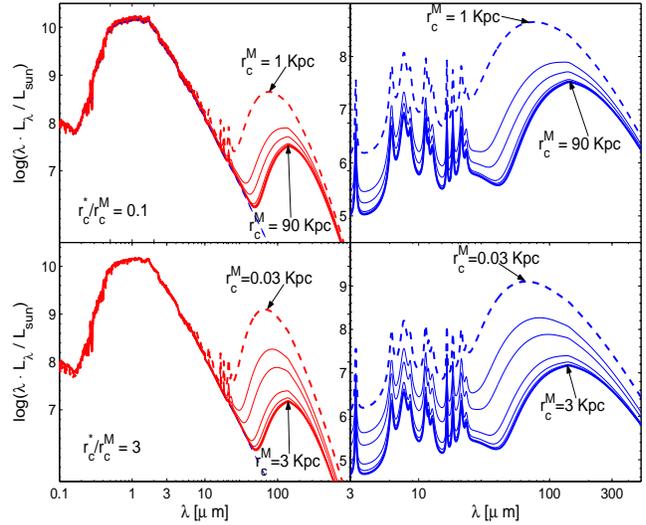,width=8.5truecm,height=7.0truecm}
\caption{In the \textbf{top-left} and \textbf{bottom-left} panels we
show the total emission of the prototype elliptical galaxy at the
age of 13 Gyr. The ratio $r_{c}^{*}/r_{c}^{M}$ between the scale
lengths is kept fixed and two values are considered:
$r_{c}^{*}/r_{c}^{M} = 0.1$ and $r_{c}^{*}/r_{c}^{M} = 3$. In both
cases $r_{c}^{*}$ is then varied from $0.1$ to $9$ kpc. The
corresponding value of $r_{c}^{M}$ is shown. In the \textbf{right
panels (top and bottom)}, the contribution to the total emission of
the diffuse ISM is shown in detail.} \label{RATIORCSTARRCCLOUD}
\end{figure}

Other geometrical parameters are the scale length $r_{c}^{*}$ and
the ratio $r_{c}^{*}/r_{c}^{M}$ that at fixed $r_{c}^{*}$ gives
$r_{c}^{M}$. We calculate a grid of models characterized by seven
values of $r_{c}^{*}$, i.e. $0.1$, $0.5$, $1$, $3$, $5$, $7$ and $9$
kpc ($R_{gal}$ is fixed at $10$ kpc), and ten values of the ratio
$r_{c}^{*}/r_{c}^{M}$, i.e. $0.005$, $0.05$, $0.1$, $0.2$, $0.4$,
$1$, $3$, $10$, $50$ and $100$, for a total of $170$ models covering
large intervals of $r_{c}^{*}$ and $r_{c}^{M}$. The results obtained
for various combinations of these parameters are presented in figs.
\ref{RCSTARVARIARCC} and \ref{RATIORCSTARRCCLOUD}. In Fig.
\ref{RCSTARVARIARCC}, we show the results obtained keeping fixed
$r_{c}^{*}$ and leaving free the ratio $r_{c}^{*}/r_{c}^{M}$, or
equivalently $r_{c}^{M}$. Four values of $r_{c}^{*}$ are considered.
For each of these on the left panel we show the total galaxy flux at
varying $r_{c}^{M}$, whereas on the right panel  we display the
corresponding contribution of the diffuse ISM to the total flux. For
the two smallest values of $r_{c}^{*}$, $0.5$ and $1$ kpc, at
varying $r_{c}^{M}$ the emission in the FIR grows, reaches a peak
value, and then declines. This  can be explained in the following
way. For $r_{c}^{M} \ll r_{c}^{*}$ and $r_{c}^{M} \gg r_{c}^{*}$,
the diffuse ISM is  concentrated either in the inner or in the outer
regions of the galaxy. In both cases the spatial distributions of
the ISM does not favors the interaction with the stellar radiation:
the FIR flux increases as $r_{c}^{M}$ grows from very small values
to $r_{c}^{M} \thickapprox r_{c}^{*}$, which represents the best
condition in which the intensity of the local radiation field is
strong  in high density  regions. Moving to $r_{c}^{M} \gg
r_{c}^{*}$ the flux decreases, because the dust is concentrated in
regions of weaker   radiation field. The emission peak shifts to
longer wavelenghts because for small $r_{c}^{*}$, as $r_{c}^{M}$
increases, the dust is confined in regions of lower radiation field
intensity. For the two highest values of $r_{c}^{*}$, $5$ and $7$
kpc, the flux simply increases as  dust goes from being concentrated
in the inner regions of the galaxy, to being shifted toward the
outer regions with higher density of stars.

In Fig. \ref{RATIORCSTARRCCLOUD} we keep fixed the ratio
$r_{c}^{*}/r_{c}^{M}$ and let  $r_{c}^{*}$  vary from $0.1$ to $9$
kpc. The result is the same for both  cases no matter whether
$r_{c}^{*}/r_{c}^{M} < 1$ or $> 1$. The FIR emission is stronger
when both  $r_{c}^{*}$ and $r_{c}^{M}$ are small, i.e. when stars
and dust occupy the same inner region of the galaxy. With high
values of the two radial scale lengths, stars and dust are more
dispersed across the galaxy, therefore the FIR emission is weaker
and the peak shifts toward longer wavelengths.

\begin{figure}
\psfig{file=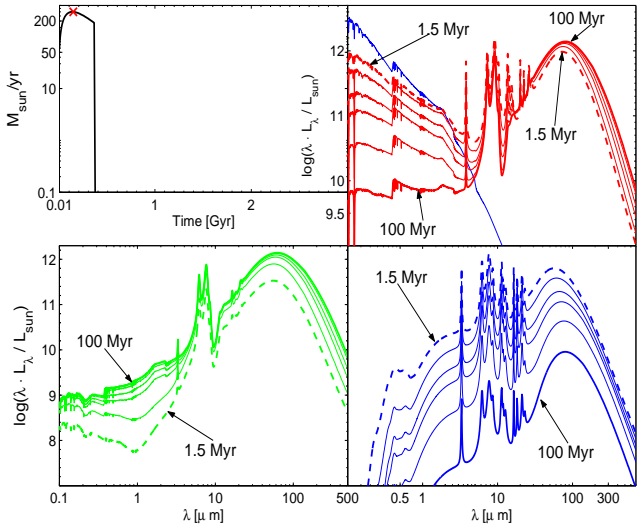,width=8.5truecm,height=7.0truecm}
\caption{\textbf{Top-left panel}: star formation history of the
prototype elliptical galaxy of $M_{L} = 10^{11}M_{\odot}$ with a
mark on the modelled age at $0.15$ Gyr. \textbf{Top-right panel}:
SEDs of ellipticals at varying the  time scale $t_{0}$ from $1.5$
Myr to $100$ Myr. The considered values are $1.5$, $4.5$, $15$,
$30$, $60$ and $100$ Myr. All the other parameters are kept fixed.
We also show the SED of the same galaxy calculated with the
classical EPS technique. \textbf{Bottom-left panel}: contribution to
the total emission  by dusty MCs at varying $t_{0}$.
\textbf{Bottom-right panel}: contribution to the total emission of
the diffuse ISM for the same set of time scales $t_{0}$.}
\label{T0ELLITTICA}
\end{figure}

To show the effects of $t_{0}$, the time scale for a young SSP to
evaporate the MC in which it is embedded, and $f_{M}$, the
fractionary gas mass in the diffuse ISM, we take the $0.15$ Gyr old
model galaxy, at the peak of star formation. In Fig.
\ref{T0ELLITTICA}, we show the SEDs for several values of $t_{0}$.
The effect of this parameter is strong and straightforward: the
longer  this time scale, the higher is the obscuration of the
UV-optical light because the energy of  newly born young and
luminous objects is long trapped by the dusty environment (top-right
panel). Simultaneously, more energy is emitted in the FIR by the
dusty MCs regions (bottom-left panel). In any case it is clear from
the top-right and the bottom-right panel that an increase  of
$t_{0}$, corresponds to a decrease of the emission by the diffuse
ISM, whose peak also shifts toward longer wavelengths. The
explanation is that increasing $t_{0}$ the UV-optical intensity of
the local radiation field  dramatically drops, leaving a cooler and
weaker mean radiation field to heat the dust.

\begin{figure}
\psfig{file=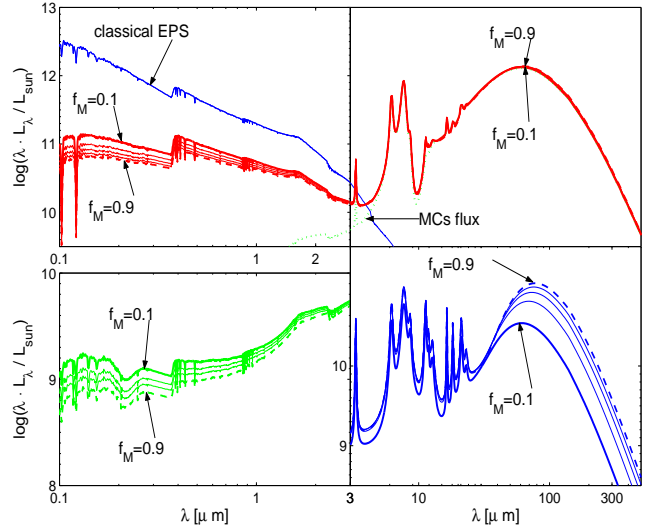,width=8.5truecm,height=7.0truecm}
\caption{\textbf{Top panels}: SEDs of test ellipticals at varying
the fraction of gas in the diffuse ISM $f_{M}$ with respect to the
total amount of gas. The range of values goes from $0.1$ to $0.9$.
All the other parameters are kept fixed. The considered values are
$0.1$, $0.3$, $0.5$, $0.7$ and $0.9$. We also show the SED of the
same galaxy calculated with the classical EPS technique.
\textbf{Bottom-left panel}: contribution to the total emission by
dusty MCs at varying $f_{M}$. \textbf{Bottom-right panel}:
contribution to the total emission of the diffuse ISM for the same
set of time scales $f_{M}$.} \label{FMCELLITTICHE}
\end{figure}

Finally, we compare in Fig. \ref{FMCELLITTICHE} the SEDs of our test
galaxy at varying the amount of gas in the diffuse ISM, i.e. $f_M$.
The remaining  gas is stored in MCs. As shown by the top panels,
increasing the amount of gas in the diffuse ISM has little influence
on the global SED. The reason  is that the galaxy is at the peak of
star formation, where the attenuation and IR emission are clearly
dominated by the star forming MCs. Similar albeit smaller effects of
$f_{M}$ are expected in starburst galaxies with high star formation
rate. In any case, as expected, increasing the amount of gas in the
diffuse ISM yields stronger attenuation of the light emitted by old
and young stars (top and bottom left panels). The emission by the
diffuse ISM (which anyhow scarcely contributes to the total flux)
increases with the amount of gas and also becomes cooler because of
the weaker local radiation field heating the grains (bottom right
panel).

\subsubsection{Late-type galaxies}\label{par_examples_late}

We consider a disk galaxy of $10^{11}\, M_{\odot}$ at the age of 13
Gyr. In our model  disk galaxy (i.e. with  infall scheme and the
Schmidt law of star formation) and likely in real disk galaxies as
well, the star forming activity never ceases. Due to the lack of
significant galactic winds in disk galaxies, the gas remain for ever
in the disk and continuously refuels star formation. This means that
our model has to take all possible components into account even at
the age of 13 Gyr: young stars just formed and still embedded in
their parental molecular cloud, bare stars of any age and diffuse
ISM. The task is more complicate than with the spheroidal models due
to the lower degree of symmetry (only the azimuthal one).

The choice of the three main parameters driving the star formation
and chemical enrichment of a disk galaxy, i.e. the infall time scale
$\tau$, the exponent $k$, and the efficiency $\nu$ of the star
formation rate, rests on the following considerations. In the star
formation rate, $k$ typically varies from 1 \citep{Schmidt63}to 2
\citep{Larson91}. In our model for a late-type galaxy the choice is
guided by the \citet{Portinari98} results for the Solar Vicinity.
Therefore we adopt $k=1$, and $\nu =0.7$, whereas for the infall
time scale we assume $\tau = 4$. The resulting star formation rate
starts small, grows to a maximum at about $3.5$ Gyr, and then gently
declines to the present day value.

\begin{figure}
\psfig{file=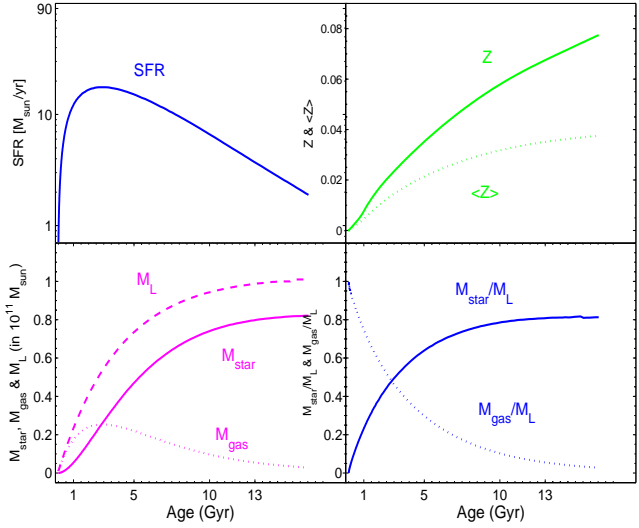,width=8.5truecm,height=7.0truecm}
\caption{Basic quantities of the chemical model for the test
late-type galaxy as function of the age: the top left panel shows
the star formation rate in $M_\odot$/yr; the top right panel
displays the maximum ($Z$, solid line) and mean metallicity
($\langle Z \rangle $, dotted line); the bottom left panel shows the
mass of living stars $M_{star}$ (solid line), the gas mass $M_{gas}$
(dotted line), and the total mass of baryons $M_{L}$ (dashed line);
finally the bottom right panel displays the ratios $M_{star} /
M_{L}$ (solid line) and $M_{gas} / M_{L}$ (dotted line). All masses
are in units of $10^{11}\,M_\odot$. Ages are in Gyr.}
\label{chimicaSpi}
\end{figure}

The temporal evolution of a few physical quantities characterizing
the model galaxy is shown in Fig. \ref{chimicaSpi}.

\begin{figure}
\psfig{file=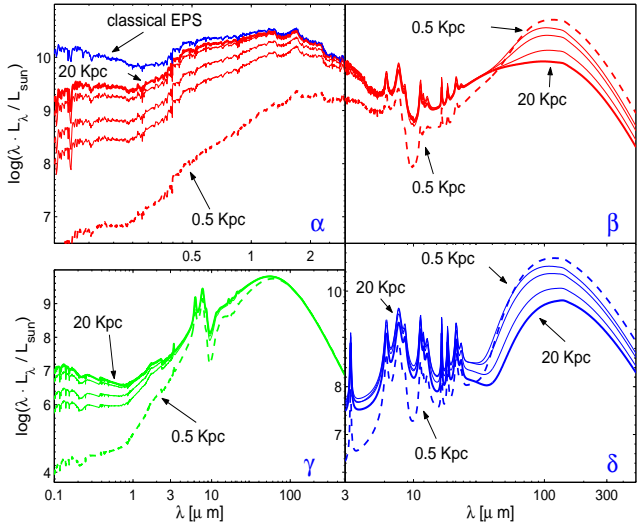,width=8.5truecm,height=7.0truecm}
\caption{\textbf{Top panels}: SEDs of prototype disk galaxies at
varying the radius of the galaxy $R_{gal}$, going from $0.5$ to $20$
Gyr. All the other parameters are kept fixed. We also show the SED
of the test galaxy calculated with the classical EPS technique.
\textbf{Bottom-left panel}: contribution to the total emission
coming from dusty MCs at varying $R_{gal}$. \textbf{Bottom-right
panel}: contribution to the total emission of the diffuse ISM for
the same set of radii $R_{gal}$.} \label{RGALSPIRALE}
\end{figure}

To proceed further other parameters must be specified. First, we fix
the geometrical parameters, i.e. the radial and vertical scale
heights, for which we make the most simple choice, i.e.
$R_{d}^{*}=R_{d}^{MC}=R_{d}^{M}=5$ kpc and
$z_{d}^{*}=z_{d}^{MC}=z_{d}^{M}=0.4$ kpc (the spatial scales for the
three components are the same) in agreement with typical values
found for local spiral galaxies such as M100 and M51
\citep{Beckman96}. Second, we need to specify the fraction $f_{M}$
of the gas present in form of diffuse ISM. The task is not easy and
to a certain extent beyond the aims of this study because it would
require a multi-component description taking into account all
heating and cooling processes that transform atomic likely hot gas
into cool molecular gas. So, as already presented in Sect.
\ref{par_chosen}, we consider $f_{M}$ as a parameter, and, for the
sake of illustration, we adopt $f_{M} = 0.5$. This means that half
of the gas is in the diffuse medium and half  in the  MCs. Third,
for the timescale $ t_{0}$ we assume $\simeq 5$ Myr which roughly
corresponds to the lifetime of the most massive stars. This means
that the heavily obscured part of young stars lifetime is rather
short. For the sake of illustration we consider the case with
inclination angle $\Theta = 60^{\circ }$.

In Table \ref{table_par}, column (4),  we summarize the set of
parameters we have used to model our test disk galaxy.

In the same way we did for the ellipticals we calculated a sequence
of models with $R_{gal}$ going from $20$ to $0.5$ kpc, from very
expanded to very compact systems. In Fig. \ref{RGALSPIRALE} we show
the resulting SEDs (top panels $\alpha$ and $\beta$) also with the
spectrum calculated with the classical EPS. In the bottom panels we
show the emission in the UV-optical region (bottom-left panel)
coming from MCs and the contribution to the total flux coming from
the diffuse ISM (bottom-right panel). The effect of the radius is
the same as for ellipticals: for smaller galaxies, higher optical
depths, the stronger will be the extinction of the UV-optical light.
For the same reason the emission of dust in the FIR will be stronger
for the smaller dimensions because of the higher density
(bottom-right panel of Fig. \ref{RGALSPIRALE}).

\begin{figure*}
\centerline{
\psfig{file=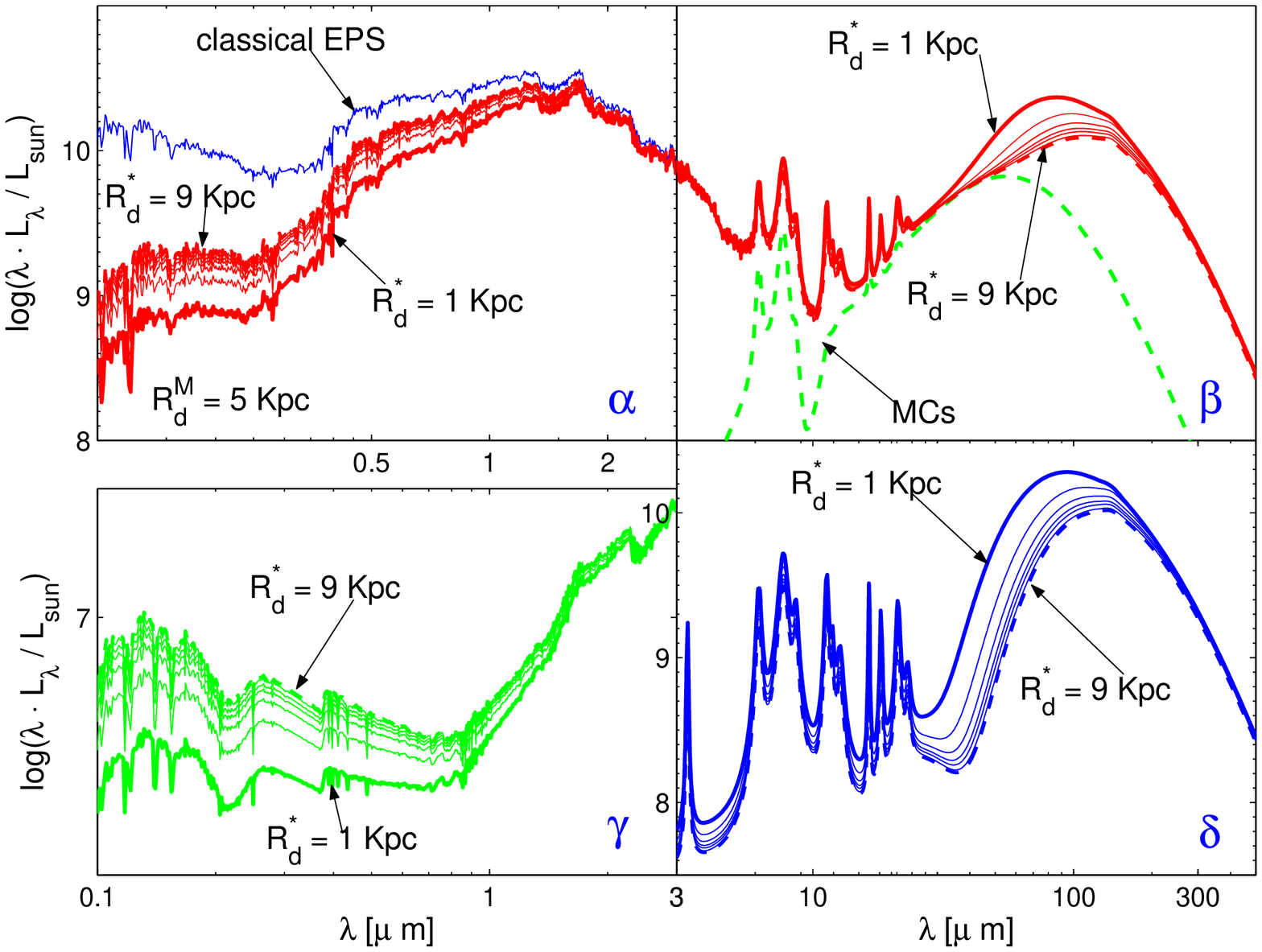,width=8.5truecm,height=7.0truecm}
\psfig{file=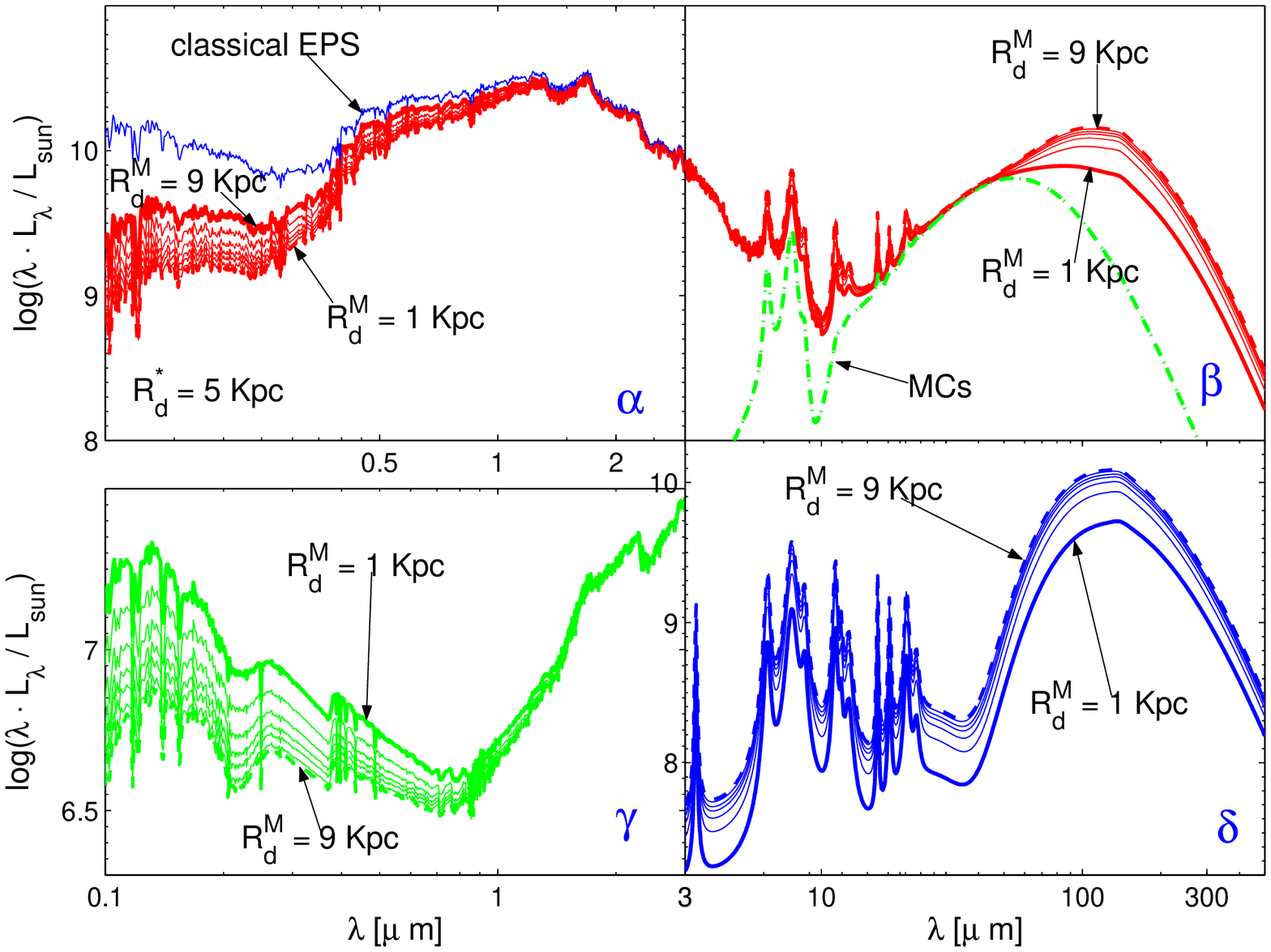,width=8.5truecm,height=7.0truecm}}
\caption{{\bf Four Left Panels}: in panels $\alpha$ and $\beta$ we
show the SEDs of the 13 Gyr test disk galaxy with fixed $R_{d}^{M} =
5$ kpc. The radial scale length $R_{d}^{*}$ is varied from $1$ to to
$9$ kpc. The following  values are considered, $1$, $2$, $3$, $4$,
$5$, $7$ and $9$ kpc. The SED obtained with the classical EPS
technique is also shown in panel $\alpha$ for the sake of
comparison. The dashed line in panel $\beta$ represents the
contribution of MCs to the total flux in the FIR. In panel $\gamma$
we show the effect of varying $R_{d}^{*}$ on the MCs flux. In
particular the uv-optical/NIR spectral region is shown. Finally, in
panel $\delta$ we plot the contribution of the diffuse ISM to the
total emission. {\bf Four Right Panels}: the same as above but for
fixed $R_{d}^{*}$ and varying $R_{d}^{M}$. $R_{d}^{M}$ goes from $1$
to $9$ kpc.} \label{SPIRALIRDSTARRDDUST}
\end{figure*}

\begin{figure*}
\centerline{
\psfig{file=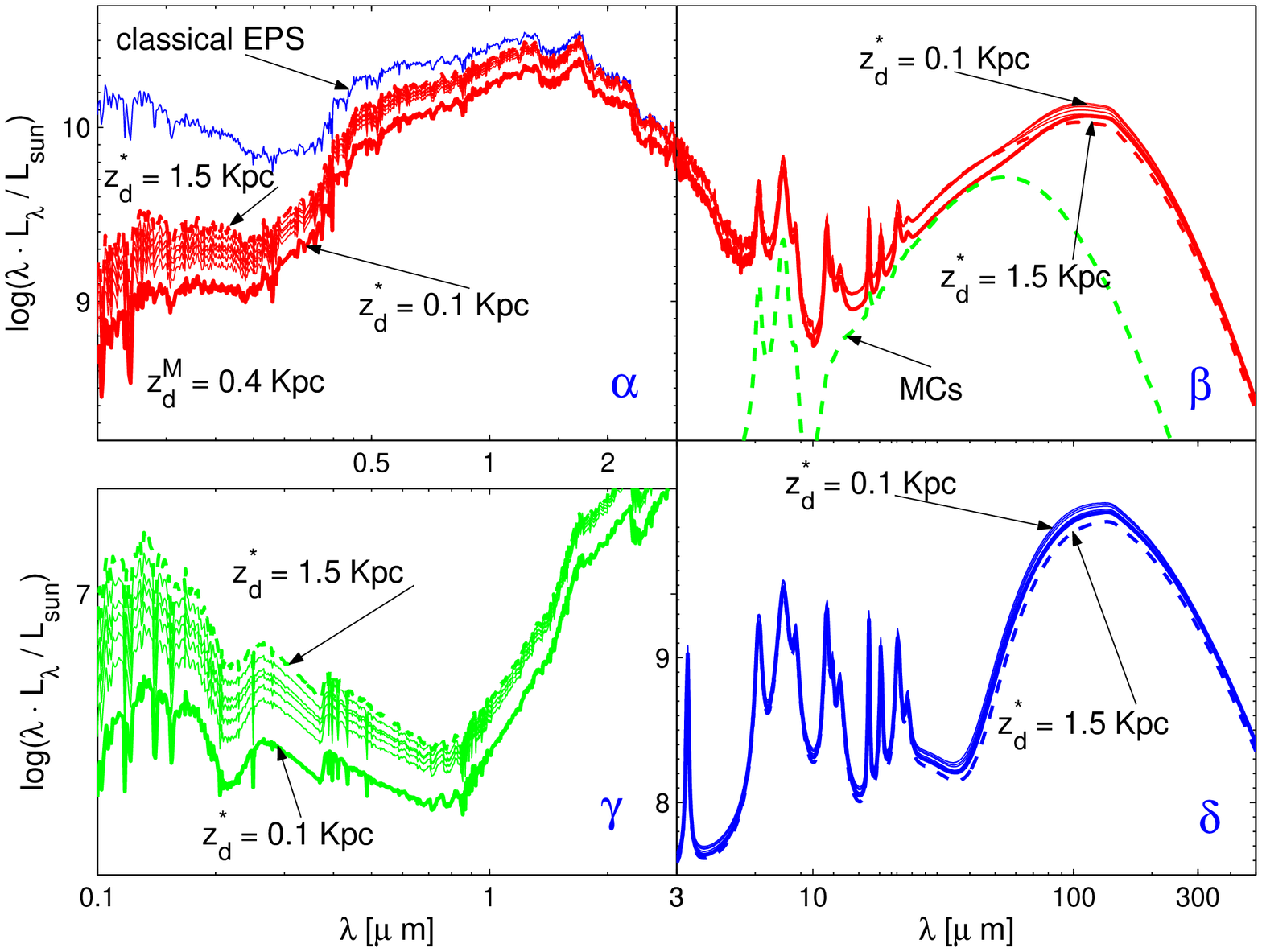,width=8.5truecm,height=7.0truecm}
\psfig{file=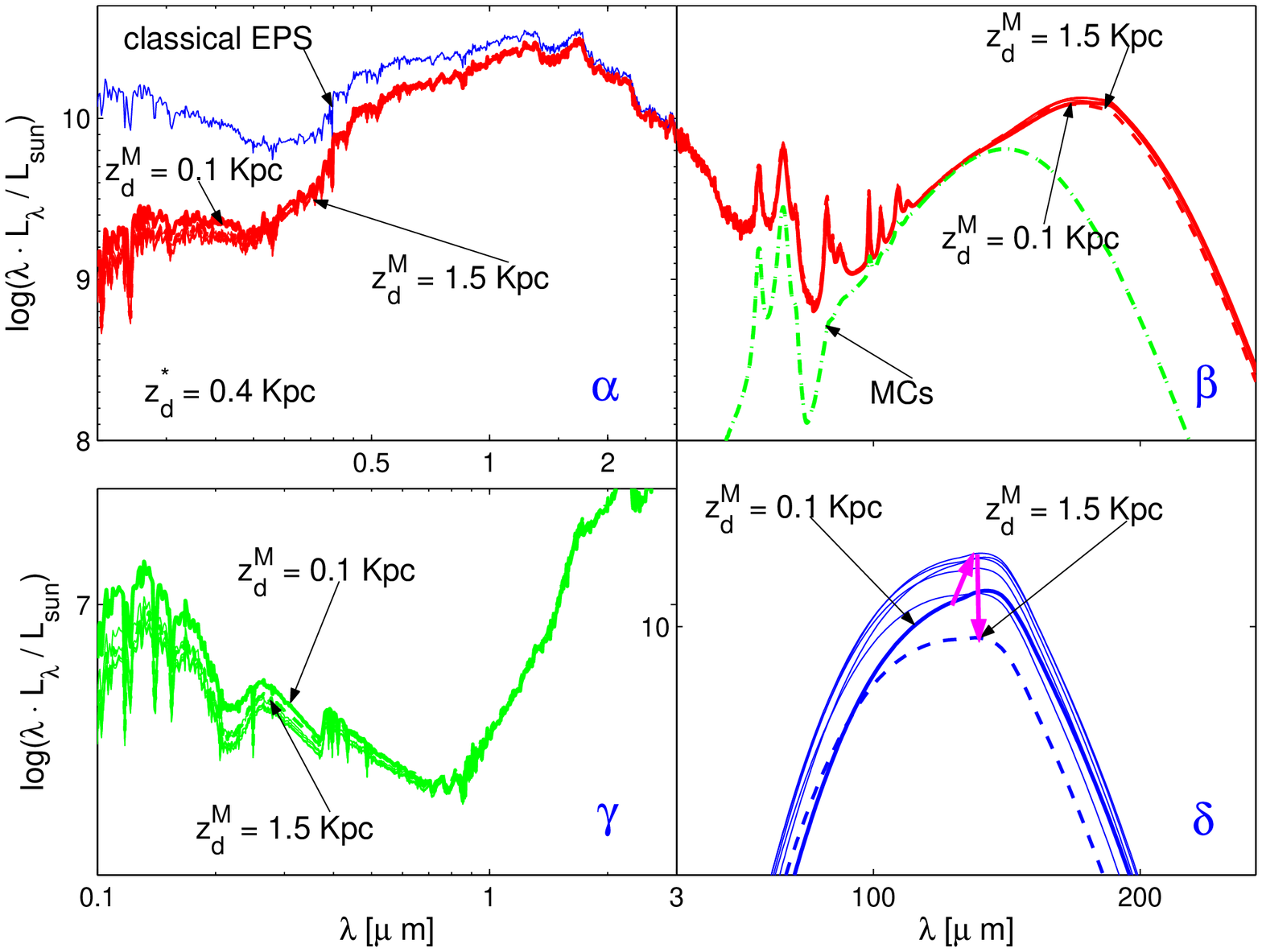,width=8.5truecm,height=7.0truecm}}
\caption{{\bf Four Left Panel}: in panels $\alpha$ and $\beta$ we
show the SEDs of the test disk galaxy of $13$ Gyr for the fixed
value of the vertical scale length $z_{d}^{M} = 0.4$ kpc. The scale
length $z_{d}^{*}$ is varied  from $0.1$ to to $1.5$ kpc. The
considered values for $z_{d}^{*}$ are $0.1$, $0.25$, $0.4$, $0.55$,
$0.7$, $1$ and $1.5$ kpc. The SED obtained with the classical EPS
technique is also shown in panel $\alpha$ for the sake of
comparison. The dashed line in panel $\beta$ represents the emission
of the MCs in the FIR. In panel $\gamma$ we show the effect of
varying $z_{d}^{*}$ on the MCs flux. In particular the
uv-optical/NIR spectral region is shown. Finally, in panel $\delta$
we plot the contribution of the ISM to the total emission. {\bf Four
Right Panels}: the same as above but for $z_{d}^{*}$ fixed and
$z_{d}^{M}$ varying from $0.1$ to $1.5$ kpc.}
\label{SPIRALIZDSTARZDDUST}
\end{figure*}

In the case of disk galaxies the degree of symmetry is lower than
for spheroidal systems as only azimuthal symmetry is conserved. In
Fig. \ref{SPIRALIRDSTARRDDUST}, we show  the SED at varying  the
radial scale lengths $R_{d}^{M}$ and $R_{d}^{*}$. In the four left
panels we keep fixed $R_{d}^{M}$ and vary $R_{d}^{*}$, the opposite
in the four right panels. Examining the four panels on the left, we
notice that once fixed $R_{d}^{M}$, the attenuation becomes weaker
going from low to high values of $R_{d}^{*}$ (panels $\alpha$ and
$\beta$). This is simply due to the fact that growing the value of
$R_{d}^{*}$, more stars are distributed in the outer regions of the
galaxy and thus they are less obscured. It is worth noticing how in
our model disk galaxy the attenuation is partially due to young MCs
and partially to the diffuse ISM. The ultimate reason of this is the
low evaporation time for dusty regions we have chosen for the model.
In panel $\gamma$ we see the effect of attenuation on the emission
of MCs: again it is stronger for $R_{d}^{*} < R_{d}^{M}$, because
for these values young dusty SSPs are confined in the inner region
$\left( R_{d}^{MC} = R_{d}^{*} \right)$. The effect of varying the
star scale on the diffuse ISM emission (panel $\delta$) is that this
emission becomes weaker and cooler at the increasing of $R_{d}^{*}$,
because more stars are distributed in the outer regions of the
galaxy weakly heating  the ISM. Passing now to the four right panels
the above effects are reversed. $R_{d}^{*}$ is fixed and the
obscuration of the stellar light by the diffuse ISM (panels $\alpha$
and $\gamma$) grows at the increasing of $R_{d}^{M}$, because the
ISM shifts toward the outer regions more and more wrapping the
stellar component. In the same way the emission of the ISM becomes
stronger at the increasing of $R_{d}^{M}$, because dust is more
evenly distributed in the regions occupied by stars.

In Fig. \ref{SPIRALIZDSTARZDDUST} we show the effect of the vertical
scale lengths. As we did for the radial scales, we consider two
cases: first we fix $z_{d}^{M}$ and  let $z_{d}^{*}$ vary, second we
fix $z_{d}^{*}$ and let $z_{d}^{M}$  change. In the four left panels
$z_{d}^{M}$ is fixed. At increasing $z_{d}^{*}$ the extinction
becomes lower, because more stars are distributed in the outer
regions of the galaxy ($\alpha$ panel). This effect of growing
attenuation is the same on young dusty SED (panel $\gamma$) that in
this model are distributed with the same vertical scale of bare
stars. The ISM emission tends to be stronger for lower values of
$z_{d}^{*}$ (panel $\delta$). In the four right panels of Fig.
\ref{SPIRALIZDSTARZDDUST} the vertical scale of stars $z_{d}^{*}$ is
kept fixed. In this case the effect of varying the other parameter
is smaller: in particular in panel $\delta$ we can see as the
emission of the diffuse ISM grows to a maximum and then decreases.
The maximum is reached when the scale lengths of stars and dust are
similar. If $z_{d}^{M} \ll z_{d}^{*}$, or $z_{d}^{M} \gg z_{d}^{*}$
the emission tends to be lower.

\begin{figure}
\psfig{file=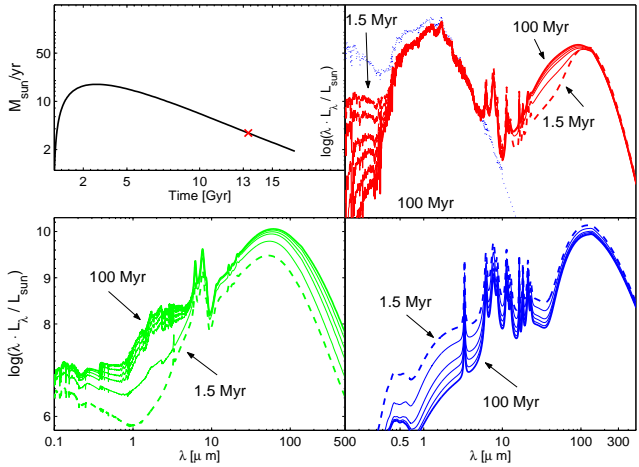,width=8.5truecm,height=6.2truecm}
\caption{\textbf{Top-left Panel}: star formation history of the
prototype disk galaxy with $M_{L} = 10^{11}M_{\odot}$. The age of
the 13 Gyr model whose SEDs are examined in detail is marked.
\textbf{Top-right Panel}: SEDs of the galaxy at varying the
evaporation time scale $t_{0}$  from $1.5$ Myr to $100$ Myr. The
considered values are  $1.5$, $4.5$, $15$, $30$, $60$ and $100$ Myr.
All the other parameters are kept fixed. We also show the SED
calculated with the classical EPS technique. It steeply declines for
wavelengths longer than $2 \mu m$. \textbf{Bottom-left Panel}:
contribution to the total emission  from dusty MCs at varying
$t_{0}$. \textbf{Bottom-right panel}: contribution to the total
emission of the diffuse ISM at varying $t_{0}$.} \label{T0SPIRALI}
\end{figure}

In Fig. \ref{T0SPIRALI}, we show the effect of varying the
evaporation time $t_{0}$ of young dusty SSPs. The longer the time
scale, the higher is the effective extinction of the UV-optical
light because the energy of newly stars is obscured for a longer
time (top-left panel), and the higher is the amount of energy
shifted toward the FIR by MCs (bottom-left panel). As for the
elliptical galaxy of $0.15$ Gyr, at increasing $t_{0}$ the emission
of the diffuse ISM decreases and becomes cooler, because the
intensity of the local radiation field is weaker.

\begin{figure}
\psfig{file=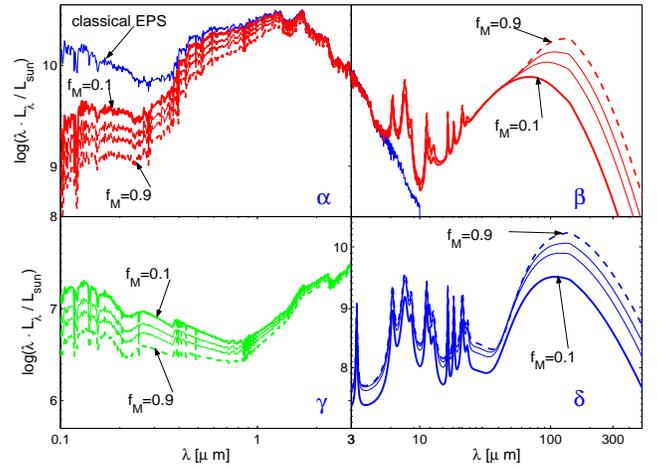,width=8.5truecm,height=6.2truecm}
\caption{\textbf{Top panels}: SEDs of model disk galaxies at varying
the fraction of gas in the diffuse ISM $f_{M}$ respect to the total
amount of gas. The range of values goes from $0.1$ to $0.9$. All the
other parameters are kept fixed. The considered values are $0.1$,
$0.3$, $0.5$, $0.7$ and $0.9$. We also show the SED of the prototype
spiral calculated with the classical EPS technique.
\textbf{Bottom-left panel}: contribution to the total emission
coming from dusty MCs at varying $f_{M}$. \textbf{Bottom-right
panel}: contribution to the total emission of the diffuse ISM for
the same set of $f_{M}$.} \label{FMCSPIRALI}
\end{figure}

Finally, we compare the SEDs of our  disk galaxy at varying the
amount of gas in the diffuse ISM with respect to that in dusty MCs,
using the parameter $f_M$. This is shown in Fig. \ref{FMCSPIRALI}.
The effect of $f_{M}$ can be easily seen in the top panels ($\alpha$
and $\beta$). Increasing $f_{M}$, i.e. the amount of gas in the
diffuse ISM, strongly affects  the  SED. The higher  the amount of
gas, the stronger is the obscuration, and the higher is the flux in
the FIR. As expected, increasing the amount of gas in the diffuse
ISM makes stronger also the effect of attenuation on the light from
young stars (bottom left panel). The emission of the diffuse ISM
becomes also cooler with more gas in the diffuse ISM, because of the
weaker radiation field heating the grains (bottom right panel).

\subsubsection{Starburst galaxies}\label{par_examples_starburst}

Starburst galaxies are objects that show a recent and transient
increase in SFR by a large factor (ten at least). The burst is often
confined to a few hundred parsecs near the nucleus, although bursts
extending to wider regions  are easy to find. The high SFR of
starburst galaxies is of great interest, because it is the local
analog of high redshift galaxies during their formation, involving
strong star formation in dust rich environments.

The light emitted by newly born stars in these star forming galaxies
is absorbed by dust and re-emitted in the FIR/MIR. One of the main
problem is to disentangle whether the dominant energy source heating
dust is a starburst or an AGN. In our models, AGNs as energy source
is not yet included. Future work is planned to include the AGN
contribution thus providing a more realistic description of these
systems. In Table \ref{table_par}, column (5), we summarize the set
of parameters we have adopted to model our prototype starburst
galaxy.

To simplify things we take a spherical model for which we adopt a
long infall time scale and relatively low star formation efficiency,
i.e. $\tau = 9$ and $\nu = 1$. In this model star formation is
maximum at about 3 Gyr, slowly declines up the the present time
($13$ Gyr) and never ceases. It is a sort of late type galaxy but
for the spherical symmetry. As already mentioned the specific shape
is irrelevant for the purposes of this experiment. At the age of
$12.95$ Gyr (not long ago) we introduce a short burst whose
intensity is  $30$ times stronger than the current SFR at the age of
$12.95$ Gyr.

Star formation in starbursters is tightly related to reddening and
obscuration which render the detailed interpretation of their
continuum and emission-lines very complicate. Starbursters are known
to possess a flat obscuration law (with no bump) as pointed out by
\citet{Calzetti94}, who considered a mix of stars and dust and took
the effects by scattering into account. This obscuration law is very
puzzling, because starbursters are typically objects of high
metallicity and so we would expect an extinction law  similar to
that of the MW. It is worth noticing, however, that the dust
obscuration in galaxies is not strictly equivalent to dust
extinction in  stars. The latter measures the optical depth of the
dust between the observer and the star, whereas the former
corresponds to a more general attenuation, to which many effects may
concur such as  those by extinction, scattering, and geometrical
distribution of the dust relative to the emitters
\citep{Calzetti01}. Because of this, there is nowadays much debate
on the origin of the flat law.  Is it due to geometrical effects
\citep{Granato00}, or peculiar distribution and the composition of
the dust in star forming environments? Furthermore, there is some
evidence that reddening by gas and stars are systematically
different \citep{Calzetti01}, as stars are on average less reddened
than ionized gas. Another point is that  both stars and ionized gas
are contrasted by  a foreground-like distribution of dust. This
could be due to dust being preferentially associated to the star
forming regions. In addition to this, the evidence of different
amounts of gas mass along different lines of sight, could mirror a
density and reddening structure. These considerations led
\citet{Calzetti01} to propose a model aimed at  explaining many
observational constraints: newly born stars form in a central high
density region immersed in a bath of older stars and dust.

In our models we can easily deal with this bath of old stars and
dust: setting $r_{c}^{*}=r_{c}^{M}$ (in spherical models) and/or
$R_{d}^{*}=R_{d}^{M}$ (in disks), the region of dust coincides with
the region in which stars are present. We can also allow for
different scales between old and newly born stars so that centrally
confined burst can be superimposed to a ''normal" galaxy, setting
$r_{c}^{*}\neq r_{c}^{MC}$ or $R_{d}^{*}\neq R_{d}^{MC}$. However,
as in all this paper, we will simply put $r_{c}^{*}=r_{c}^{MC}$ or
$R_{d}^{*}=R_{d}^{MC}$, leaving this point aside because it would
require a deep investigation beyond the purposes of this study.

\begin{figure}
\psfig{file=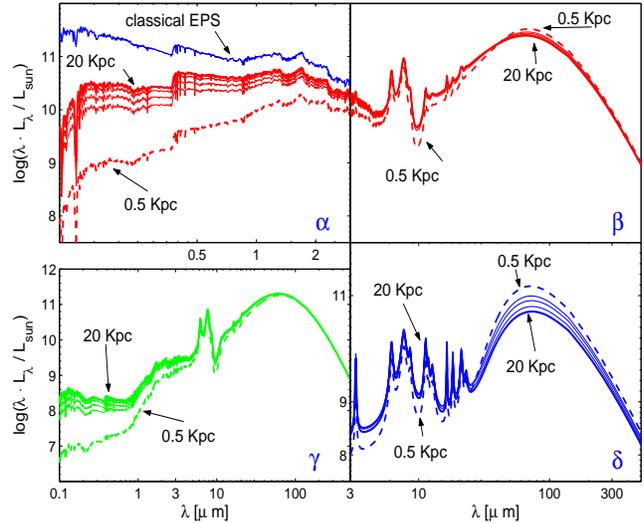,width=8.5truecm,height=7.0truecm}
\caption{\textbf{Top panels}: SEDs of the model star-burst galaxy at
varying $R_{gal}$ from $0.5$ to $20$ Gyr. The values are the same as
for the model elliptical and spiral. All the other parameters are
kept fixed. The SED derived with the classical EPS is also shown.
\textbf{Bottom-left panel}: contribution to the total emission by
dusty MCs at varying $R_{gal}$. \textbf{Bottom-right panel}:
contribution to the total emission by the diffuse ISM at varying
$R_{gal}$.} \label{RGALSTARBURST}
\end{figure}

As for our starburster galaxy we have adopted the spherical
symmetry, all scale lengths are the same as in model for early-type
galaxies. In Fig. \ref{RGALSTARBURST} we show the effect on the
total SED of varying the  galaxy radius $R_{gal}$. The effect is the
same as already found for the model elliptical and disk galaxies:
the smaller  the galaxy, the higher is the optical depth and the
stronger is the attenuation of the UV-optical light. The emission of
dust in the FIR is stronger for compact objects because of the
higher density (bottom-right panel of Fig. \ref{RGALSTARBURST}.

\begin{figure*}
\centerline{
\psfig{file=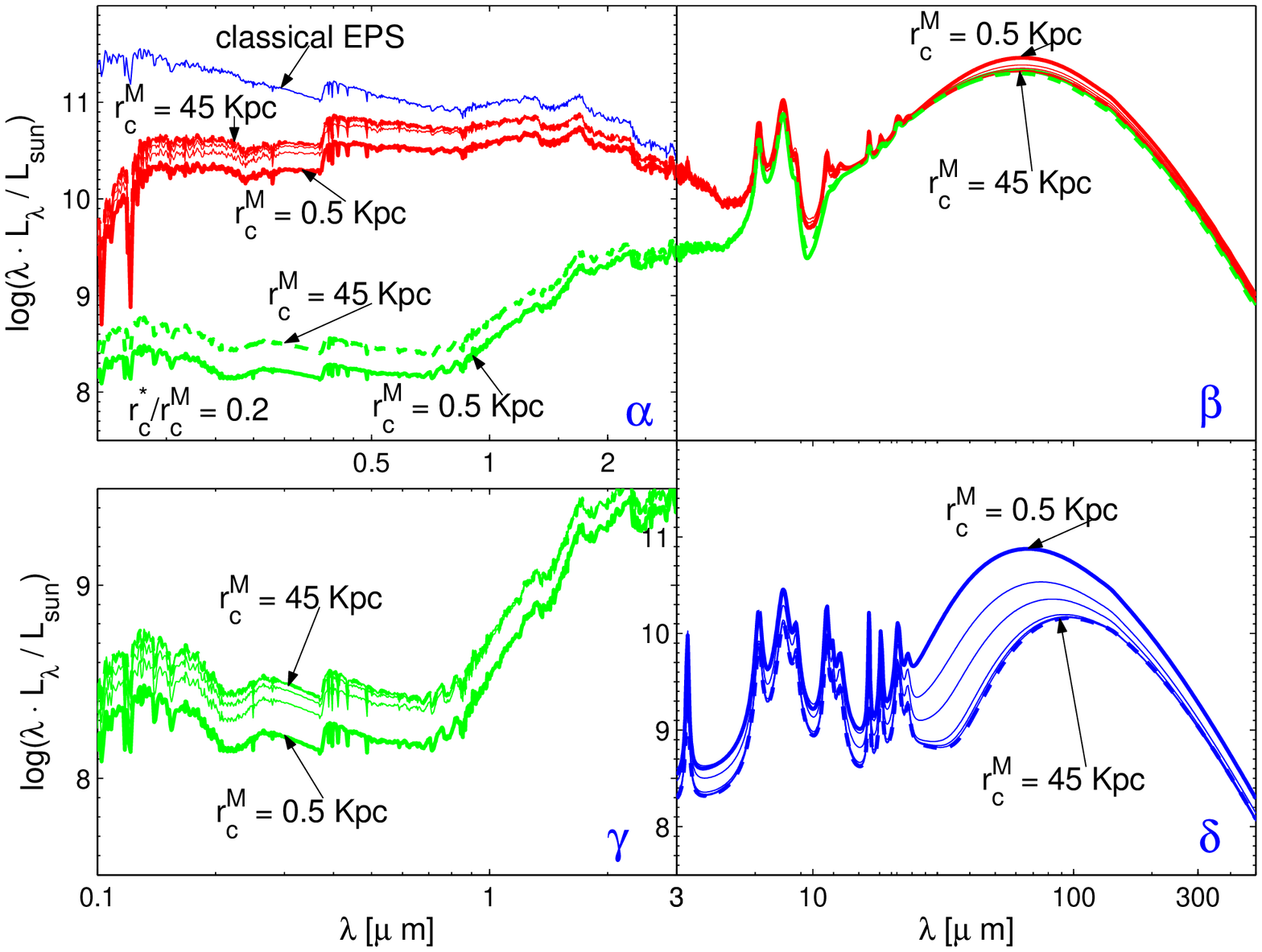,width=8.5truecm,height=7.0truecm}
\psfig{file=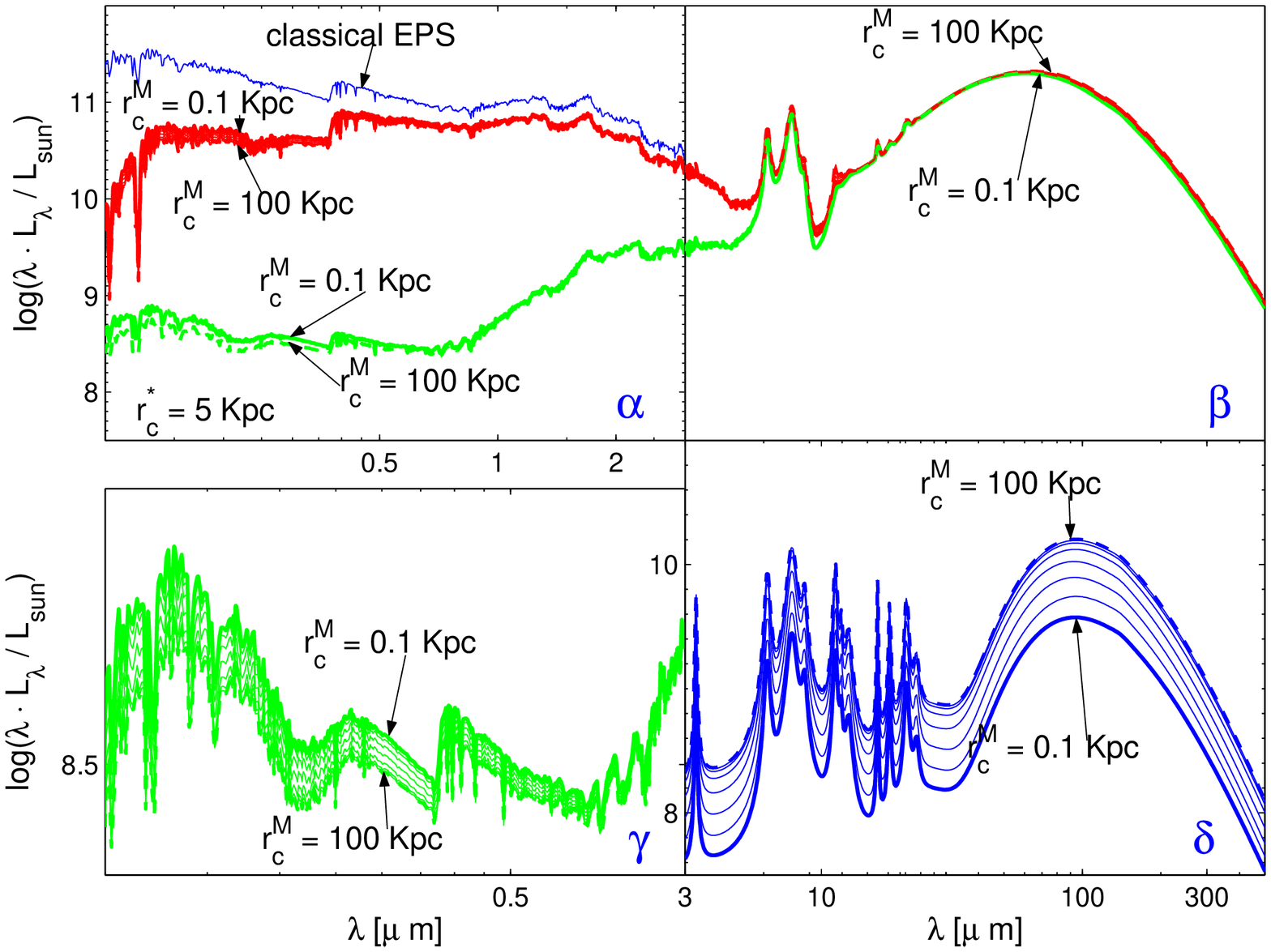,width=8.5truecm,height=7.0truecm}}
\caption{{\bf Four Left Panels}: in this group panels $\alpha$ and
$\beta$ show the SEDs of the model starburst at the age of $13$ Gyr
for fixed value of the ratio $r_{c}^{*}/r_{c}^{M}$=0.2. The radial
scale length $r_{c}^{*}$ is let varied from $0.1$ to to $9$ kpc. The
extreme values are marked with the corresponding $r_{c}^{M}$. The
SED obtained with the classical EPS is also shown for the sake of
comparison. The contribution of MCs to the total flux is also shown
for the extreme values. In panel $\gamma$ we show the MC flux in the
UV-optical/NIR and in panel $\delta$ we plot the contribution of the
diffuse ISM to the total emission. {\bf Four Right Panels}: the same
as above but for $r_{c}^{*}$ fixed and varying $r_{c}^{M}$.}
\label{RCSTARBURST}
\end{figure*}

In Fig. \ref{RCSTARBURST} we show how the total SED changes at
varying the scale radii. Their effects on  the emission of diffuse
ISM and young dusty MCs are also highlighted. Two cases are
considered. First, the ratio $r_{c}^{*}/r_{c}^{M}$ is fixed and
$r_{c}^{*}$ is let vary over a wide range of values (four left
panels). Second, $r_{c}^{*}$ is fixed and  $r_{c}^{M}$ is let change
(four right panels). In both groups of four, panels $\alpha$ and
$\beta$ display the total SED, the old SED without dust, the
contribution of MCs for the extreme values of the parameters,
whereas panels $\gamma$ and $\delta$ show the contribution of MCs
and diffuse ISM to the total flux. In the four left panels of Fig.
\ref{RCSTARBURST} we notice that the attenuation is stronger when
$r_{c}^{*}$ and $r_{c}^{M}$ are the lowest, both for stars (panel
$\alpha$) and MCs (panel $\gamma$). The emission of diffuse ISM
grows for more compact systems, because the radiation field heating
the grains is stronger in the central regions of high density (panel
$\delta$). In the four right panels where $r_{c}^{*}$ is fixed,
there is a small effect of varying $r_{c}^{M}$, because the emission
is clearly dominated from the MCs component. For higher values of
$r_{c}^{M}$, the system is more obscured (panel $\gamma$), because
there is more dust in the outer regions to screen the stellar light.

\begin{figure}
\psfig{file=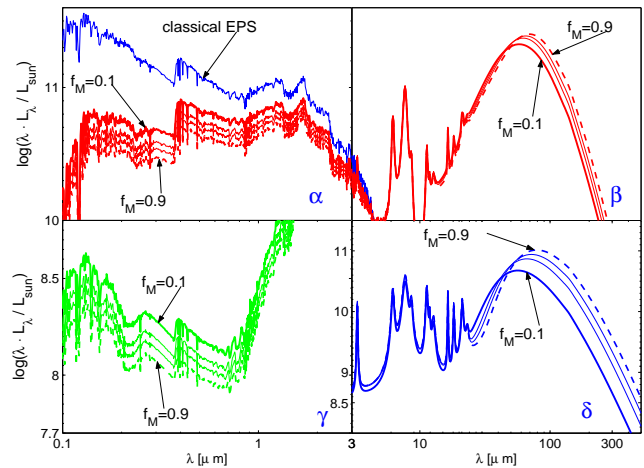,width=8.5truecm,height=6.2truecm}
\caption{\textbf{Top panels}: SEDs of the model star-burst galaxy at
varying the fraction of gas in the diffuse ISM $f_{M}$ with respect
to the total amount of gas. All the other parameters are kept fixed.
The considered values for $f_M$ are $0.1$, $0.3$, $0.5$, $0.7$ and
$0.9$. We also show the SED  calculated with the classical EPS
technique. \textbf{Bottom-left panel}: contribution to the total
emission by dusty MCs at varying $f_{M}$. \textbf{Bottom-right
panel}: contribution to the total emission by the diffuse ISM at
varying $f_{M}$.} \label{FMCSTARBURST}
\end{figure}

We compare the SEDs of our model starburster at varying the amount
of gas in the diffuse ISM. The remaining part of the gas is
distributed among the dusty MCs. In Fig. \ref{FMCSTARBURST} we
compare  SEDs obtained with different amounts of gas in the diffuse
ISM, varying the parameter $f_{M}$. Increasing the amount of gas in
the diffuse ISM  makes stronger the extinction of the light from old
and young stars, whereas it makes higher and cooler the FIR  flux,
because of the higher density and the weaker radiation field.

\begin{figure}
\psfig{file=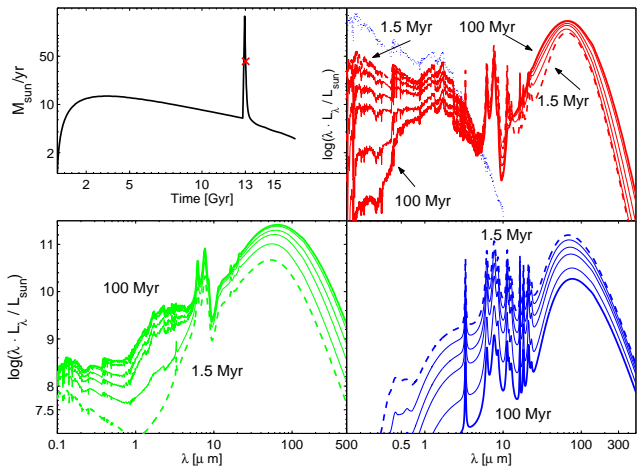,width=8.5truecm,height=6.2truecm}
\caption{\textbf{Top-left panel}: star formation rate of the model
star-burst galaxy of $M_{L} = 10^{11}M_{\odot}$ on which the age of
$13$ Gyr is marked. \textbf{Top-right panel}: SEDs  at varying the
evaporation time scale $t_{0}$, namely $1.5$, $4.5$, $15$, $30$,
$60$ and $100$ Myr. All the other parameters are kept fixed. We also
show the classical SED  (dotted line). \textbf{Bottom-left panel}:
contribution to the total emission by dusty MCs at varying $t_{0}$.
\textbf{Bottom-right panel}: contribution to the total emission by
the diffuse ISM at varying $t_{0}$.} \label{T0STARBURST}
\end{figure}

Finally, in Fig. \ref{T0STARBURST}, we show the effect of varying
the evaporation time scale of young dusty SSPs. The effect is the
same as in our model elliptical of $0.15$ Gyr (see Fig.
\ref{FMCELLITTICHE}) For longer time scales, the attenuation of the
UV-optical light becomes stronger (top-right panel), and the amount
of energy shifted toward the MIR/FIR by MCs becomes higher
(bottom-left panel).

\section{Late-type galaxies of the Local Universe} \label{spiral_local}

In this section, using our models we seek to reproduce the SEDs of
two late-type galaxies of the Local Universe, namely M$100$
and NGC$6946$.

In Table \ref{table_par}, columns (6) and (7), we summarize the
parameters characterizing the models that best match the properties
of the two galaxies under consideration. Part of the parameters are
based on observational hints, the others are suitably varied to get
agreement between observational and theoretical SEDs. Specifically,
the geometrical parameters and distances are from literature and
kept constant. The gas mass in the ISM and MCs is let vary within
the range indicated by current observational data. The star
formation efficiency $\nu$ and infall time scale $\tau$ are also let
vary around the typical values currently estimated for spiral
galaxies. Finally the evaporation timescale $t_0$ and the set of
dusty SSPs in our library are free parameters.

\textbf{M$100$}. This Sbc spiral galaxy is one of the most important
members of the Virgo Cluster, characterized by two huge and luminous
spiral arms and many other smaller ones. The redshift of the galaxy,
taken from NED\footnote{\textit{http://nedwww.ipac.caltech.edu/}.}
(Nasa/Ipac Extragalactic Database) is $z = 0.00524$. Using the
Hubble constant $H_{0} = 72$ km/s/Mpc, it corresponds to a distance
of $21.8$ Mpc. \citet{Shapley01}, who performed a thorough search in
literature of the distances for the spiral galaxies of their sample,
report for NGC 4321 the distance of $16.1$ Mpc, based on Cepheid
distance by \citet{Freedman94} and $H_{0} = 75$ km/s/Mpc. We adopt
here for the distance the average of the two estimates, i.e. $19$
Mpc. The major and minor diameters of the galaxy in arcmin (NED) are
$7.4^{\prime} \times 6.3^{\prime}$. Taking the average dimension of
$7^{\prime}$, we obtain the linear radius $R_{gal} = 19$ Kpc.
Following \citet{Gnedin95}, \citet{Knapen96} and \citet{Beckman96}
we adopt the inclination angle $\Theta = 27^{\circ}$ and we assume
that both stars and dust have the same radial and vertical scale
lengths of $5$ and $0.5$ kpc, respectively.

\begin{figure}
\psfig{file=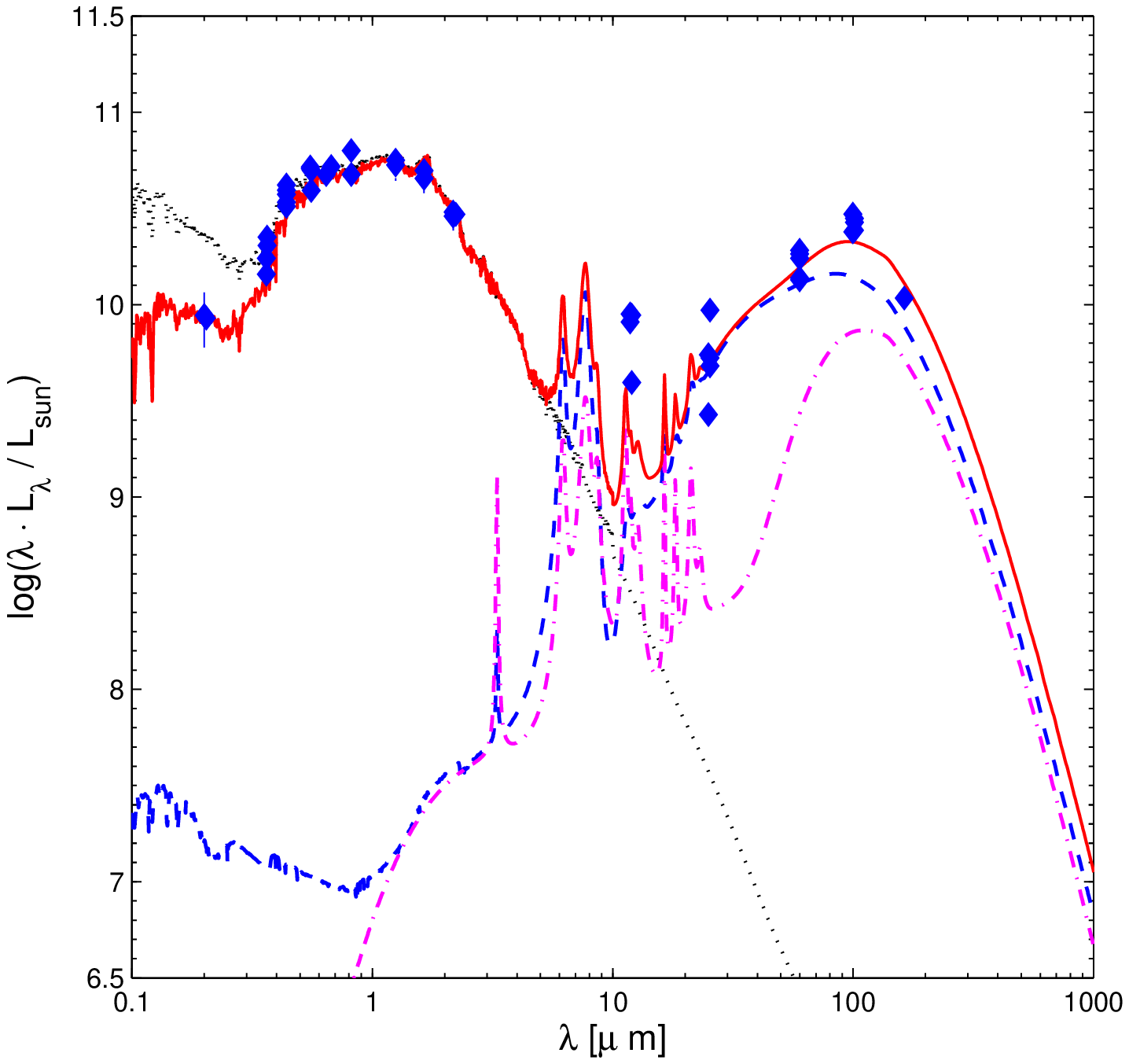,width=8.5truecm,height=7.0truecm}
\caption{SED of the modelled Sbc spiral galaxy M100 at the age of 13
Gyr (continuous line). We represented also the old SED with
classical EPS (dotted line), the emission of the diffuse ISM
(dot-dashed line) and the emission from young dusty SSPs (dashed
line). Data for M$100$ are taken from \citet[UV]{Buat89},
\citet[UV]{Donas87}, RC3 catalogue of
\citet[UBVRI]{deVaucouleurs92}, \citet[BVRIHK]{deJong94}, 2MASS
\citep[JHK]{Jarrett03}, IRAS \citep{Soifer89,Moshir90} and, finally,
\citet[FIR]{Stark89}. } \label{M100fit}
\end{figure}

The star formation history of M$100$ is derived assuming  the total
baryonic mass of $2 \cdot 10^{11} M_{\odot}$,   $\nu = 0.7$ and
$\tau = 4$ Gyr. With this choice, the star formation never stops: it
starts small, grows to a maximum and then  smoothly declines. We
remain with the following parameters to adjust: the gas mass in the
diffuse ISM and in MCs, the evaporation  time $t_{0}$ and the
library of young dusty SSPs. We may also slightly vary the radial
and vertical scale lengths with respect to current estimates. For
the gas masses  we adopt the values given by \citet{Young89a} based
on $HI$ and $H_{2}$, According to \citet{Young89a}, $M_{H_{2}}=0.77
\cdot M_{g}$, where $M_{g}$ is the total gas mass. It means that
this galaxy is dominated by the molecular component. A good fit of
the observational SED is obtained adopting the fractionary gas mass
in MCs   $f_{M} =0.35$. For the evaporation time we find $t_{0} = 5
Myr$, consistent with a dust-poor star forming medium in which the
first supernovae explosions in the newly born stellar populations
evaporate the  parental gas. The library of young SSPs best suited
to  M$100$  (and the other spiral as well) is characterized by the
optical depth $\tau = 35$, $R = 5$, and high $b_{c}$ abundance.
Nothing can be said for the ionization state of PAHs.  We adopt the
case  with  the complete treatment of the ionization state. Only the
detailed spectrum in the MIR region may help clarifying the issue.
It is worth noticing that for spirals, more extended and cooler MCs
with $R = 5$ fit the FIR emission better than  MCs with $R = 1$. In
Fig. \ref{M100fit} we show our best fit of the observational data.
The result is remarkably good.

\textbf{NGC $6946$} is a Sc/Scd nearby galaxy, highly obscured by
the interstellar matter of our galaxy, as it is quite close to the
Galactic plane and it is seen nearly face-on. However, the
inclination angle is uncertain due to the global asymmetry
\citep{Blais-Ouellette04}. We adopt  $\Theta = 35^{\circ}$
\citep[see also][]{Bonnarel88,Carignan90,Oey90}. The distance to NGC
$6946$ is uncertain and going  from $5$ \citep{deVaucouleurs79} to
$10$ Mpc \citep{Rogstad73}. Our distance to this galaxy has been
taken from \citet{Shapley01} who give $5.5$ Mpc  as a mean value
based on previous studies. From the NED online catalogue, the
diameters of the galaxy in arcmin are $11.5^{\prime} \times
9.8^{\prime}$. Adopting the average dimension of $10.5^{\prime}$, we
obtain  the radius $R_{gal} = 10$ Kpc. According to
\citet{Tacconi86}, the galaxy has some evidence of a spiral
structure extending beyond   $20$ kpc with HI emission out to $30$
kpc. However they adopted the distance of $10.1$ Mpc, that is two
times longer than the recent value proposed by \citet{Shapley01}
that we have adopted here. Taking $5.5$ Mpc, the radius is  $R_{gal}
= 13$ Kpc consistent with the \citet{Shapley01} results and the NED
diameters. The radial scale lengths of stars and gas are also taken
from \citet{Tacconi86}, however rescaled to our shorter value for
the distance to the galaxy. We adopt the radial scale length of $5$
kpc, and for the scale height we use the same value of 1 kpc
proposed by \citet{Silva98}. As far as the fraction of  total gas
embedded in young MCs is concerned, the question is controversial.
\citet{Young89a} reports $M_{HI}/M_{H_{2}} = 2.18$: this means that
only $1/3$ of the gas is in molecular form, in good agreement with
the results by \citet{Young89b} on  the molecular to atomic gas
ratio for the morphological types Sc/Scd. \citet{Tuffs96} found that
the bulk of the FIR luminosity arises from a diffuse disk component.
However, \citet{Devereux93} reports NGC $6946$ as a galaxy where the
molecular gas dominates the interstellar medium and the thermal
emission is explained mainly with a warm dust component heated from
young massive stars, a result confirmed in \citet{Malhotra96}. A
recent work by \citet{Walsh02} confirms the importance of the
molecular component in NGC $6946$, almost as massive as the atomic
one, giving the ratio $M_{H_{2}}/M_{HI} = 0.57$, which is high with
respect to other galaxies of the same morphological type. We find
that the observational SED is best reproduced if the molecular
component is about as massive as the atomic one. The star formation
history of NGC $6946$, that exhibits a moderate starburst activity,
has been calculated with a baryonic mass similar to M$100$ ($1.2
\cdot 10^{11} M_{\odot}$, $\nu = 0.7$ and $\tau = 5$).

\begin{figure}
\psfig{file=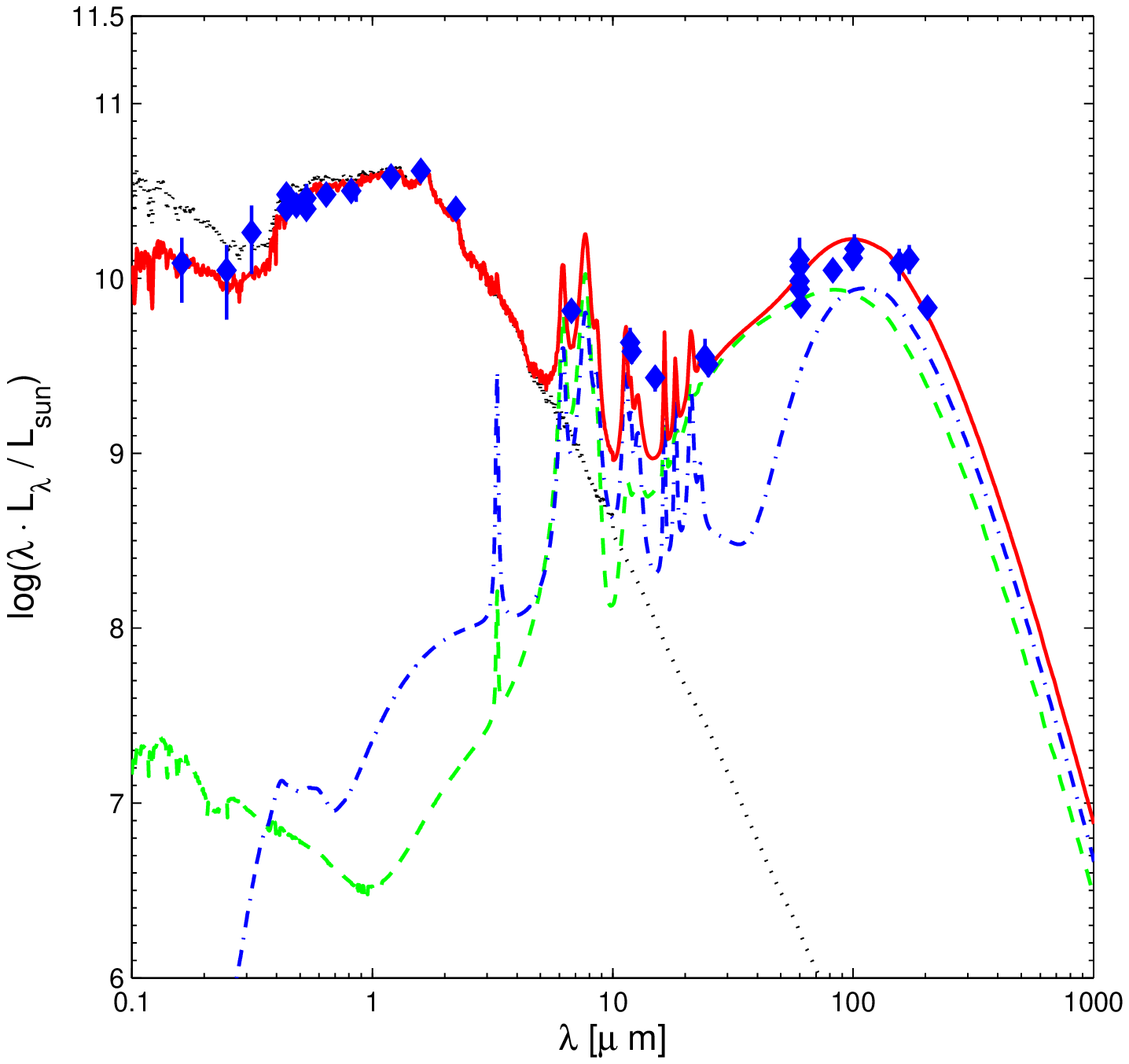,width=8.5truecm,height=6.8truecm}
\caption{SED of the modelled Scd spiral galaxy NGC$6946$ at the age
of 13 Gyr (continuous line). We represented also the old SED with
classical EPS (dotted line), the emission of the diffuse ISM
(dot-dashed line) and the emission from young dusty SSPs (dashed
line). Data for NGC $6946$ are taken from \citet[UV]{Rifatto95},
\citet[BV]{deVaucouleurs92}, ISO \citep[MIR]{Roussel01}, IRAS
\citet[MIR and FIR]{Rice88}, \citet[FIR]{Engargiola91},
\citet[FIR]{Devereux93}, \citet[FIR]{Tuffs96,Tuffs03} and
\citet{Silva99}.} \label{NGC6496fit}
\end{figure}

In Fig. \ref{NGC6496fit}, we show the result of our fit to the data
for NGC $6946$. The evaporation time $t_{0}$ is very short, $3$ Myr,
and the library of dusty SSPs is the same as for  M$100$.
As compared to M$100$,  the agreement with the MIR is
better, even if  MIR spectra would allow a better comparison
between theory and observations.

\section{Early-type galaxies of the local universe} \label{early_local}

In this section we present two old early-type galaxies of the Local
Universe, namely NGC 2768 and NGC 4491, and our attempts to
reproduce their SEDs. In Table \ref{table_par}, columns (8) and (9),
we summarize the values of the parameters we have chosen. The number
of parameters is smaller than for disk galaxies. First because of
the higher degree of symmetry, and second because parameters like
$t_{0}$ and $f_{M}$ have no role. The star formation ceased long ago
after the onset of the galactic wind, no young stars in dusty MCs
are present $\left(t_{0}=0\right)$ and all gas (if any) is in the
diffuse ISM $\left(f_{M}=1\right)$. There is however, a new
parameter to consider, i.e. the fraction $f_r$ of the gas
continuously ejected by dying and evolved stars (mostly RHB) that is
still retained in the galaxy (see Sect. \ref{par_examples_early} for
details).

\begin{figure}
\psfig{file=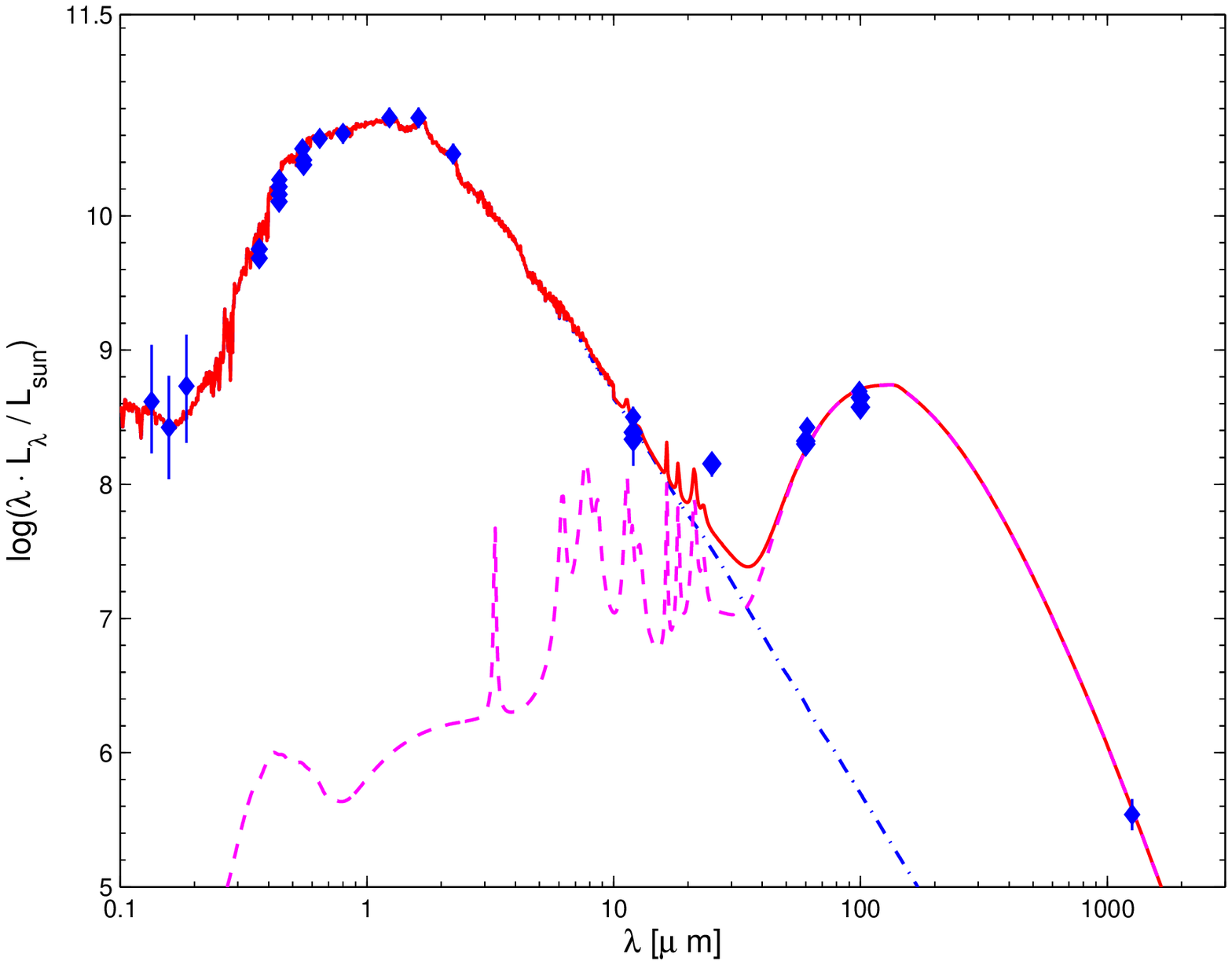,width=8.5truecm,height=6.8truecm}
\caption{SED of the modelled E6 elliptical galaxy NGC$2768$ at the
age of 13 Gyr (continuous line). We represented also the old SED
with classical EPS (dot-dashed line) and the emission of the diffuse
ISM (dashed line). Data for NGC $2768$ are taken from
\citet[far-UV]{Longo91}, \citep[UBV]{deVaucouleurs92}, NED database,
\citet[JHK]{Frogel78}, IRAS \citet[MIR and FIR]{Moshir90} and NED
database and, finally, \citet[sub-mm]{Wiklind95}. }
\label{NGC2768fit}
\end{figure}

\textbf{NGC $2768$} is an elliptical galaxy of
morphological type E$6$ at the distance of $21.5$ Mpc
\citep{Takagi03a}. As noticed by \citet{Takagi03a}, this galaxy has
been detected in the sub-mm range of wavelengths thus providing   a
better constraint on the dust emission. Data for this galaxy have
been taken from the literature as listed in Fig. \ref{NGC2768fit}.
As in \citet{Takagi03a}, all the data have been corrected where it
was necessary to be consistent with the galaxy observed as a whole,
even if there is a certain amount of error for the IUE data of
\citet{Longo91} that correspond to a small aperture in the center of
the galaxy. The diameters of the galaxy in arcmin (NED) are
$8.1^{\prime} \times 4.3^{\prime}$ and using an average dimension of
$6^{\prime}$, we obtain a radius of the galaxy of about $R_{gal} =
20$ Kpc.

\begin{figure}
\psfig{file=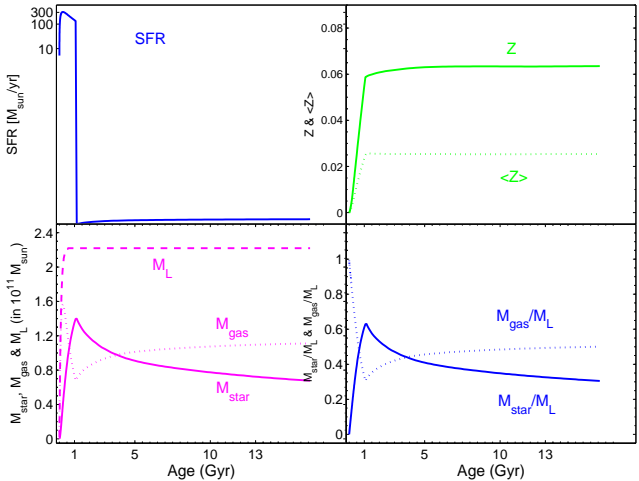,width=8.5truecm,height=6.4truecm}
\caption{Basic quantities of the chemical model for the elliptical
galaxy NGC2768 as function of the age: the top left panel shows the
star formation rate in $M_\odot$/yr; the top right panel displays
the maximum ($Z$, solid line) and mean metallicity ($\langle Z
\rangle $); the bottom left panel shows the mass of living stars
$M_{star}$ (solid line), the gas mass $M_{gas}$ (dotted line), and
the total mass of baryons $M_{L}$ (dashed line); finally the bottom
right panel displays the ratios $M_{star} / M_{L}$ (solid line) and
$M_{gas} / M_{L}$ (dotted line). All masses are in units of
$10^{11}\,M_\odot$. Ages are in Gyr.} \label{NGC2768chimica}
\end{figure}

The parameters chosen to model the SFH are as follows: the infall
time scale is $\tau$=0.1 Gyr, the exponent for the star formation
law is $k$=1 as in \citet{Tantalo96,Tantalo98a}, whereas the
efficiency is $\nu$=2. In Fig. \ref{NGC2768chimica} we show the SFH
with other quantities derived by the chemical code. The onset of the
galactic wind takes place at about $1.1$ Gyr. The scale lengths of
stars and gas have been fixed as for the prototype elliptical of
Sect. \ref{par_examples_early}.

Finally, in Fig. \ref{NGC2768fit}, we show the theoretical SED and
its comparison with the observational data. Agreement is excellent
from the UV to  FIR. Note the strong reduction factor for the gas
content.

\textbf{NGC $4494$} is a roughly spherical elliptical galaxy of the
morphological type E$1$. For this galaxy the NED catalog reports the
redshift $z = 0.00451$ corresponding to a distance of about $19$
kpc, using $H_{0} = 72$ km/s/Mpc. \citet{Temi04} report from the
LEDA catalog a distance of about $21.28$ Mpc. We choose the average
distance between the two determinations, i.e. $20$ Mpc. The
diameters of the galaxy in arcmin (NED) are $4.8^{\prime} \times
3.5^{\prime}$ and using an average dimension of $4.2^{\prime}$, we
obtain a dimension of about $R_{gal} = 12$ Kpc.

\begin{figure}
\psfig{file=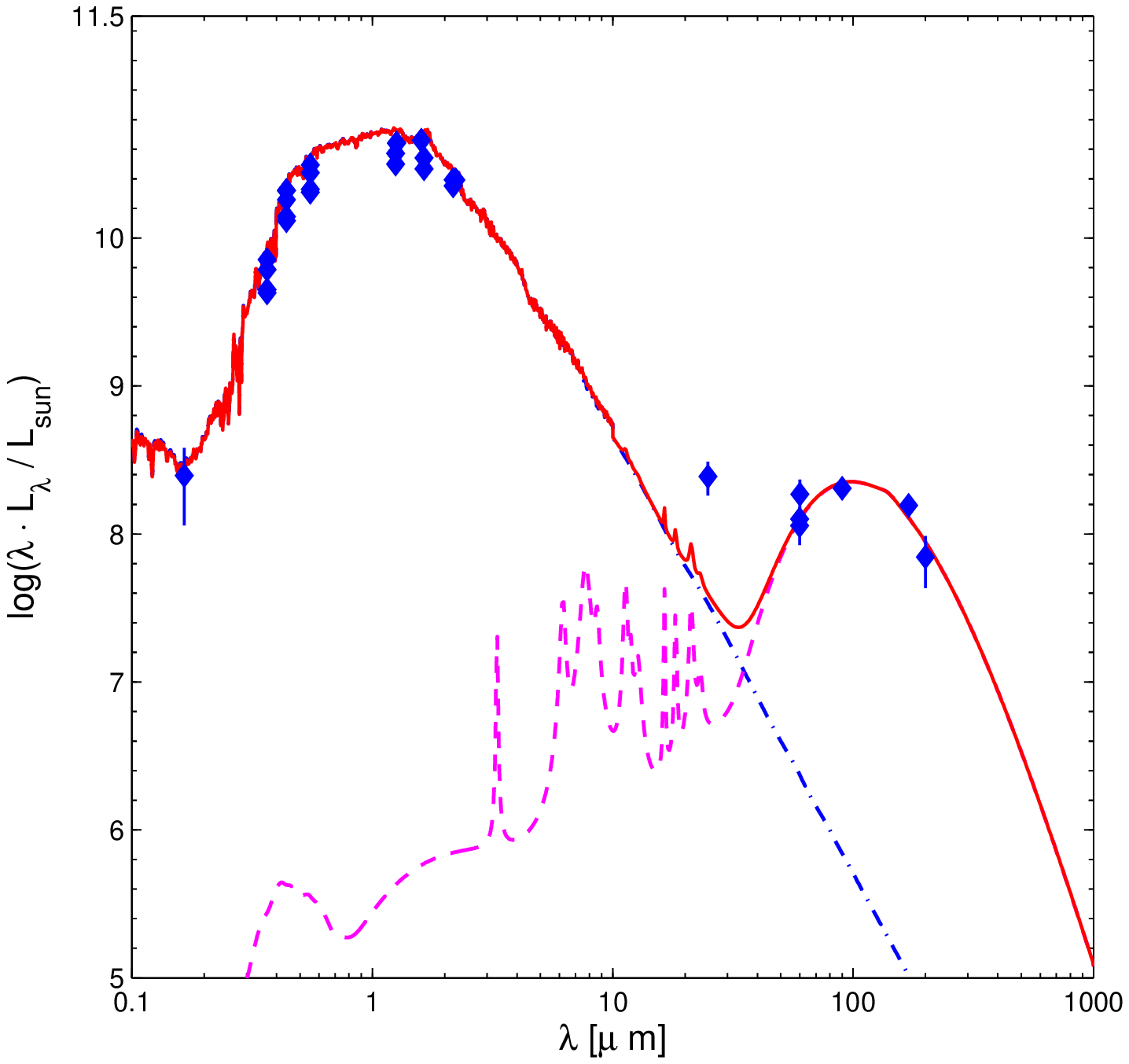,width=8.5truecm,height=6.4truecm}
\caption{SED of the modelled E6 elliptical galaxy NGC$4494$ at the
age of 13 Gyr (continuous line). We represented also the old SED
with classical EPS (dot-dashed line) and the emission of the diffuse
ISM (dashed line). Data for NGC$4494$ have been taken from
\citet[far-UV]{Rifatto95}, \citet[UBV]{deVaucouleurs92}, NED
database, \citet[UBVJHK]{Gavazzi96}, 2MASS \citep[JHK]{Jarrett03},
IRAS \citep[MIR and FIR]{Moshir90} and ISO \citep{Temi04}.}
\label{NGC4494fit}
\end{figure}

The SFH and the evolution in metallicity of NGC$4494$ are calculated
in the same way as for NGC$2768$, only with a slightly different
value of the baryonic mass. The scale lengths are the same as for
the other elliptical. In Fig. \ref{NGC4494fit} we show the result of
our fit. The agreement is good. The main reason of uncertainty is
the correction we made for the IUE UV-data  of \citet{Rifatto95}. As
pointed in \citet{Takagi03a}, only knowing the true UV profile it is
possible to properly correct these IUE data that cover only a small
region in the central region of the galaxy. Finally, the same remark
on the gas content made for the other elliptical can be made also
here.

\section{Starburst galaxies}\label{starburst_local}

Finally, we present models for two well known and thoroughly studied
star-burst galaxies of the local universe, namely Arp220 and M82. In
Table \ref{table_par}, columns (10) and (11),
we summarize the values of the parameters we
chose to reproduce the SEDs of these galaxies. We adopt the
spherical geometry to describe both objects. As for ellipticals, the
number of parameters to deal with is much lower than for disk
galaxies. The situation, however, is different from the case of
ellipticals because now the parameters $f_{M}$ and particular
$t_{0}$ play in the fit of the observational data.

\textbf{Arp$220$}. Arp $220$ is the brightest object in the Local
Universe. There is nowadays the general consensus that Arp $220$ is
a starburst-dominated galaxy and not an AGN-dominated object
\citep{Lutz96,Genzel98,Lutz99,Rigopoulou99,Tran01}. Recently,
\citet{Spoon04} re-analyzing the ISO MIR spectrum of Arp $220$
confirm the starburst-dominated hypothesis, suggesting that the IR
luminosity should be be probably powered by the starburst activity
in extremely dense regions, even if the AGN contribution cannot be
definitely ruled out because of the high extinction.

The redshift of the galaxy, taken from the NED database is $z =
0.01813$. Using the Hubble constant $H_{0} = 72$ km/s/Mpc, it
corresponds to a distance of about $76$ Mpc. This value is fully
consistent with the distances proposed in \citet{Soifer87} and in
\citet{Spoon04}. The diameters of the galaxy in arcmin (NED) are
$1.5^{\prime} \times 1.2^{\prime}$. Adopting the mean value of
$1.35^{\prime}$, the radius of the galaxy is of about $R_{gal} =
16-17$ Kpc. \citet{Wynn-Williams93} showed that almost all the MIR
flux of Arp220 comes from a small central region of about $5^{"}$
aperture and this concentration of the MIR emission has been
confirmed by \citet{Soifer99} comparing the fluxes at a fixed MIR
wavelength and varying the beam. For this reason we can quite safely
use the MIR data from small apertures even modelling the whole
galaxy.

\begin{figure}
\psfig{file=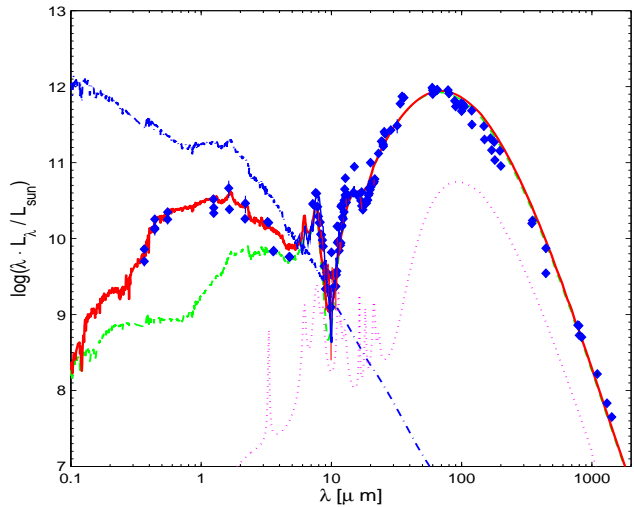,width=8.5truecm,height=6.8truecm}
\caption{SED of the model for the starburst galaxy Arp$220$ at the
age of 13 Gyr (continuous line). We represent also the old SED with
classical EPS (dot-dashed line), the emission of the diffuse ISM
(dotted line) and, finally, the emission of dusty MCs (dashed line).
Optical and near IR data are taken from RC3 catalog
\citep[UBV]{deVaucouleurs92} and \citet[2MASS - JHK]{Jarrett03}. The
MIR fluxes are from \citet[MIR]{Spinoglio95}, \citet[MIR]{Smith89},
\citet[from MIR to FIR]{Klaas97}, \citet[MIR]{Wynn-Williams93},
\citet[5-16 $\mu$m ISOCAM-CVF spectrum]{Tran01}. FIR and radio data
are from \citet[sub-mm]{Rigopoulou96}, \citet[UKIRT
sub-mm]{Eales89}, \citet[SCUBA sub-mm]{Dunne00}, \citet[SCUBA
sub-mm]{Dunne01}. Other MIR/FIR data are taken from
\citet{Spoon04}.} \label{Arp220fit}
\end{figure}

The mass of molecular gas $H_{2}$ in Arp$220$ according to CO
estimates is about $3 \cdot 10^{10} M_{\odot}$
\citep{Solomon97,Scoville97}. \citet{Mundell01} estimated the
nuclear column densities of $H_{2}$ to be of the order of $N_{H_{2}}
= 2-4 \cdot 10^{22} cm^{-2}$. These values are comparable to the
mean $HI$ column densities that are of the order of $N_{H} = 1.5
\cdot 10^{20}T_{s}$, with the spin temperature between $100$ and
$200$ K. These column densities  are likely lower limits because of
the uncertainty in the values of $T_{s}$ \citep{Kulkarni88}, in the
abundance of $CO$ \citep{Frerking82}, in the excitation conditions
and optical depth, which all suggest us to adopt similar contents
for both atomic and molecular gas. We fix $f_{M} = 0.5$. The SFH
history of Arp$220$ is modelled adding to the current SFH (which
peaked in past and ever since decreased) a very strong and short
burst. The burst is obtained by increasing the star formation of a
factor of $60$. In Fig. \ref{Arp220fit} we show our best fit of
Arp$220$, the agreement is good. However, there are two points that
need to be clarified. First, we have no data of Arp$220$ in the near
and far UV. As outlined by \citet{Takagi03a}, the data by
\citet{Goldader02} are puzzling because, owing to the large angular
size of Arp 220 filling the field of view of the instrument, the
flux level of the sky (to be subtracted) is highly uncertain.
Second, to get satisfactory agreement with observational data we had
to extend the parameter space, namely the scale radius has been
lowered to $R = 0.5$ and the $\tau_{V}=40$, this one in agreement
with \citet{Sturm96} and \citet{Genzel98}. The ionization of PAHs is
fully considered, but it is worth noticing that the ionization model
does not bear very much on the extinction curve we have adopted
thanks to the low carbon abundance and low contribution of PAHs.

There is another interesting point to note. Arp$220$ is modelled as
a $13$ Gyr old galaxy with a strong burst superimposed to a
spiral-like SFH that reached a maximum in the past and gently
declines. This long history of star formation slowly increases the
metallicity well above the solar value. For these high metallicities
values, one should (has to) use a MW-like extinction curve for dense
regions to calculate the SED of young SSPs emerging from dusty
regions. This type of extinction curve is characterized by a high
abundance of carbon. We tried to fit the observational SED of
Arp$220$ using the lowest available value of $b_{c}$, but the MIR
emission was always too high and the $10 \mu m$ absorption features
of silicates not as deep as required. To obtain a good fit we had to
use for young dusty SSPs the extinction curve of the SMC, which is
poor of carbon. Similar result was found by \citet{Takagi03a}. There
seems to be a point of contradiction, because the carbon-poor
extinction curve of the SMC is characteristic of a low metallicity
environment, whereas here we are dealing with a high metallicity
one. Although this point deserves careful future investigation, a
plausible explanation could be that the high rate of type II
supernovae due to the strong burst alters the composition of dust,
growing the amount of silicates with respect to that of carbonaceous
grains. Another hypothesis is that this high energy output  of the
star forming process destroys molecules and small grains thus
altering the distribution of the grains. Finally the geometry of the
system could play a role.

\begin{figure}
\psfig{file=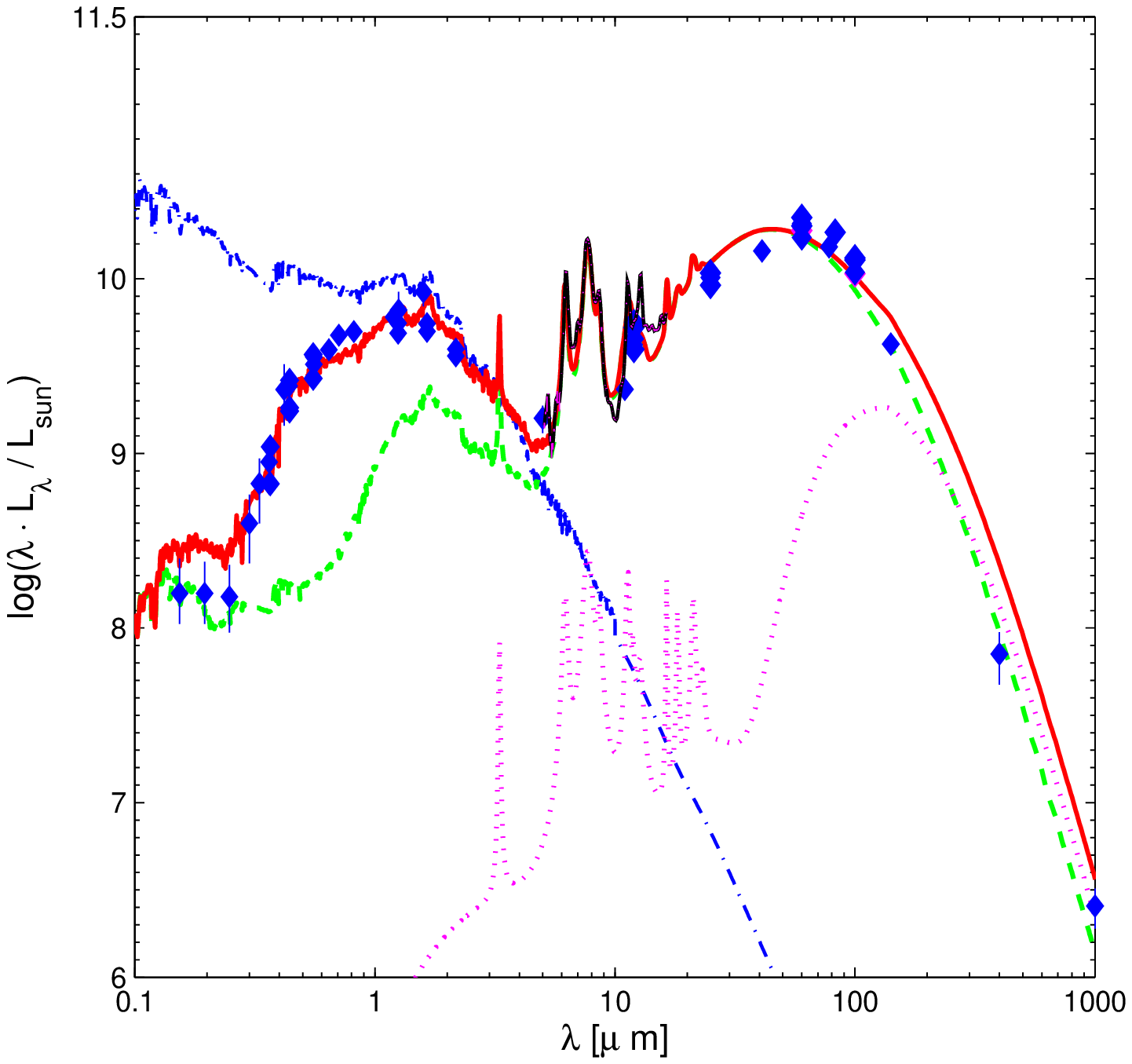,width=8.5truecm,height=6.8truecm}
\caption{SED of the model for the starburst galaxy M$82$ at the age
of 13 Gyr (continuous line). We represent also the old SED with
classical EPS (dot-dashed line), the emission of the diffuse ISM
(dotted line) and the emission of the dusty MCs (dashed line). Data
are from \citet[UBV]{deVaucouleurs92}, \citet[VRIJKL]{Johnson66},
\citet[2MASS - JHK]{Jarrett03}, \citet[IRAS]{Soifer87},
\citet[IRAS]{Golombek88}, \citet[IRAS]{Rice88}, \citet[FIR and
sub-mm]{Klein88}, and \citet{ForsterSchreiber03}.} \label{M82fit}
\end{figure}

\textbf{M$82$}. The central region of the irregular starburst galaxy
M$82$ seems to have suffered a strong gravitational interaction
about $10^{8}$ years ago \citep{ForsterSchreiber03} with its
companion M$81$ and it shows a remarkable burst of star formation
activity. M$82$ is still strongly interacting with $M81$ as proved
by the common envelope of neutral hydrogen \citep{Yun93,Ichikawa95}.
The classical distance to M$82$ is $3.25 \pm 0.20$ Mpc
\citep{Tammann68} that makes of M$82$ the nearest and most studied
starburst galaxy \citep[see][and references therein]{Shopbell98}.
For M$82$ data for the emission of PAHs in the MIR region are
available. The ISO-SWS data by \citet[][ private
communication]{Sturm00} cover a wide region going from $2.4$ to $45
\mu m$ with different parts of a SWS full grating scan that are
observed with different aperture sizes, going from $14^{"} \times
20^{"}$ to $20^{"}\times 33^{"}$. The problem is that with these
small apertures it is not possible to observe the whole galaxy, but
only part of it. In our case the south-western star formation lobe.
Recently, \citet[][private communication]{ForsterSchreiber03}
observed M$82$ in the MIR range between $5.0$ and $16 \mu m$ with
ISOCAM  on board of ISO. The total field of view is $96^{"}\times
96^{"}$. It covers almost entirely the MIR sources of M$82$ and it
is more suitable to be used to model M$82$ as a whole. The mass of
molecular gas for M$82$ is of the order of $1-2 \cdot 10^{8}
M_{\odot}$, as derived from CO determinations
\citep{Lo87,Wild92,ForsterSchreiber01}. The estimates of the total
mass of gas are of the order of $10^{9} M_{\odot}$
\citep{Solinger77}. For this reason we use a parameter $f_{M} \sim
0.8-0.9$, with almost all the gas in the diffuse ISM. In Fig.
\ref{M82fit} we show the best fit obtained for M$82$. The agreement
with the data is good and the main features of PAHs are well
reproduced. The young dusty SSPs used for this fit are characterized
by the following parameters: $R=1$, $\tau_{V}=8$, $b_{c}=2 \cdot
10^{-5}$ and MW average ionization of PAHs. An interesting point is
that we found it difficult to reach the low level of flux in the UV
region. Even if data are highly uncertain, this could be due to the
model of MCs  with uniform distribution of dust, $\eta = 1000$
\citep[see][]{Piovan06a}, we have adopted. It might be possible that
for M$82$, MCs with a shell-like distribution of dust with  more
dust in the outer regions and therefore with stronger attenuation of
the flux from young stars, are better suited to obtain a good fit in
that spectral region.

\section{Discussion and conclusions}\label{dis_concl5}

In this paper, improving upon the standard EPS technique, we have
developed theoretical SEDs of galaxies, whose morphology goes from
disk to spherical structures, in presence of dust in the ISM.
Properly accounting for the effects of dust on the SED of a galaxy
increases the complexity of the problem with respect to the standard
EPS theory because it is necessary to consider the distribution of
the energy sources (the stars) inside the ISM absorbing and
re-emitting the stellar flux. This means that the geometry and
morphological type of the galaxy become important and unavoidable
ingredients of the whole problem, together with the transfer of
radiation from one region to another. The emergent SEDs of our model
galaxies have been shown to reproduce very well, even in details,
the observational data for a few test galaxies of different
morphological type. The model is versatile and applicable to a large
range of objects of astrophysical interest at varying the star
formation and chemical enrichment histories, the geometrical shape
or morphology of the galaxies and the amounts of gas and dust
contained in their ISM.

Before concluding, it is worth mentioning a few points of weakness
that could be improved . First, the chemical models we have adopted
are from \citet{Tantalo96,Tantalo98a}, whereas the chemical yields
are from \citet{Portinari98}. These models are state-of-the-art in
the study of the chemical evolution of galaxies. However, they  do
not include a proper description of the formation/destruction of
dust as for instance  recently developed  by \citet{Dwek98,Dwek05}.
Even if the dust content can be related to the metallicity of the
galaxy (see Sect. \ref{par_chosen}), the relative proportions of the
various components of the dust would require the detailed  study of
the evolution of the dusty environment and the complete information
on the dust yields \citep{Dwek98,Dwek05,Galliano05}. This would lead
to a better and more physically sounded correlation between the
composition of dust and the star formation and chemical enrichment
history of the galaxy itself. All this is missing in most galaxy
models in which dust is considered. The problem may be particularly
severe for high metallicity environments.

Second, the models for disk galaxies with central bulge need to be
tested against SEDs of local galaxies of intermediate type going
from $S0$ to $Sb/Sc$, trying to match some observational
constraints, like the UV-optical average colours
\citep{Buzzoni02,Buzzoni05}.

Finally, many other physical ingredients can be improved and/or
considered. Just to mention a few, the inclusion of the recent
models of thermally pulsing AGB stars with varying molecular
opacities in the outer layers \citet{Marigo02}, the extension of the
SED to the radio range, and the simulation of the nebular emission.

\section*{Acknowledgements}

We would like to deeply thank A. Weiss for the many stimulating
discussions and for showing much interest in our work. L.P. is
pleased to acknowledge the hospitality and stimulating environment
provided by Max-Planck-Institut f\"ur Astrophysik in Garching where
part of the work described in this paper has been made during his
visit as EARA fellow on leave from the Department of Astronomy of
the Padua University. This study has been financed by the Italian
Ministry of Education, University, and Research (MIUR), and the
University of Padua.

\appendix

\appendix

\section{Matter in the cylinder between $V$ and $V'$}
\label{AppendixA}

To calculate the number of H atoms contained in the cylinder of
matter between two generic volumes $V=V\left( i,j,k\right)$ and
$V^{\prime}=V\left( i^{\prime },j^{\prime },k^{\prime }\right)$ one
has to know the total gas mass in the cylinder

\begin{equation}
\int_{V}^{{V}^{\prime}}\rho _{M}\left( l\right) dl \label{A1}
\end{equation}

\noindent where $\ \rho_{M}$  is the density profile of the diffuse
medium, given by the eqns. (\ref{rho_stars_disk}) or
(\ref{rhostar_ell}) or both, according to whether we are dealing
with a disk galaxy, a spherical galaxy or a disk plus bulge one.

Normalizing the density to the central value, i.e.
$\rho'=\rho_{M}/\rho_{0M}$, and the radial distance to the radial
scale length for disks or core radius for spheroidal galaxies as
appropriate, i.e. $x=r/R_{d}^{M}$ or $x=r/r_{c}^{M}$, we may express
the density profiles by means of dimensionless quantities. For a
disk galaxy we have $\rho'(x)=\exp \left( -x\sin \theta \right) \exp
\left( -x\left| \cos \theta \right| \cdot R_{d}^{M}/z_{d}^{M}
\right)$ which depends only on  the ratio  $R_{d}^{M}$ and
$z_{d}^{M}$, whereas for a  spheroidal galaxy we have
$\rho'(x)=\left[1+\left( x\right)^{2}\right]^{-\gamma_{M}}$.

\noindent In this way it is possible to calculate once for all the
mass contained in the cylinders between any two generic volume
elements of the galaxy, independently from the mass of the galaxy,
given the coordinate grid $\left( nr,n\theta ,n\Phi \right) $ and
ratio $R_{d}^{M}/z_{d}^{M}$ (for disk galaxies only) or exponent
$\gamma_{M}$ (for spheroidal galaxies). The integral of eqn.
\ref{A1} is then numerically evaluated.

\section{The distance $r^{2}$} \label{AppendixB}

Following \citet{Silva99} the distance $r^{2}\left( i,j,k,i^{\prime
},j^{\prime },k^{\prime }\right)$ shortly indicated as
$r^{2}\left(V,V^{\prime}\right)$, which is defined as volume
averaged value of the square of the distance between $V\left(
i^{\prime },j^{\prime },k^{\prime }\right) $ and all the points
belonging to $V\left( i,j,k\right) $, is $r^{2}\left(
V,V^{\prime}\right) = \displaystyle \int \nolimits \int \nolimits
\int \nolimits_{V} d^{2}\left( V,V^{\prime}\right) r^{2}\sin \theta
d\theta d\Phi dr/V$ where $d^{2}\left( V,V^{\prime}\right) $ is
given by $d^{2}\left(V,V^{\prime}\right) =\left( r_{i^{\prime }}\sin
\theta _{i^{\prime }}-r\sin \theta \cos \Phi \right) ^{2}+
 \left( r\sin \theta \sin \Phi \right)^{2}+\left(
r_{i^{\prime }}\cos \theta _{i^{\prime }}-r\cos \theta \right) ^{2}$
and the volume $V$ by $V = \int \nolimits \int \nolimits \int
\nolimits_{V} r^{2} \sin \theta d \theta d \Phi dr$.

\section{Matter in the cylinder between $V$ and the galactic edge}
\label{AppendixC}

The matter contained in the cylinder between a generic element $V$
and the edge of the galaxy extinguishes the radiation emerging from
the volume element, travelling across the galaxy up to its edge,
from which it escapes toward an external observer. To determine this
mass we proceed as follows. Firstly, since the emission
$j^{TOT}\left( \lambda ,V\right)$  of eqn. (\ref{Geitot}) has been
defined per unit volume  and  the problem has  azimuthal symmetry,
we resize the grid of azimuthal coordinates by taking many equally
spaced bins. Using this resized grid, we introduce the Cartesian
coordinates $\left( X_{V},Y_{V},Z_{V}\right)$ of the center of the
volume element by means of the transformations $X_V=x_{iV} \sin
\theta_{jV} \cos \Phi_{kV}$, $Y_V=x_{iV} \sin \theta_{jV} \sin
\Phi_{kV}$ and $Z_V=x_{iV} \cos \theta_{jV}$, where
$\left(x_{iV},\theta_{jV}, \Phi_{kV}\right)$ are the polar
coordinates of centre of the volume $V\left(i,j,k\right)$. The
starting reference system is in spherical coordinates
$\left(O,x,\theta,\Phi\right)$ and corresponds to a Cartesian system
$\left(O,X,Y,Z\right)$. Let us now consider another reference system
$\left(O^{\prime},x^{\prime},\theta^{\prime},\Phi^{\prime}\right) $
to which another Cartesian system $\left(O^{\prime}, X^{\prime},
Y^{\prime},Z^{\prime}\right)$ would correspond. The transformations
from one Cartesian system to the other is given by $X=X^{\prime
}+X_{O^{\prime}}, Y=Y^{\prime}+Y_{O^{\prime}}$ and $Z=Z^{\prime} +
Z_{O^{\prime}}$, where
$\left(X_{O^{\prime}},Y_{O^{\prime}},Z_{O^{\prime}}\right)$ are the
coordinates of the centre $O^{\prime}$ of the new system in the old
one. We take the new reference system $\left(O^{\prime}, X^{\prime},
Y^{\prime},Z^{\prime}\right)$ with  the origin $O^{\prime}$ in
$\left(X_{O^{\prime}},Y_{O^{\prime}},Z_{O^{\prime}}\right)=\left(0,x_{iV}\sin
\theta_{jV}\sin \Phi_{kV},0\right)$,
 where $\overline{OO^{\prime}}=x_{iV}\sin \theta_{jV}\sin
\Phi_{kV}$. This represents a translation of the origin $O$ along
the Y-axis to a new origin $O^{\prime}$ so that the centre of the
volume lies in the plane $\left( O^{\prime}, X_{O^{\prime}},
Z_{O^{\prime}} \right)$. The Cartesian coordinates of the volume
centre in the new system will be $X_{V}^{\prime}=X_{V}$,
$Y_{V}^{\prime}=Y_{V}-Y_{O^{\prime}}=0$ and $Z_{V}^{\prime}=Z_{V}$.
\noindent Applying the inverse relationships to pass from Cartesian
to spherical coordinates, the $x^{\prime}$ and $\theta^{\prime}$ of
the volume centre are $x_{iV}^{\prime } = \sqrt{\left(
X_{V}^{\prime}\right)^{2}+\left( Y_{V}^{\prime }\right) ^{2}+\left(
Z_{V}^{\prime}\right)^{2}} = x_{iV}\sqrt{\left( \sin \theta_{jV}\cos
\Phi_{kV}\right) ^{2}+\left( \cos \theta_{jV}\right)^{2}}$ and
$\theta_{jV}^{^{\prime}}=\arccos \left( Z_{V}^{\prime}/\sqrt{\left(
X_{V}^{\prime}\right)^{2}+\left( Y_{V}^{\prime }\right) ^{2}+\left(
Z_{V}^{\prime}\right)^{2}} \right) =\arccos \left(\cos \theta
_{jV}/\sqrt{\left( \sin \theta_{jV}\cos \Phi_{kV}\right) ^{2}+\left(
\cos \theta_{jV}\right)^{2}}\right)$.

\noindent Obviously the new coordinate $\phi_{k}^{\prime} $ is equal
to $0$ or $\pi $, because the volume belong to the plane $\left(
O^{\prime}, X_{O^{\prime}}, Z_{O^{\prime}} \right)$. For
$x_{i}^{\prime}<0$ $\phi_{k}^{\prime}=\pi$, whereas for
$x_{i}^{\prime}>0$ $\phi_{k}^{\prime}=0$. To avoid useless
complications,  the grid of azimuthal coordinates has been redefined
in such a way that $x_{i}^{\prime}$ is always different from zero.
We proceed now to determine the radius of the circular section of
the galaxy coincident with the plane $\left( O^{\prime},
X_{O^{\prime}}, Z_{O^{\prime}} \right)$. This  will depend on the
type of galaxy under consideration. We get $x_{G}^{\prime} =
\sqrt{\left( R_{gal}/R^{M}_{G}\right) ^{2}-Y_{O^{\prime}}^{2}}
\nonumber = \sqrt{\left(R_{gal}/R^{M}_{G}\right)^{2}-\left(
x_{iV}\sin \theta_{jV}\sin \Phi_{kV}\right)^{2}}$.

\noindent where $R_{gal}$ is the galactic radius and $R^{M}_{G}$ is
equal to $R_{d}^{M}$ for disks and to $r_{c}^{M}$ for spherical
galaxies. Let us now consider an observer looking at the galaxy from
the view angle $\Theta$ with respect to the equatorial plane of the
galaxy. For the sake of simplicity we place the observer on the
plane $\left( O^{\prime}, X_{O^{\prime}}, Z_{O^{\prime}} \right)$.
Therefore, $\Theta =0$ corresponds to a galaxy seen edge-on, whereas
$\Theta =\pi /2$ to the case face-on. Thanks to the azimuthal and
equatorial symmetries we make take $\Phi=0$ and $Z^{\prime} \leq 0$.

Associated to the volume center with coordinates $\left(
x_{i}^{\prime},\theta_{j}^{^{\prime}},\Phi_{k}^{\prime}\right)$
there will be a point  $P$ located on the galaxy edge with
coordinates $P\left( x_{G}^{\prime
},\theta_{G,}^{\prime}\Phi_{G}^{\prime}\right)$ where
$\theta_{G}^{\prime }$ and $\Phi_{G}^{\prime }$ are still unknown.
 Let us first calculate  $\Phi_{G}^{\prime}$. If
$\Theta=0$ all the points at the galactic edge  have
$\Phi_{G}^{\prime}=0$. If $\Theta =\pi /2$  two cases are
possible: for $x_{i}^{\prime}<0\Rightarrow \Phi_{G}^{\prime
}=\pi$, whereas for $x_{i}^{\prime}>0\Rightarrow
\Phi_{G}^{\prime}=0$. In the general case with $0<\Theta <\pi /2$
we have to calculate the equation of the straight line with
angular coefficient $m=tg\left( \pi -\Theta \right)$ passing
through the point $Q\left( X^{\prime},Z^{\prime}\right) =Q\left(
0,-R_{G}^{^{\prime }}\right) $. We obtain
$Z^{\prime}=-X^{\prime}\tan\Theta -x_{G}^{\prime}$. We calculate
now $\left(Z_{V}^{\prime}+X_{V}^{\prime}\tan\Theta
+x_{G}^{\prime}\right)$ where $\left(
X_{V}^{\prime},Z_{V}^{\prime}\right)$ are the coordinates of the
 volume center. If
$\left(Z_{V}^{\prime}+X_{V}^{\prime}\tan\Theta
+x_{G}^{\prime}\right) >0$ we have $\Phi_{G}^{\prime}=0$, whereas
if $\left( Z_{V}^{\prime}+X_{V}^{\prime}\tan\Theta
+x_{G}^{\prime}\right) <0$ we get $\Phi_{G}^{\prime}=\pi $.

The derivation of $\theta_{G}^{\prime}$ is slightly more complicate
because five cases are possible. In any case, it is a matter of
lengthy  trigonometrical manipulations. The line for the center of
the plane $\left(O^{\prime}X^{\prime}Z^{\prime}\right) $ with
inclination $\Theta$ has equation $Z^{\prime} =
-X^{\prime}\tan\Theta$. This leads us to define the parameter
$\Delta$ to check whether  the volume $V$ in the plane
$\left(O^{\prime}X^{\prime}Z^{\prime}\right) $ falls above or below
the line $Z^{\prime} = -X^{\prime}\tan\Theta$. If $\Theta =\pi
/2\Rightarrow \Delta =0,$ whereas  if $0\leq \Theta <\pi /2$
$\Rightarrow \Delta =Z_{V}^{\prime }+X_{V}^{\prime }\tan\Theta $.
Finally, let us introduce   the  ratio $\chi = x_{iV}^{\prime
}/x_{G}^{\prime }$. The following cases are then possible. If
$X_{V}^{\prime }>0 $ and $ \Delta
>0 \Rightarrow $ then $\theta _{G}^{\prime }=\pi/2+\Theta -\arcsin \left[ \chi
\sin \left( \frac{\pi }{2}-\Theta +\theta _{jV}^{^{\prime }}\right)
\right]$. If $ X_{V}^{\prime }>0 $ and $ \Delta <0 \Rightarrow $
then $\theta_{G}^{\prime}=\pi/2+\Theta +\arcsin \left[ \chi \sin
\left(\theta_{jV}^{\prime} -\pi/2 -\Theta \right) \right]$. If $
X_{V}^{\prime }>0 $ and $ \Delta =0 \Rightarrow $ then
$\theta_{G}^{\prime }=\pi/2+\Theta$. If $ X_{V}^{\prime }<0 $ and $
\Delta
>0 \Rightarrow $ then $\theta_{G}^{\prime }=\pi/2+\Theta -\arcsin \left[ \chi
\sin \left( \pi/2 -\Theta -\theta_{jV}^{^{\prime }}\right) \right]$.
If $ X_{V}^{\prime }<0 $ and $ \Delta <0 $ we have three solutions:
$\theta_{G}^{\prime }=\pi/2+\Theta +\arcsin \left[ \chi \sin \left(
\Theta-\pi/2 +\theta _{jV}^{^{\prime }}\right) \right]$,
$\theta_{G}^{\prime }=\frac{3}{2}\pi -\Theta -\arcsin \left[ \chi
\sin \left(\Theta -\pi/2+\theta _{jV}^{^{\prime }}\right) \right]$
and $\theta_{G}^{\prime }=\pi$, depending on whether
$Z_{V}^{\prime}+X_{V}^{\prime}\tan\Theta +x_{G}^{\prime}$ is
greater, smaller or equal to $0$, respectively.

Once determined the spherical coordinates of $P\left(
x_{G}^{\prime},\theta_{G,}^{\prime}\Phi_{G}^{\prime}\right) $ in the
translated system, by means of an inverse transformation of
coordinates we can obtain the Cartesian coordinates $P\left(X_{P},
Y_{P}, Z_{P}\right)$ in the old system of coordinates:
$X_{P}=x_{G}^{\prime }\sin \theta_{G}^{\prime }\cos \Phi
_{G}^{\prime}$, $Y_{P}=x_{G}^{\prime }\sin \theta_{G}^{\prime }\sin
\Phi _{G}^{\prime }+x_{iV}\sin \theta_{jV}\cos \Phi _{kV}$ and
$Z_{P}=x_{G}^{\prime }\cos \theta_{G}^{\prime}$. Having eventually
derived the Cartesian coordinates of the point $P\left(X_{P}, Y_{P},
Z_{P}\right)$ on the galactic edge and those of volume center
$\left(X_{V}, Y_{V}, Z_{V}\right)$,  the calculation of the mass in
the cylinder  comprised between $V$ and the galaxy edge is trivial
and can be straightforwardly performed as described in Appendix A.

\begin{footnotesize}
\bibliographystyle{mn2e}                          
\bibliography{mnemonic,MF1070rv}    
\end{footnotesize}

\end{document}